\documentclass[a4paper,nobind]{ociamthesis}
\usepackage{wrapfig}
\usepackage{graphicx}
\usepackage{eqnarray,amsmath}
\usepackage{color}
\usepackage{comment}
\usepackage[makeroom]{cancel}
\usepackage{cancel}
\newcommand\scalemath[2]{\scalebox{#1}{\mbox{\ensuremath{\displaystyle #2}}}}

\title{
\begin{center}
The Gr\"uneisen parameter applied to critical phenomena and experimental investigations of correlated phenomena in molecular conductors
\end{center}
}

\author{Lucas Cesar Gomes Squillante}
\college{Advisor: Prof.\,Dr.\,\emph{phil.\,nat.}\,Mariano de Souza}

\degree{}
\degreedate{Rio Claro, SP - Brazil - 2023}

\begin{document}

\selectlanguage{english}

\thispagestyle{empty}
\begin{center}
%
%

\parbox{\textwidth}{\Huge{\textbf{%
}\vspace{3.8cm}\linebreak \LARGE  The Gr\"uneisen parameter applied to critical phenomena and experimental investigations of correlated phenomena in molecular conductors}}\linebreak \vspace{5cm}


\large Lucas Cesar Gomes Squillante \linebreak
Ph.D. Advisor: Prof.\,Dr.\,Mariano de Souza\linebreak


\vspace{4.1cm}


\large{São Paulo State University - 2023}

\vspace{1.7cm}

\end{center}

\setlength{\textbaselineskip}{22pt plus2pt}

\setlength{\frontmatterbaselineskip}{17pt plus1pt minus1pt}

\setlength{\baselineskip}{\textbaselineskip}


\setcounter{secnumdepth}{2}
\setcounter{tocdepth}{2}


\begin{romanpages}




\chapter*{Acknowledgements}

\minitoc
I would like to express my profound gratitude to Prof.\,Dr.\,Mariano de Souza for the immense support during my entire scientific career and for proposing the research topics investigated in this Thesis, as well as for all his advice and guidance throughout the entire progress of this work. He supported me beyond the academic activities always encouraging me to reach the best version of myself and to flourish my humane personal side in the academic activities and in life. He always saw an inner potential in me that even I didn't know it existed. In a world full of hollowness, superficiality, and lacking true academic mentors, I feel honored and proud to have him as my advisor. For those who know me since I started my undergraduate studies, I once was a non-fully committed student. Today I feel totally different, committed with the academic path and looking forward to become a professor of Physics in a Brazilian university in a near future. In the Talmud it is written that ``Whoever saves one life, saves the world entirely'' and I feel academically saved! My profound gratitude for you, Prof.\,Dr.\,Mariano de Souza!

I cannot express my love and gratitude for my fianc\'e Mariana Alvarinho Lorizola. She always encouraged me to pursue my dreams and to work hard to achieve my goals. She was also very patient, kind, and lovely during the entire period of the developing and writing of this Thesis. She is the best life partner I would ever imagine having in my life. All the love and affection that came from our relation are somehow spread throughout all chapters of this Thesis in a lovely fashion. Arriving home everyday to find you with a bright smile and our baby cat Frida asking to be petted make me feel loved and my heart warm. I love you, thank you for believing in me! At this point, I recall Robin Willians quote in the role of John Keating in the classical movie ``Dead Poets Society'' (1989):

\begin{quote}
``...We don't read and write poetry because it's cute. We read and write poetry because we are members of the human race. And the human race is filled with passion. And medicine, law, business, engineering, these are noble pursuits and necessary to sustain life. But poetry, beauty, romance, love, these are what we stay alive for. To quote from Whitman, "O me! O life!... of the questions of these recurring; of the endless trains of the faithless... of cities filled with the foolish; what good amid these, O me, O life?" Answer: that you are here - that life exists, and identity; that the powerful play goes on and you may contribute a verse. That the powerful play goes on and you may contribute a verse. What will your verse be? ...''
\end{quote}
Thank you for contributing a verse with me in this beautiful journey, my love!

I also would like to thank you my family, my mother Lindalva Gomes and my sister Isis Gomes Marques, for always supporting me in the best way they could, even from a distance. I wish this Thesis will spark love and affection in their lives! If my mother hadn't registered me in English classes when I was a kid, I wouldn't be writing this Thesis in English. Thank you, you are very special to me and I love you both!

I acknowledge all members of the Solid State Physics group, with special thanks to my Ph.D. colleague Isys Mello for all her support throughout the years. Also, I thank everyone from the Physics department at Unesp who supported us during the Ph.D., with special thanks to Leandro Xavier Moreno and Elizabete Pereira das Neves. I also thanks Maria from \emph{Klips Papelaria \& Copiadora} who printed all the posters I have presented in conferences for always caring too much about the exceptional printing quality of the posters, contributing to the best poster awards during the Ph.D.

I thank my great friends Renan Vieira Barreto and Vinicius Cavassano Zampier for the time we have lived together. Our home was always filled with nice discussions, celebrations, and a lot of hard work. I remember that we were always working in our manuscripts and projects and it really helped me during the first years of my Ph.D. to live with other people committed with their studies. Special thanks to Renan who really teached me everything I know about training exercises and helped me to get a better shape during the pandemics. Renan always told me what I needed to hear instead of what I wanted to hear, which was very important to me.

Special thanks to one special friend André Kyoshi Fujii Ferrazo who always kindly supported me and encouraged me to do great things. Thank you for all endless philosophical talks, walks at night, and emotional support as well.

I profoundly acknowledge Carlos Eduardo Ortolani Prado de Moura and Cristina Rosa Campos for the amazing support during this Ph.D. period and for making, in particular, this time of my life much lighter and smoother. Thank you both for the enlightenments and emotional support.

I also acknowledge all professors who did not gave me an opportunity and did not believe in my potential. If it weren't by them, I wouldn't be in this lovely academic path I am today.

This study was financed in part by the Coordenação de Aperfeiçoamento de Pessoal de Nível Superior - Brasil (CAPES) - Finance Code 001.

Finally, I thank everyone who directly or indirectly supported me in this Thesis.


\begin{abstract}
	Critical phenomena are of great interest to the scientific community and can be widely extended to various fields of research, such as classical and quantum phase transitions, caloric effects, and even Biology in brain functionality, for instance. As broadly discussed in the literature, the \emph{smoking-gun} physical quantity to experimentally access both finite-temperature and quantum critical points is the so-called Gr\"uineisen ratio. In this Ph.D. Thesis, a systematic review is performed on the derivation and generalization of the Gr\"uneisen parameter followed by its unprecedented applications to several distinct scenarios, such as magnetic model systems, zero-field quantum phase transitions, the maximization of caloric effects close to \emph{any} critical-end point based on entropy arguments, the here-proposed adiabatic magnetization of a paramagnetic salt, as well as for Cosmology in the frame of the universe expansion. Since this Ph.D. Thesis is a symbiosis between theoretical and experimental results, an experimental investigation of correlated phenomena was carried out for molecular conductors of the (TMTTF)$_2$X family, where TMTTF is the base molecule tetramethyltetrathiafulvalene and X a monovalent counter-anion such as PF$_6$, SbF$_6$, or AsF$_6$. Such strongly correlated electron systems are considered suitable ones for the exploration of Mott insulating phase, charge-ordering, spin-Peierls, and superconductivity. In particular, the investigation of a possible multiferroic character in these salts was performed via quasi-static (low-frequency) dielectric constant $\varepsilon'$ measurements as a function of temperature where a maximum in $\varepsilon'$ as a function of temperature was observed at the corresponding charge-ordering temperature for both hydrogenated and 97.5\% deuterated (TMTTF)$_2$SbF$_6$ salts. Furthermore, Raman measurements were performed on the 97.5\% deuterated (TMTTF)$_2$PF$_6$, showing a possible magneto-optical effect on the $\nu_4(a_g)$ vibrational mode of the TMTTF molecule. Yet, fluorescence measurements demonstrated that the fully-hydrogenated (TMTTF)$_2$AsF$_6$ presents an expressive fluorescence background, which is roughly five orders of magnitude lower than that for the 97.5\% deuterated variant of (TMTTF)$_2$PF$_6$.
\newline\newline
Keywords: critical phenomena; Gr\"uneisen parameter; strongly correlated electron systems.
\end{abstract}\clearpage


\dominitoc 

\flushbottom

\tableofcontents




\end{romanpages}

\flushbottom
\begin{savequote}[8cm]
“The graveyard is the richest place on Earth, because it is there that you will find all the hopes and dreams that were never fulfilled, the books that were never written, the songs that were never sung, the inventions that were never shared, the cures that were never discovered, all because someone was too afraid to take that first step, keep with the problem, or determined to carry out their dream.”
  \qauthor{--- Les Brown (1945 -)}
\end{savequote}

\chapter{\label{ch:1-intro}Introduction}

\minitoc

Thermodynamics plays an unquestionable crucial role in Physics covering mathematical descriptions from heat engines \cite{ralph,callen} to black holes \cite{hawking}, being considered as one of the most relevant fields in Physics in the last centuries \cite{peterwinberg}. Upon browsing classical Thermodynamics textbooks \cite{ralph,callen}, it is easily noticed that most of its applications are based on industrial machines and this is not a mere coincidence. Indeed, Thermodynamics was the driving intellectual force behind the industrial revolution and all its applications that emerged from this process in the beginning of the nineteenth century \cite{peterwinberg}. Only in the last decades, Thermodynamics started to play an increasingly key-role in other fields, such as Biology. A few examples lies, for instance, in the description of cell compartmentalization \cite{klosin} or brain criticality in neural networks \cite{Munoz2013}. Afterwards, Thermodynamics was employed in modern technology associated with eco-friendly cooling, the so-called caloric effects \cite{mrb}, which refers to the temperature decrease in response to an adiabatic removal of a tuning parameter, which can be pressure or magnetic field, for example. In this context, the \emph{smoking-gun} to explore critical phenomena and caloric effects is the so-called Gr\"uneisen ratio $\Gamma$ \cite{zhu,EJP2016}, which is defined as the ration between thermal expansivity and the heat capacity at constant pressure. Also, such physical quantity is the appropriate experimental tool to probe quantum critical points \cite{gegenwartgruneisen}. Hence, it becomes evident that $\Gamma$ is a key quantity in Thermodynamics. This Ph.D. Thesis is based in two main chapters: a theoretical and an experimental one. At this point, it is worth mentioning that all theoretical topics covered in this Thesis, including the elastocaloric effect, adiabatic magnetization, among others, were possible to be investigated only because of the intense experimental activities and the robust experimental background of the Solid State Physics Laboratory at Unesp - Rio Claro, SP. Thus, in the frame of this Ph.D. Thesis, a theoretical background is reviewed followed by the obtained results regarding modern and cutting-edge applications of $\Gamma$.

In the last decades, molecular conductors have been recognized as an appropriate playground for the exploration of correlated phenomena in low-dimensions \cite{reviewpouget,reviewlorentz}. A class of systems of particular interest is the (TMTTF)$_2$$X$, where TMTTF is the base molecule tetramethyltetrathiafulvalene and $X$ a monovalent centrosymmetric counter-anion, such as PF$_6$, AsF$_6$, and SbF$_6$. This electronic system enables the investigation of contemporary aspects in the field of correlated phenomena, like the metal-Mott insulator transition \cite{reviewlorentz} and the charge-ordered phase \cite{reviewpouget}. It is possible to change the physical properties of such systems either by employing chemical substitution in the base molecules or by applying external hydrostatic pressure \cite{monceau,mds,dressel}. The samples of such systems are single-crystals, which enables the investigation of correlated phenomena in a \emph{clean} environment since no doping is made, i.e., absence of disorder via doping, being only the replacement of the counter-anions $X$ in the electrochemical synthesis performed. In 2018, the so-called Mott-Hubbard ferroelectric phase was investigated in the Fabre salts \cite{prbrioclaro}. A proper comparison between the physical properties of the fully-hydrogenated and the 97.5\% deuterated variants of the system (TMTTF)$_2$PF$_6$ was performed via systematic quasi-static dielectric constant measurements. It was theoretically predicted that the Mott-Hubbard ferroelectric phase would present magnetoelectric effects \cite{giovaneti}. Hence, the investigations regarding possible multiferroic/magnetoelectric effects in these systems is a hot topic nowadays, which is the main focus of the experimental part of this Ph.D. Thesis. After this brief introduction, it is worth mentioning that this Ph.D. Thesis is divided in the following way:

\begin{itemize}
  \item \textbf{Chapter 2}: this chapter is focused on the obtained theoretical results. In the beginning, a theoretical background on the detailed derivation of the effective Gr\"uneisen parameter and its relation with the Gr\"uneisen ratio is presented. After, a brief recapitulation of the fundamental aspects of the Brillouin-like paramagnet is performed, followed by the theoretical results obtained in this Ph.D. regarding zero-field quantum criticality, the relation between the canonical definition of temperature and the magnetic Grüneisen parameter, and caloric effects. Also, a discussion about the Friedmann equations and the Einstein field equations are also presented, as well as the application of the Gr\"uneisen parameter to Cosmology (submitted manuscript). In this context, the following manuscripts were published with the obtained results discussed in this chapter: Physical Review B \textbf{100}, 054446 (2019), Scientific Reports \textbf{10}, 7981 (2020), Scientific Reports \textbf{11}, 9431 (2021), Materials Research Bulletin \textbf{142}, 111413 (2021).
  \item \textbf{Chapter 3}: this chapter is dedicated to the experimental results obtained in this Ph.D. Thesis regarding the molecular conductors of the (TMTTF)$_2$X salts. A brief introduction, motivation, and state-of-the art are presented. Before discussing the results, a compilation of the technical aspects regarding the operation and use of the Teslatron-PT cryostat is included. The experimental results are then presented followed by its discussion.
  \item \textbf{Chapter 4}: here, the summary and conclusions of this Thesis are provided.

  \item \textbf{Chapter 5}: perspectives and outlook are briefly discussed.

  \item Some appendices are included in the end of the Thesis regarding the non-perfect alinement between the magnetic moment and the external magnetic field, the impossibility of connecting adiabatic deformations and the concept of negative temperatures, power down during experiments, technical aspects like the fixing of a SMB connector, participation in joint reports during the Ph.D. degree, tutorships, attendance to scientific events, prizes and awards, participation as the graduate student representative, and the full texts of the manuscripts that were published in the frame of this Thesis.
\end{itemize}

\begin{savequote}[8cm]
``Take a good rest, small bird,'' he said. \\
``Then go in and take your chance like any man or bird or fish.''
  \qauthor{--- Ernest Hemingway, The Old Man and the Sea.}
\end{savequote}

\chapter{Theoretical investigations employing the Gr\"uneisen parameter}

\minitoc

In this chapter, several distinct physical scenarios that were explored using the various definitions of the Gr\"uneisen parameter, namely effective, magnetic, electric, elastic, and the Gr\"uneisen ratio, are presented, which essentially covers magnetism at ultra low-temperatures, finite-temperature critical points, quantum critical points, caloric effects, and the dark energy problem in modern Cosmology. The obtained results for each topic of research are presented, as well as the published manuscripts that originated from these investigations. Key equations are boxed throughout the text.

\section{State of the art}

During the last four decades, there have been an enormous interest to explore the behaviour of the thermodynamical quantities in the critical regime, i.e., close to a finite-temperature critical point (CP). Particular examples include brain criticality in the frame of neural networks \cite{Munoz2013} and the onset of unconventional superconductivity \cite{putzke,huang}. Besides finite-temperature critical points, investigations on quantum critical points have been growing continuously since they can be related to superconductivity \cite{Flouquet1989,schenck} or even to quantum Griffiths phases \cite{Coldea2021}, just to mention a few key examples. In this context, the so-called Gr\"uneisen ratio $\Gamma$ \cite{EJP2016, gegenwartgruneisen}, which is given by the ratio between thermal expansivity and the specific heat, plays a fundamental role in exploring critical phenomena, since it quantifies the entropy variation in terms of temperature and the tuning parameter, which can be pressure $p$ or external magnetic field $B$, for instance. It is well known that upon approaching a finite-temperature CP, $\Gamma$ is enhanced close to the critical parameters \cite{MI1}. Over the years, $\Gamma$ has been employed not only to explore finite-temperature CP \cite{MI1}, but also magnetic-field induced quantum CP \cite{schenck} and zero-field quantum CP \cite{sakai2016,matsumoto2011,deguchi2012}. When the tuning parameter is $B$, the magnetic Gr\"uneisen parameter $\Gamma_{mag}$ is the proper quantity to probe, for instance, quantum criticality \cite{gegenwartgruneisen}. Also, $\Gamma_{mag}$ quantifies the so-called magnetocaloric effect \cite{zhu}, which is the temperature decrease due to an adiabatic removal of $B$ \cite{ralph}. Hence, this chapter focus on the obtained theoretical results regarding critical phenomena employing the various definitions of $\Gamma$.

\section{Theoretical background}\label{theoreticalbackground}

Before starting the discussions of this chapter followed by the obtained results, a theoretical background based on the discussion on Ref.\,\cite{earthinterior} regarding the effective and magnetic Gr\"uneisen parameters, the Gr\"uneisen ratio, and the well-known Brillouin-like paramagnet are provided.

\subsection{Detailed deduction of the effective Gr\"uneisen parameter}\label{deductionofgamma}

First, it is considered a system where an infinitesimal pressure and temperature variation, namely $dp$ and $dT$, respectively, can take place as a result of both an infinitesimal entropy variation $dS$ and an infinitesimal strain variation $d\varepsilon$, so that the coupled equations for the differentials can be written as \cite{earthinterior}:
\begin{equation}
dp = \left(\frac{\partial p}{\partial\varepsilon}\right)_S d\varepsilon + \left(\frac{\partial p}{\partial S}\right)_{\varepsilon} dS,
\label{dp}
\end{equation}
\begin{equation}
dT = \left(\frac{\partial T}{\partial\varepsilon}\right)_S d\varepsilon + \left(\frac{\partial T}{\partial S}\right)_{\varepsilon} dS.
\label{dt}
\end{equation}\newline
Also, an isotropic compression is considered, so that $\varepsilon = \Delta v/v_0 = (v - v_0)/v_0$, being $v$ the system's volume and $v_0$ its initial volume. Dividing both sides of Eq.\,\ref{dp} by $d\varepsilon$, it reads:
\begin{equation}
\frac{dp}{d\varepsilon} = \left(\frac{\partial p}{\partial\varepsilon}\right)_S + \left(\frac{\partial p}{\partial S}\right)_{\varepsilon} \frac{dS}{d\varepsilon},
\end{equation}
\begin{equation}
\frac{dp}{d\varepsilon} = \left(\frac{\partial p}{\partial\varepsilon}\right)_S + \left(\frac{\partial p}{\partial S}\right)_{\varepsilon} \frac{dS}{d\varepsilon}.
\label{dpdepsilon}
\end{equation}\newline
Assuming an isothermal case, i.e., $dT = 0$, Eq.\,\ref{dt} becomes:
\begin{equation}
0 = \left(\frac{\partial T}{\partial\varepsilon}\right)_S d\varepsilon + \left(\frac{\partial T}{\partial S}\right)_{\varepsilon} dS,
\end{equation}
\begin{equation}
-\left(\frac{\partial T}{\partial S}\right)_{\varepsilon} dS = \left(\frac{\partial T}{\partial\varepsilon}\right)_S d\varepsilon,
\end{equation}
\begin{equation}
-dS = \frac{\left(\frac{\partial T}{\partial\varepsilon}\right)_S d\varepsilon}{\left(\frac{\partial T}{\partial S}\right)_{\varepsilon}},
\end{equation}
\begin{equation}
\frac{dS}{d\varepsilon} = -\frac{\left(\frac{\partial T}{\partial \varepsilon}\right)_S}{\left(\frac{\partial T}{\partial S}\right)_{\varepsilon}}.
\label{dsdepsilon}
\end{equation}
Replacing Eq.\,\ref{dsdepsilon} in Eq.\,\ref{dpdepsilon}, it reads:
\begin{equation}
\frac{dp}{d\varepsilon} = \left(\frac{\partial p}{\partial\varepsilon}\right)_S + \left(\frac{\partial p}{\partial S}\right)_{\varepsilon}\left[-\frac{\left(\frac{\partial T}{\partial\varepsilon}\right)_S}{\left(\frac{\partial T}{\partial S}\right)_{\varepsilon}}\right],
\end{equation}
\begin{equation}
\frac{dp}{d\varepsilon} = \left(\frac{\partial p}{\partial\varepsilon}\right)_S - \left(\frac{\partial p}{\partial S}\right)_{\varepsilon}\left(\frac{\partial S}{\partial T}\right)_{\varepsilon}\left(\frac{\partial T}{\partial\varepsilon}\right)_S.
\label{differentialrelation0}
\end{equation}
Considering the generalized differential relation for the state functions given by \cite{Stanleybook}:
\begin{equation}
\left(\frac{\partial x}{\partial z}\right)_g = \left(\frac{\partial x}{\partial y}\right)_g\left(\frac{\partial y}{\partial z}\right)_g.
\label{differentialrelation}
\end{equation}
Assuming $x = p$, $y = S$, $z = T$, and $g = \varepsilon$ in Eq.\,\ref{differentialrelation}, it reads:
\begin{equation}
\left(\frac{\partial p}{\partial T}\right)_{\varepsilon} = \left(\frac{\partial p}{\partial S}\right)_{\varepsilon}\left(\frac{\partial S}{\partial T}\right)_{\varepsilon}.
\label{differentialrelation2}
\end{equation}
Replacing Eq.\,\ref{differentialrelation2} into Eq.\,\ref{differentialrelation0}:
\begin{equation}
\frac{dp}{d\varepsilon} = \left(\frac{\partial p}{\partial\varepsilon}\right)_S - \left(\frac{\partial p}{\partial T}\right)_{\varepsilon}\left(\frac{\partial T}{\partial\varepsilon}\right)_S.
\label{differentialrelation3}
\end{equation}
Assuming that $dp = (\partial p/\partial\varepsilon)_{T}d\varepsilon$, Eq.\ref{differentialrelation3} becomes:
\begin{equation}
\left(\frac{\partial p}{\partial\varepsilon}\right)_T = \left(\frac{\partial p}{\partial\varepsilon}\right)_S - \left(\frac{\partial p}{\partial T}\right)_{\varepsilon}\left(\frac{\partial T}{\partial\varepsilon}\right)_S,
\end{equation}
\begin{equation}
\left(\frac{\partial p}{\partial\varepsilon}\right)_{T} = \left(\frac{\partial p}{\partial\varepsilon}\right)_S - \left(\frac{\partial p}{\partial T}\right)_{\varepsilon}\left(\frac{\partial T}{\partial\varepsilon}\right)_S.
\label{differentialrelation4}
\end{equation}
The term $(\partial p/\partial\varepsilon)_T$ in Eq.\,\ref{differentialrelation4} can be rewritten based on the definition of $\varepsilon = \Delta v/v_0 = (v - v_0)/v_0$, so that:
\begin{equation}
\left(\frac{\partial p}{\partial\varepsilon}\right)_T = \left[\frac{\partial p}{\partial(\Delta v/v_0)}\right]_T = \left[\frac{\partial p}{\frac{1}{v_0}\partial(\Delta v)}\right]_T = v_0\left(\frac{\partial p}{\partial\Delta v}\right)_T = v_0\left[\frac{\partial p}{\partial(v-v_0)}\right]_T.
\label{bulk}
\end{equation}
Considering the case of a compression, $v$ is always lower than $v_0$ so that $\partial[v-v_0] = -\partial v$ and Eq.\,\ref{bulk} reads:
\begin{equation}
\left(\frac{\partial p}{\partial\varepsilon}\right)_T = v_0\left(\frac{\partial p}{-\partial v}\right)_T = -v_0\left(\frac{\partial p}{\partial v}\right)_T.
\label{bulk2}
\end{equation}
The right side of Eq.\,\ref{bulk2} is the definition of the isothermal bulk modulus $\mathcal{B}_T$ \cite{Stanleybook}, so that:
\begin{equation}
\left(\frac{\partial p}{\partial\varepsilon}\right)_T = -v_0\left(\frac{\partial p}{\partial v}\right)_T = \mathcal{B}_T.
\label{isothermal}
\end{equation}
The same analysis can be made in terms of $(\partial p/\partial\varepsilon)_S$ in Eq.\,\ref{differentialrelation4}, so that:
\begin{equation}
\left(\frac{\partial p}{\partial\varepsilon}\right)_S = \mathcal{B}_S,
\label{adiabatic}
\end{equation}
where $\mathcal{B}_S$ is the adiabatic bulk modulus \cite{Stanleybook}. The term $(\partial T/\partial\varepsilon)_S$ in Eq.\,\ref{differentialrelation4} can also be rewritten in terms of $\varepsilon = \Delta v/v_0 = (v - v_0)/v_0$, so that:
\begin{equation}
\left(\frac{\partial T}{\partial\varepsilon}\right)_S = \left[\frac{\partial T}{\partial(\Delta v/v_0)}\right]_S = \left[\frac{\partial T}{\frac{1}{v_0}\partial(\Delta v)}\right]_S = v_0\left[\frac{\partial T}{\partial(\Delta v)}\right]_S = v_0\left[\frac{\partial T}{\partial(v-v_0)}\right]_S,
\end{equation}
so that:
\begin{equation}
\left(\frac{\partial T}{\partial\varepsilon}\right)_S = v_0\left(\frac{\partial T}{-\partial v}\right)_S = -v_0\left(\frac{\partial T}{\partial v}\right)_S.
\label{(G)}
\end{equation}
Now, replacing Eqs.\,\ref{isothermal}, \ref{adiabatic}, and \ref{(G)} into Eq.\,\ref{differentialrelation4}, it reads:
\begin{equation}
\mathcal{B}_T = \mathcal{B}_S - \left(\frac{\partial p}{\partial T}\right)_{\varepsilon}\left[-v_0\left(\frac{\partial T}{\partial v}\right)_S\right],
\end{equation}
\begin{equation}
(\mathcal{B}_T - \mathcal{B}_S) = \left(\frac{\partial p}{\partial T}\right)_{\varepsilon}\left(\frac{\partial T}{\partial v}\right)_S v_0.
\label{(H)}
\end{equation}
Since the condition of a constant $v$ strictly implies in a constant $\varepsilon$ as well:
\begin{equation}
\left(\frac{\partial p}{\partial T}\right)_{\varepsilon} = \left(\frac{\partial p}{\partial T}\right)_v.
\end{equation}
The term $(\partial p/\partial T)_v$ can be rewritten in terms of the well-known relation \cite{Stanleybook}:
\begin{equation}
\left(\frac{\partial v}{\partial p}\right)_T \left(\frac{\partial p}{\partial T}\right)_v \left(\frac{\partial T}{\partial v}\right)_p = -1,
\end{equation}
so that:
\begin{equation}
\left(\frac{\partial p}{\partial T}\right)_v = -\frac{1}{\left(\frac{\partial v}{\partial p}\right)_T \left(\frac{\partial T}{\partial v}\right)_p}.
\label{(I)}
\end{equation}
The term $(\partial T/\partial v)_p$ is related to the definition of the thermal expansion coefficient at constant pressure $\alpha_p$ by $(\partial v/\partial T)_p = \alpha_p v_0$ \cite{Stanleybook}, so that $(\partial T/\partial v)_p = 1/(\alpha_p v_0)$ and Eq.\,\ref{(I)} reads:
\begin{equation}
\left(\frac{\partial p}{\partial T}\right)_v = -\frac{1}{\left(\frac{\partial v}{\partial p}\right)_T \frac{1}{\alpha_p v_0}} = -\frac{\alpha_p v_0}{\left(\frac{\partial v}{\partial p}\right)_T}.
\label{(J)}
\end{equation}
Following the definition of $\mathcal{B}_T$, the term in Eq.\,\ref{(J)} $(\partial v/\partial p)_T = -{v_0}/\mathcal{B}_T$ and thus Eq.\,\ref{(J)} becomes:
\begin{equation}
\left(\frac{\partial p}{\partial T}\right)_v = -\frac{\alpha_p \cancel{v_0}}{-\frac{\cancel{v_0}}{\mathcal{B}_T}} = \alpha_p \mathcal{B}_T.
\label{alphabeta}
\end{equation}
Now, the term $(\partial T/\partial v)_S$ in Eq.\,\ref{(H)} can be rewritten employing the relation of Eq.\,\ref{differentialrelation} considering $x = T$, $y = p$, $z = v$, and $g = S$, so that:
\begin{equation}
\left(\frac{\partial T}{\partial v}\right)_S = \left(\frac{\partial T}{\partial p}\right)_S \left(\frac{\partial p}{\partial v}\right)_S.
\label{(K)}
\end{equation}
The term $(\partial T/\partial p)_S$ can be rewritten following the relation \cite{Stanleybook}:
\begin{equation}
\left(\frac{\partial T}{\partial p}\right)_S \left(\frac{\partial S}{\partial T}\right)_p \left(\frac{\partial p}{\partial S}\right)_T = -1,
\label{generalization2}
\end{equation}
\begin{equation}
\left(\frac{\partial T}{\partial p}\right)_S =  -\frac{1}{\left(\frac{\partial S}{\partial T}\right)_p \left(\frac{\partial p}{\partial S}\right)_T}.
\label{(L)}
\end{equation}
Plugging Eq.\,\ref{(L)} into Eq.\,\ref{(K)}:
\begin{equation}
\left(\frac{\partial T}{\partial v}\right)_S = -\frac{1}{\left(\frac{\partial S}{\partial T}\right)_p\left(\frac{\partial p}{\partial S}\right)_T}\left(\frac{\partial p}{\partial v}\right)_S = -\frac{\left(\frac{\partial S}{\partial p}\right)_T}{\left(\frac{\partial S}{\partial T}\right)_p}\left(\frac{\partial p}{\partial v}\right)_S.
\label{(M)}
\end{equation}
Given the fact that $\alpha_p = (1/v_0)(\partial v/\partial T)_p = (1/v_0)(-\partial S/\partial p)_T$ and the heat capacity at constant pressure $c_p = T(\partial S/\partial T)_p$ \cite{Stanleybook}, Eq.\,\ref{(M)} becomes:
\begin{equation}
\left(\frac{\partial T}{\partial v}\right)_S = \frac{\alpha_p v_0 T}{c_p}\left(\frac{\partial p}{\partial v}\right)_S.
\label{(N)}
\end{equation}
Given the definition of $\mathcal{B}_S = -v_0(\partial p/\partial v)_S$ \cite{Stanleybook}, the term in Eq.\,\ref{(N)} $(\partial p/\partial v)_S = -\mathcal{B}_S/v_0$, so that Eq.\,\ref{(N)} reads:
\begin{equation}
\left(\frac{\partial T}{\partial v}\right)_S = \frac{\alpha_p \cancel{v_0} T}{c_p}\left(-\frac{\mathcal{B}_S}{\cancel{v_0}}\right) = -\frac{\alpha_p T \mathcal{B}_S}{c_p}.
\label{(O)}
\end{equation}
Finally, inserting Eqs.\,\ref{alphabeta} and \ref{(O)} in Eq.\,\ref{(H)}, it reads:
\begin{equation}
(\mathcal{B}_T - \mathcal{B}_S) = \alpha_p\mathcal{B}_T\left(-\frac{\alpha_p\mathcal{B}_S T}{c_p}\right) v_0 = -\alpha_p\mathcal{B}_TT\left(\frac{\alpha_p\mathcal{B}_Sv_0}{c_p}\right).
\label{(P)}
\end{equation}
Multiplying both sides of Eq.\,\ref{(P)} by $-1$, it reads:
\begin{equation}
(\mathcal{B}_S - \mathcal{B}_T) = \alpha_p \mathcal{B}_T T \left(\frac{\alpha_p\mathcal{B}_Sv_0}{c_p}\right),
\end{equation}
\begin{equation}
\mathcal{B}_S = \mathcal{B}_T + \alpha_p \mathcal{B}_T T \left(\frac{\alpha_p\mathcal{B}_Sv_0}{c_p}\right) = \mathcal{B}_T\left[1 + \alpha_pT\left(\frac{\alpha_p\mathcal{B}_Sv_0}{c_p}\right)\right],
\end{equation}
\begin{equation}
\frac{\mathcal{B}_S}{\mathcal{B}_T} = 1 + \alpha_p T \left(\frac{\alpha_p\mathcal{B}_Sv_0}{c_p}\right).
\label{(Q)}
\end{equation}
Employing the relation for the partial derivatives in Eq.\,\ref{differentialrelation} assuming $x = p$, $y = T$, $z = S$, and $g = v$, it reads:
\begin{equation}
\left(\frac{\partial p}{\partial S}\right)_v = \left(\frac{\partial p}{\partial T}\right)_v \left(\frac{\partial T}{\partial S}\right)_v.
\label{(R)}
\end{equation}
Employing the Maxwell relation $(\partial p/\partial S)_v = -(\partial T/\partial v)_S$ \cite{Stanleybook} and the term $(\partial T/\partial v)_S$ computed in Eq.\,\ref{(O)}, so that, $(\partial p/\partial S)_v = \alpha_p\mathcal{B}_ST/c_p$. The latter can be replaced in Eq.\,\ref{(R)} together with Eq.\,\ref{alphabeta} and the definition of the heat capacity at constant volume $c_v = T(\partial S/\partial T)_v$ \cite{Stanleybook}, i.e., $(\partial T/\partial S)_v = T/c_v$, so that Eq.\,\ref{(R)} now reads:
\begin{equation}
\frac{\cancel{\alpha_p} \mathcal{B}_S \cancel{T}}{c_p} = \cancel{\alpha_p} \mathcal{B}_T \frac{\cancel{T}}{c_v},
\end{equation}
\begin{equation}
\frac{\mathcal{B}_S}{c_p} = \frac{\mathcal{B}_T}{c_v},
\end{equation}
which can be rewritten simply as:
\begin{equation}
\frac{\mathcal{B}_S}{\mathcal{B}_T} = \frac{c_p}{c_v}.
\label{(S)}
\end{equation}
Replacing Eq.\,\ref{(S)} into Eq.\,\ref{(Q)}, it reads:
\begin{equation}
\frac{c_p}{c_v} = 1 + \alpha_p T \left(\frac{\alpha_pv_0}{c_p}\mathcal{B}_S\right) = 1 + \alpha_p T \left(\frac{\alpha_pv_0}{\cancel{c_p}}\frac{\cancel{c_p} \mathcal{B}_T}{c_v}\right) =  1 + \alpha_p T\overbrace{\left(\frac{\alpha_p v_0 \mathcal{B}_T}{c_v}\right)}^{\Gamma_{eff}}.
\label{(T)}
\end{equation}
The dimensionless physical quantity between parenthesis in Eq.\,\ref{(T)} is the effective Gr\"uneisen parameter \cite{gruneisen,EJP2016}:
\begin{equation}
\boxed{\Gamma_{eff} = \frac{\alpha_p v_0 \mathcal{B}_T}{c_v}}.
\label{gammaeffective}
\end{equation}\vspace{0.3cm}
Note that $\Gamma_{eff}$ is also related to the specific heat ratio, namely $\Gamma_{eff} = (c_p/c_v) - 1$ \cite{vdw}. It is worth mentioning that $\Gamma_{eff}$ can be rewritten in terms of the internal energy $U$. Employing the definition of $\Gamma_{eff}$ in Eq.\,\ref{gammaeffective} and, considering $\alpha_p \mathcal{B}_T = (\partial p/\partial T)_v$ \cite{earthinterior} and $c_v = (\partial U/\partial T)_v$, we have:\vspace{0.3cm}
\begin{equation}
\Gamma_{eff} = v_0\frac{\left(\frac{\partial p}{\partial T}\right)_v}{\left(\frac{\partial U}{\partial T}\right)_v} = v_0\left(\frac{\partial p}{\partial T}\right)_v \left(\frac{\partial U}{\partial T}\right)_v^{-1}.
\label{gammaeff2}
\end{equation}\vspace{0.3cm}
Employing the relation from Eq.\,\ref{differentialrelation}, one can write:\vspace{0.3cm}
\begin{equation}
\left(\frac{\partial p}{\partial U}\right)_v = \left(\frac{\partial p}{\partial T}\right)_v \left(\frac{\partial T}{\partial U}\right)_v = \left(\frac{\partial p}{\partial T}\right)_v {\left(\frac{\partial U}{\partial T}\right)_v}^{-1}.
\label{gammaeff3}
\end{equation}\vspace{0.3cm}
Replacing Eq.\,\ref{gammaeff3} into Eq.\,\ref{gammaeff2}, it reads \cite{EJP2016,wallace}:\vspace{0.3cm}
\begin{equation}
\boxed{\Gamma_{eff} = v_0\left(\frac{\partial p}{\partial U}\right)_v}.
\label{dpdegru}
\end{equation}
Strictly speaking, $\Gamma_{eff}$ must be computed upon taking into account all the entropy contributions to the system, namely phononic $S_{ph}$ and electronic $S_{el}$, so that $S_{tot} = S_{ph} + S_{el}$. Rewriting $\Gamma_{eff}$ upon considering that $(\partial S/\partial \ln{v})_T = v_0\alpha_p\mathcal{B}_T$ and $(\partial S/\partial \ln{T})_v = c_v$, $\Gamma_{eff}$ reads \cite{wallace}:
\begin{equation}
\Gamma_{eff} = \frac{\left(\frac{\partial S}{\partial \ln{v}}\right)_T}{\left(\frac{\partial S}{\partial \ln{T}}\right)_v}.
\label{gammaeff4}
\end{equation}\vspace{0.3cm}
Replacing the expression for $S_{tot}$ in Eq.\,\ref{gammaeff4}, we have:
\begin{equation}
\Gamma_{eff} = \frac{\left[\frac{\partial (S_{ph} + S_{el})}{\partial \ln{v}}\right]_T}{\left[\frac{\partial (S_{ph} + S_{el})}{\partial \ln{T}}\right]_v},
\end{equation}
\begin{equation}
\Gamma_{eff} = \frac{\left(\frac{\partial S_{ph}}{\partial\ln{v}}\right)_T + \left(\frac{\partial S_{el}}{\partial\ln{v}}\right)_T}{\left(\frac{\partial S_{ph}}{\partial \ln{T}}\right)_v + \left(\frac{\partial S_{el}}{\partial \ln{T}}\right)_v}.
\label{gammaeff5}
\end{equation}\vspace{0.3cm}
Employing Eq.\,\ref{gammaeff4} into Eq.\,\ref{gammaeff5}, it reads:
\begin{equation}
\Gamma_{eff} = \frac{\Gamma_{eff}^{ph}c_v^{ph} + \Gamma_{eff}^{el}c_v^{el}}{c_v^{ph} + c_v^{el}},
\label{gammaeff6}
\end{equation}
where the phononic and electronic contributions are taken into account both in $\Gamma_{eff}$ and in $c_v$. Therefore, Eq.\,\ref{gammaeff6} can be written in its final form \cite{wallace}:
\begin{equation}
\boxed{\Gamma_{eff} = \frac{\sum_i \Gamma_{eff}^{i}c_{v}^{i}}{\sum_i c_{v}^{i}}}.
\label{gammaeff7}
\end{equation}
Equation\,\ref{gammaeff7} shows that although $S$ is an additive quantity in a system, $\Gamma_{eff}$ is not. After almost 100 years that Eduard Gr\"uneisen (1877-1949) proposed the relation shown in Eq.\,\ref{gammaeffective}, it was proposed in the literature that a singular portion of $\Gamma_{eff}$ can be employed to explore the vicinities of a pressure-induced quantum critical point \cite{zhu}. Equation\,\ref{gammaeffective} can be rewritten as \cite{zhu}:
\begin{equation}
\Gamma_{eff} = \frac{\alpha_p}{c_p} \frac{c_p}{c_v} v_0\mathcal{B}_T = \Gamma \frac{c_p}{c_v} v_0\mathcal{B}_T,
\end{equation}
where $\Gamma$ is the so-called Gr\"uneisen ratio \cite{zhu}. Since $\alpha_p$ is more singular than $c_p$ close to a pressure-induced quantum critical point \cite{zhu}, $\Gamma$ is the sensitive quantity that presents a divergent-like behaviour in the vicinity of the quantum critical point. The reason $\alpha_p$ is more pronounced than $c_p$ close to a QCP, or a finite-$T$ critical end point, is that usually $T$ is fixed and the tuning parameter is varied, making the $S$ variation more pronounced upon varying the tuning parameter. Another way of writing $\Gamma$ is employing the definition of $\alpha_p$ and the molar heat capacity at constant pressure $c_p = T/N(\partial S/\partial T)_p$ \cite{Stanleybook}, where $N$ is the number of particles in the system, so that:
\begin{equation}
\Gamma = \frac{\alpha_p}{c_p} = \frac{1}{v_0}\left(\frac{\partial v}{\partial T}\right)_p \left[\frac{1}{\frac{T}{N}\left(\frac{\partial S}{\partial T}\right)_p}\right] = \frac{1}{v_m T}\frac{\left(\frac{\partial v}{\partial T}\right)_p}{\left(\frac{\partial S}{\partial T}\right)_p},
\end{equation}
where $v_m = v/N$ is the molar volume. Employing the Maxwell relation $(\partial v/\partial T)_p = -(\partial S/\partial p)_T$ \cite{Stanleybook}, $\Gamma$ reads \cite{zhu}:
\begin{equation}
\Gamma = -\frac{1}{v_m T}\frac{\left(\frac{\partial S}{\partial p}\right)_T}{\left(\frac{\partial S}{\partial T}\right)_p}.
\label{gamma2}
\end{equation}
Using the relation \cite{Stanleybook}:
\begin{equation}
\left(\frac{\partial S}{\partial T}\right)_p \left(\frac{\partial T}{\partial p}\right)_S \left(\frac{\partial p}{\partial S}\right)_T = -1,
\end{equation}
\begin{equation}
\left(\frac{\partial T}{\partial p}\right)_S = -\frac{1}{\left(\frac{\partial S}{\partial T}\right)_p \left(\frac{\partial p}{\partial S}\right)_T} = -\frac{\left(\frac{\partial S}{\partial p}\right)_T}{\left(\frac{\partial S}{\partial T}\right)_p}.
\label{(U)}
\end{equation}
Replacing Eq.\,\ref{(U)} in Eq.\,\ref{gamma2}, $\Gamma$ can be written as:\vspace{0.3cm}
\begin{equation}
\boxed{\Gamma = \frac{1}{v_m T}\left(\frac{\partial T}{\partial p}\right)_S}.
\label{gamma3}
\end{equation}
Equation\,\ref{gamma3} shows that $\Gamma$ quantifies the temperature change upon adiabatically varying $p$, i.e., the so-called barocaloric effect. A couple of years after the proposal that $\Gamma$ is the singular contribution of $\Gamma_{eff}$ in the vicinities of quantum critical points \cite{zhu}, it was also proposed by Prof. Mariano and collaborators that $\Gamma$ can be also employed to explore finite-$T$ critical points as well \cite{MI1}. The authors of Ref.\,\cite{zhu} also proposed that $\Gamma$ can be rewritten for a magnetic-field-induced quantum critical point, so that \cite{zhu}:\vspace{0.3cm}
\begin{equation}
\boxed{\Gamma_{mag} = -\frac{\left(\frac{\partial M}{\partial T}\right)_B}{c_B} = -\frac{1}{T}\frac{\left(\frac{\partial S}{\partial B}\right)_T}{\left(\frac{\partial S}{\partial T}\right)_B}},
\label{mgp}
\end{equation}
where $M$ is the modulus of the magnetization, $c_B$ the heat capacity at constant magnetic field, $\Gamma_{mag}$ the magnetic Gr\"uneisen parameter, and $B$ the modulus of the external magnetic field. Employing the relation \cite{Stanleybook}:\vspace{0.3cm}
\begin{equation}
\left(\frac{\partial S}{\partial T}\right)_B \left(\frac{\partial T}{\partial B}\right)_S \left(\frac{\partial B}{\partial S}\right)_T = -1,
\end{equation}
\begin{equation}
-\left(\frac{\partial T}{\partial B}\right)_S = \frac{1}{\left(\frac{\partial S}{\partial T}\right)_B \left(\frac{\partial B}{\partial S}\right)_T} = \frac{\left(\frac{\partial S}{\partial B}\right)_T}{\left(\frac{\partial S}{\partial T}\right)_B}.
\label{(V)}
\end{equation}\newline
Replacing Eq.\,\ref{(V)} in Eq.\,\ref{mgp}, it reads:\vspace{0.3cm}
\begin{equation}
\Gamma_{mag} = -\frac{1}{T}\left[-\left(\frac{\partial T}{\partial B}\right)_S\right],
\end{equation}\vspace{0.3cm}
\begin{equation}
\boxed{\Gamma_{mag} = \frac{1}{T}\left(\frac{\partial T}{\partial B}\right)_S}.
\label{mgp2}
\end{equation}
Equation\,\ref{mgp2} shows that $\Gamma_{mag}$ quantifies the so-called magnetocaloric effect \cite{ralph}, i.e., the temperature decrease due to an adiabatic removal of the external magnetic field at a starting temperature $T$, which is accounted by the pre-factor $1/T$. Hence, it becomes evident that $\Gamma$ and $\Gamma_{mag}$ are key-parameters in probing both classical and quantum criticality, as well as in the quantification of caloric effects.

\subsection{The Brillouin-like paramagnet}

\begin{figure}[h!]
\centering
\includegraphics[width=0.8\textwidth]{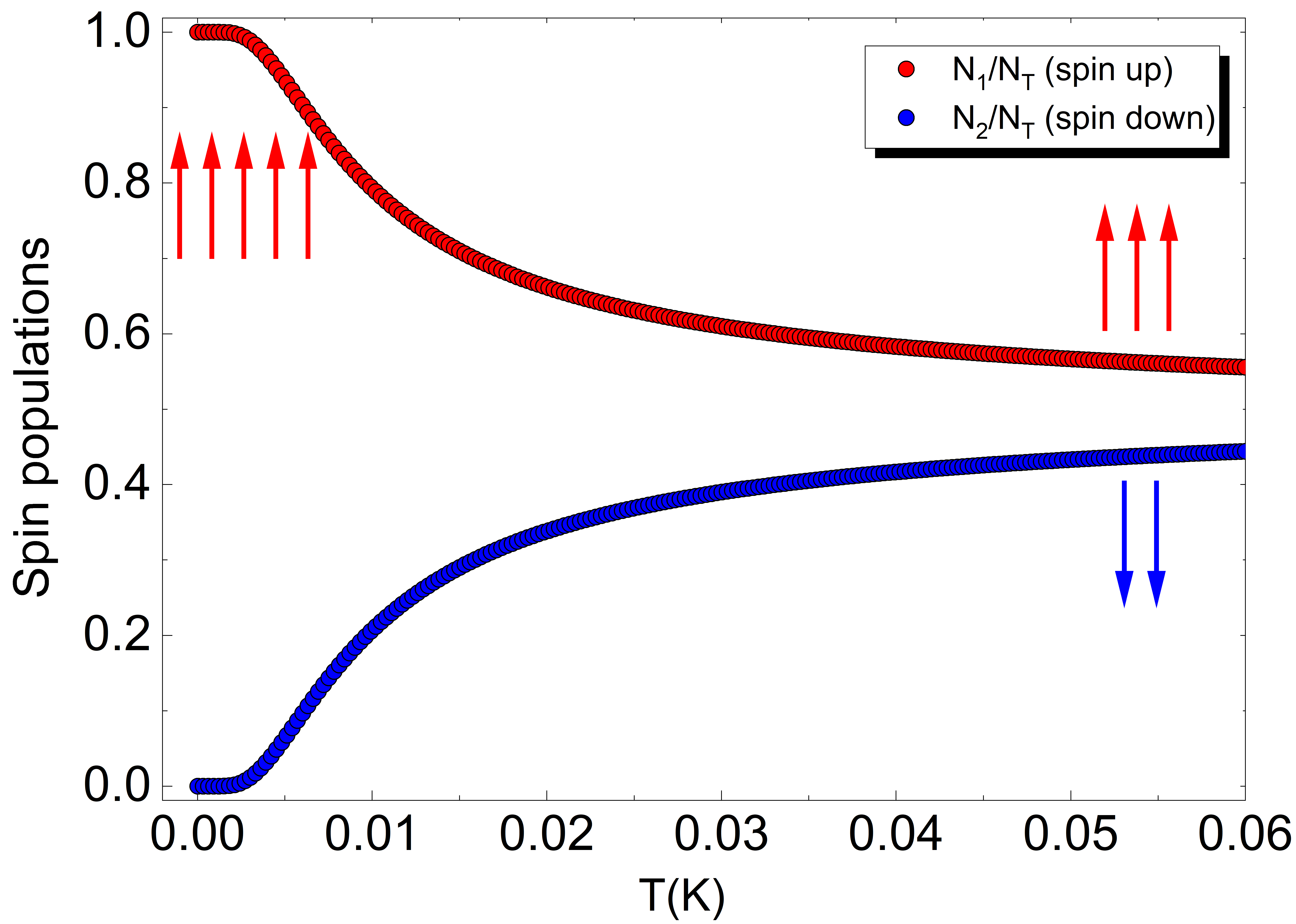}
\caption{\footnotesize Spin populations $N_1/N_T$ (up) and $N_2/N_T$ (down) represented by the red and blue color data, respectively. Note that when $T \rightarrow 0$\,K, $N_1/N_T \rightarrow 1$ and $N_2/N_T \rightarrow 0$, which means that the ground-state of the Brillouin-type paramagnet is a ferromagnetic phase. In the regime of high temperatures, i.e., $T \rightarrow \infty$, both $N_1/N_T$ and $N_2/N_T$ go to 0.5, which means that both configurations, namely spin up and down configurations are equally probable [$S = k_B\ln{(2)}]$. The arrows illustrate the spin populations in two distinct temperatures.}
\label{spinpopulations}
\end{figure}

Before discussing the results, for the sake of completeness, the fundamental physical aspects of the Brillouin-type paramagnet are briefly recalled, since those are connected with some of the obtained theoretical results obtained. Just to mention, the modulus of many vectorial physical quantities discussed in this Ph.D. Thesis, such as the external magnetic field, magnetization, and related quantities, are considered. The well-known from textbooks Brillouin-type paramagnet is a non-interacting model that describes the behaviour of a paramagnetic insulating system \cite{Blundell}. The spins are localized and the system has only two eigenstates, i.e., there is a probability that the magnetic moment is either aligned in a parallel or an anti-parallel configuration with respect to $B$, being hereafter defined as $P_{par}$ and $P_{anti-par}$, respectively. Such probabilities are given by \cite{ralph}:
\begin{equation}
P_{par} (T,B) = \frac{e^{\mu_B B/k_B T}}{2\cosh{(\mu_B B/k_B T)}},
\label{par}
\end{equation}
\begin{equation}
P_{anti-par} (T,B) = \frac{e^{-\mu_B B/k_B T}}{2\cosh{(\mu_B B/k_B T)}},
\label{anti-par}
\end{equation}\newline
where $k_B$ is Boltzmann constant (1.38$\times$10$^{-23}$)\,J/K and $\mu_B$ the Bohr magneton (9.27$\times$10$^{-24}$\,J/T). Recalling that the entropy $S$ can be computed employing \cite{ralph}:
\begin{equation}
S = -k_B \sum_j P(\psi_j)\ln{(P(\psi_j))},
\label{canonicalentropy}
\end{equation}
where $\psi_j$ represents the $j^{th}$ eigenstates. Since for the case of an insulating paramagnet there is only the probabilities $P_{par}$ and $P_{anti-par}$, it is then written:
\begin{equation}
S = -k_B[P_{par}\ln{(P_{par})} + P_{anti-par}\ln{(P_{anti-par})}].
\label{canonicalentropy2}
\end{equation}
Inserting Eqs.\,\ref{par} and \ref{anti-par} into Eq.\,\ref{canonicalentropy2}, it is straightforward to write \cite{ralph}:
\begin{equation}
S(T, B) = -\frac{B \mu_B}{T}\tanh{\left(\frac{B \mu_B}{k_B T}\right)} + k_B\textmd{ln}\left[2\cosh{\left(\frac{B \mu_B}{k_B T}\right)}\right],
\label{entropyBrillouin}
\end{equation}
Upon analysing the high-temperature limit for Eq.\,\ref{entropyBrillouin} at a fixed $B$, it is obtained $S (T\rightarrow\infty, B = \textmd{constant}) = k_B\ln{(2)}$, which indicates that the two eigenstates are equally probable. In the opposite side, at a fixed $B$, when $T \rightarrow 0$ $\Rightarrow$ $S \rightarrow 0$, as expected by the third law of Thermodynamics \cite{ralph}. The latter indicates that the ground-state of a Brillouin-type paramagnet is a ferromagnetic phase. This can be analysed in terms of the two spin populations $N_1$ and $N_2$, which refer, respectively, to the number of spins aligned in parallel or anti-parallel with respect to $B$. The expressions for $N_1$ and $N_2$ with respect to the total spin population $N_T = (N_1 + N_2)$ are well-known in textbooks \cite{kittel}, being given by:
\begin{equation}
\frac{N_1}{N_T} = \frac{\exp(\mu_B B/k_B T)}{\exp(\mu_B B/k_B T)\,+\,\exp(-\mu_B B/k_B T)}\label{popu1}
\end{equation}
and
\begin{equation}
\frac{N_2}{N_T} = \frac{\exp(-\mu_B B/k_B T)}{\exp(\mu_B B/k_B T)\,+\,\exp(-\mu_B B/k_B T)}.\label{popu2}
\end{equation}
Figure\,\ref{spinpopulations} depicts that, on one hand, when $T \rightarrow 0$\,K, all magnetic moments are in a parallel alignment with respect to $B$. On the other hand, at high temperatures, i.e., $T \rightarrow \infty$, the populations $N_1/N_T$ and $N_2/N_T$ are equally probable.\newline
Also, the heat capacity at constant magnetic field $c_B$ can be computed employing the expression $c_B = T(\partial S/\partial T)_B$ and Eq.\,\ref{entropyBrillouin}, so that:
\begin{equation}
c_B\,(T,B) = T\left(\frac{\partial S}{\partial T}\right)_B = \frac{B^2 {\mu_B}^2}{k_B T^2}{\textmd{sech}^2\left(\frac{\mu_B B}{k_B T}\right)}.
\end{equation}
\begin{figure}[h!]
\centering
\includegraphics[width=0.8\textwidth]{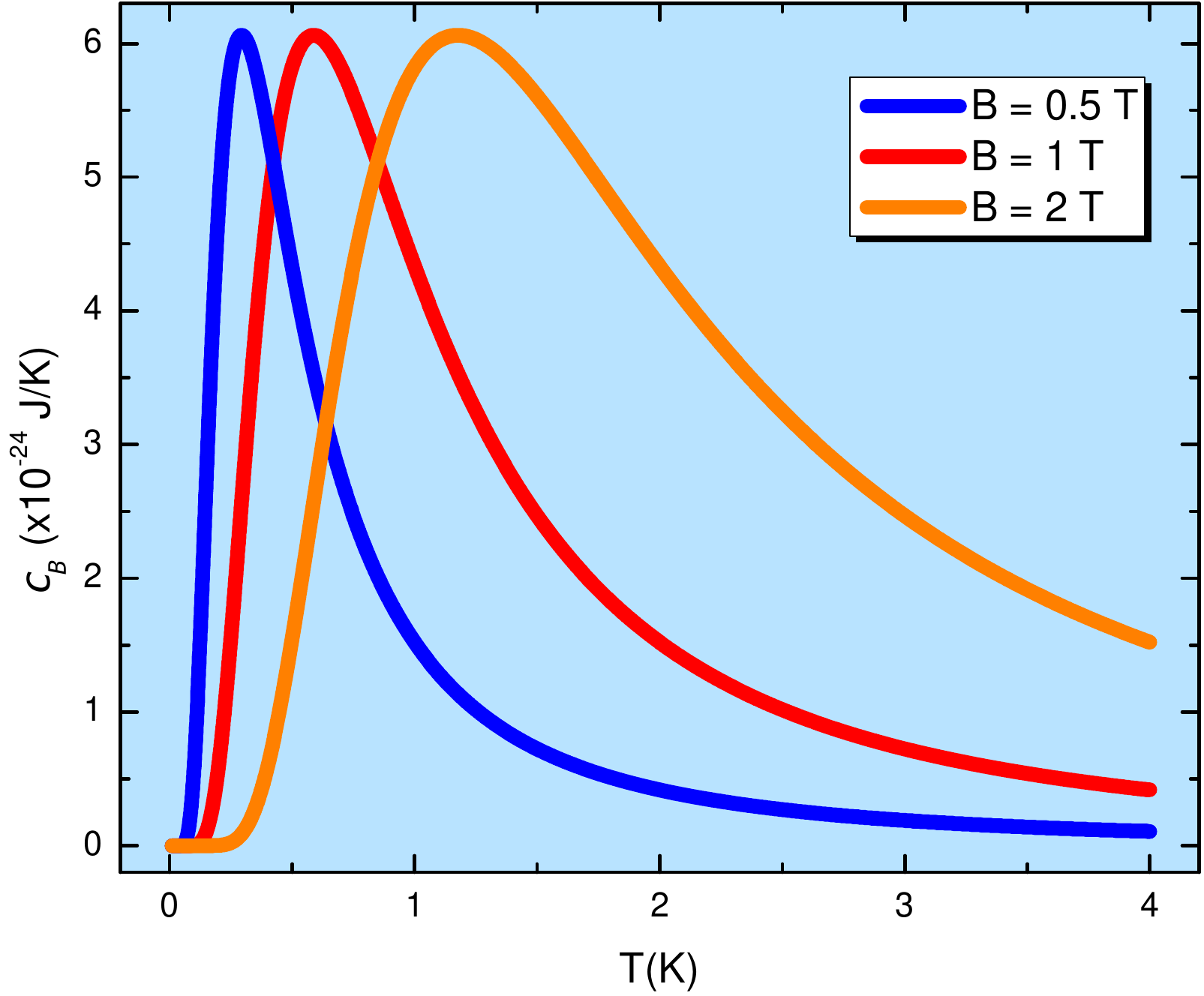}
\caption{\footnotesize Heat capacity at constant magnetic field $c_B$ as a function of $T$ for $B = 0.5$\,T (blue line), $B = 1$\,T (red line), and $B = 2$\,T (orange color line). The maximum on $c_B$ is associated with the Schottky anomaly for two-level systems \cite{pathria,tari}. More details in the main text.}
\label{schottky}
\end{figure}
The maximum in $c_B$ shown in Fig.\,\ref{schottky} for various fixed values of $B$ is associated with the famous Schottky anomaly \cite{pathria,tari}. Such an anomaly is a fingerprint of two-level systems and it shows that, upon lowering $T$, there is a particular condition in which entropy is dramatically varied as a consequence of the ordering of the system when $T \rightarrow 0$\,K. Upon approaching such a condition, it becomes energetically more favorable for the system to order in a long-range fashion and thus the population $N_1/N_T$ is favored, which in turn dramatically affects the available eigenstates of the system and, as a consequence, the entropy. As well-known from textbooks, such a maximum in $c_B$ takes place when ($k_B T$)/($\mu_B B$) $\simeq$ 0.834 \cite{pathria,tari,multilevel}, i.e., when the thermal energy matches exactly the value of the energy between the two levels $\Delta$. Hence, as depicted in Fig.\,\ref{schottky}, upon increasing the fixed value of $B$, the Schottky anomaly will take place at higher temperatures. This happens because when a higher $B$ is applied, $\Delta$ is increased so that the condition for the Schottky anomaly to take place is thus reached at higher temperatures. In other words, given that ($k_B T$)/($\mu_B B$) $\simeq$ 0.834 \cite{pathria,tari}, if the magnetic energy increases then the thermal energy has to be increased as well in order to satisfy such a condition, which in turn makes the Schottky anomaly to be reached at higher temperatures upon increasing $B$. At very high temperatures, all curves shown in Fig.\,\ref{schottky} converge to the same value of $c_B$, since $S$ does not vary much with increasing $T$, namely $S \rightarrow k_B\ln{(2)}$. Regarding $S$ for the Brillouin-type paramagnet, for $B = 0$\,T $\Rightarrow$ $S = k_B\ln{(2)}$ which is a non-temperature dependent $S$ and thus it violates the third law of Thermodynamics, i.e., $S \rightarrow 0$ as $T \rightarrow 0$ \cite{ralph}. Hence, to solve this issue, the mutual interactions between neighboring magnetic moments must be considered \cite{ash123,ralph}. At this point, the magnetic energy $E$ associated with the dipolar interactions between adjacent spins is recalled, namely \cite{Blundell}:
\begin{equation}
E = \frac{\mu_0}{4\pi |\,\vec{r}\,|^3}\left[\vec{\mu_1}\cdot\vec{\mu_2}-\frac{3}{|\,\vec{r}\,|^2}(\vec{\mu_1}\cdot \vec{r})(\vec{\mu_2}\cdot \vec{r})\right],
\label{magneticenergy}
\end{equation}
where $\mu_0$ is the vacuum permeability (4$\pi\times$10$^{-7}$)\,T$\cdot$m/A, $\vec{r}$ the distance between neighbouring spins, and $\vec{\mu_1}$ and $\vec{\mu_2}$ two arbitrary neighbouring magnetic moments. Considering no applied $B$, it can be written $|E| = \mu_B B_{loc}$, where $B_{loc}$ is the modulus of the effective local magnetic field that emerges into the system as a result of the dipolar interactions between neighboring moments. Then, Eq.\,\ref{magneticenergy} becomes:
\begin{equation}
\mu_B B_{loc} = \frac{\mu_0}{4\pi |\,\vec{r}\,|^3}\left[\vec{\mu_1}\cdot\vec{\mu_2}-\frac{3}{|\,\vec{r}\,|^2}(\vec{\mu_1}\cdot \vec{r})(\vec{\mu_2}\cdot \vec{r})\right].
\label{magneticenergy2}
\end{equation}
Assuming $|\vec{\mu_1}| = |\vec{\mu_2}| = \mu_B = (9.27\times10^{-24})$\,J/T and a typical distance between spins given by $|\,\vec{r}\,| = (5\AA = 5\times10^{-10})\,$\,m, it is straightforward to notice that the second term on the right side of the brackets does not significantly contribute to $E$, so that Eq.\,\ref{magneticenergy2} can be approximated to:
\begin{equation}
B_{loc} \simeq \frac{\mu_0 \mu}{4\pi |\,\vec{r}\,|^3} \simeq \frac{(4\pi\times10^{-7})(9.27\times10^{-24})}{4\pi(5\times10^{-10})^3} \simeq 0.01\,T.
\label{magneticenergy3}
\end{equation}
Equation\,\ref{magneticenergy3} enables us to conclude that $B_{loc} = 0.01$\,T is intrinsically present in real paramagnets, since interactions between magnetic moments are always present. Therefore, since both $B$ and $B_{loc}$ are acting on the system, it is natural to assume that the paramagnetic system is under the influence of a resultant magnetic field, being its modulus $B_r$, based on the angle $\theta$ between $\vec{B}$ and $\vec{B}_{loc}$, which, employing the cosine rule, reads:
\begin{equation}
{B_r}^{2} = B^2 + {B_{loc}}^2 - 2BB_{loc}\cos{(\theta)}.
\end{equation}
For the sake of simplicity, it is assumed that $\theta \approx 90^{\circ}$, so that:
\begin{equation}
{B_r}^{2} = B^2 + {B_{loc}}^2 - \overbrace{2BB_{loc}\cos{(\theta)}}^{\approx\,\,0}.
\end{equation}
Hence, $B_r$ takes the form:
\begin{equation}
B_r \simeq \sqrt{B^2 + {B_{loc}}^2}.
\label{resultantb}
\end{equation}
Now, replacing $B$ for $B_r$ in Eq.\,\ref{entropyBrillouin}, it reads:\newline
\begin{equation}
\scalemath{0.91}{\boxed{S(T,B) = -\frac{\mu_B\sqrt{B^2+{B_{loc}}^2} }{T}\tanh{ \mu_B\left(\frac{\sqrt{B^2+{B_{loc}}^2}}{k_B T}\right)} + k_B\textmd{ln}\left[2\cosh{\left(\frac{\mu_B\sqrt{B^2+{B_{loc}}^2} }{k_B T}\right)}\right]}}.
\label{S1}
\end{equation}\newline
Note that, even when $B = 0$\,T, $B_{loc}$ is still non-zero and thus $S$ assumes a temperature-dependent mathematical form, which is now in perfect agreement with the third law of Thermodynamics \cite{ash123}. Thus, although the Brillouin-type paramagnet model is non-interacting, the effective dipolar interactions between adjacent spins must be considered. It is worth mentioning that the modulus of the magnetic energy $E_{mag}$ associated with $B_{loc}$ is given by $E_{mag} = \mu_B B_{loc} = (9.27\times10^{-24})(0.01) = (9.27\times10^{-26})$\,J. Comparing such a result with the thermal energy $T \simeq (9.27\times10^{-26})/k_B \simeq (9.27\times10^{-26})/(1.38\times10^{-23}) \simeq 6.7$\,mK. Such a result indicates that the relevance of the mutual interactions takes place on temperatures $T \leq 6.7$\,mK. In Section\,\ref{resultssection}, the obtained results regarding the Brillouin-type paramagnet and an investigation of the role played by $B_{loc}$ are discussed in a broader context.

\subsection{The Friedman equations}\label{friedmanequations}

Before discussing the results obtained in the frame of this Thesis regarding the application of both $\Gamma$ and $\Gamma_{eff}$ to Cosmology, a Newtonian-based derivation of the Friedman equations \cite{friedman} without the cosmological constant $\Lambda$ is properly given based on discussions presented in Ref.\,\cite{barbararyden}. Note that Friedman has derived his equations in 1922 using Einstein field equations to describe a spatially homogeneous and isotropic expanding universe, so that a full derivation of Friedman equations including all general relativistic effects is not covered here. The latter is properly reported in Ref.\,\cite{joan}. Friedman has published his equations 7 years before Hubble reported what is known today as the Hubble law \cite{hubble}, showing that galaxies are moving away from each other and the further away they are, the faster they move away. Hubble proposed a linear relationship between the radial velocity $v_R$ of galaxies and the distance to a particular galaxy $d$, so that $ H = v_Rd$ \cite{hubble}, where $H$ is the Hubble parameter broadly discussed in Cosmology. The Hubble law is considered one of the most important relations in Cosmology \cite{barbararyden}. Just to mention, the first experimental evidence that the universe is expanding was reported in 1912 by V. M. Slipher \cite{slipher}, while the observational evidence that the expansion of the universe was accelerated was only reported in 1998, cf.\,Ref.\,\cite{riess}. Back to the Newtonian-based derivation of the Friedman equations, first it is considered that the universe is a homogeneous sphere of matter with a mass $M_s$ that does not vary in time. Such a sphere can be either contracting or expanding isotropically over time, so that its radius $R_s(t)$ is time-dependent. Considering a test mass $m$ at the surface of the sphere, the modulus of the gravitational force $F$ experienced by the test mass considering Newton's law of gravity is given by \cite{alonsofinn}:
\begin{equation}
F = -\frac{GM_s m}{{R_s}^2},
\end{equation}
where $G$ is the universal gravitational constant. Considering Newton's second law, $F$ can be written in terms of the gravitational acceleration, which is given by:
\begin{equation}
F = ma = m\frac{d^2 R_s}{dt^2},
\end{equation}
which reads:
\begin{equation}
\cancel{m}\frac{d^2 R_s}{dt^2} = -\frac{GM_s\cancel{m}}{{R_s}^2},
\end{equation}
\begin{equation}
\frac{d^2 R_s}{dt^2} = -\frac{GM_s}{{R_s}^2}.
\label{F1}
\end{equation}
Multiplying both sides of Eq.\,\ref{F1} by $dR_s/dt$, it reads:
\begin{equation}
\frac{dR_s}{dt}\frac{d}{dt}\left(\frac{dR_s}{dt}\right) = -\frac{GM_s}{{R_s}^2}\frac{dR_s}{dt}.
\end{equation}
Integrating both sides:
\begin{equation}
\int \left(\frac{dR_s}{dt}\right)\frac{d}{dt}\left(\frac{dR_s}{dt}\right) = \int -\frac{GM_s}{{R_s}^2}\left(\frac{dR_s}{dt}\right) = GM_s\int-\frac{1}{{R_s}^2}\left(\frac{dR_s}{dt}\right).
\label{trickymath}
\end{equation}
Upon analysing the term $(dR_s/dt)d/dt(dR_s/dt)$, one can notice employing the chain rule that such a term represents the time derivative of the function $1/2(dR_s/dt)^2$, i.e., $d/dt[1/2(dR_s/dt)^2] = (dR_s/dt)d/dt(dR_s/dt)$. Hence, Eq.\,\ref{trickymath} is rewritten as:
\begin{equation}
\int \frac{d}{dt}\left[\frac{1}{2}\left(\frac{dR_s}{dt}\right)^2\right] = GM_s\int-\frac{1}{{R_s}^2}\left(\frac{dR_s}{dt}\right).
\end{equation}
Employing basic Calculus, the integral of the derivative of a function yields in the function itself, so that:
\begin{equation}
\frac{1}{2}\left(\frac{dR_s}{dt}\right)^2 + c_1 = GM_s\int-\frac{1}{{R_s}^2}\left(\frac{dR_s}{dt}\right).
\end{equation}
Now, the same analysis is employed for the term $-1/{R_s}^2(dR_s/dt)$, i.e., employing the chain rule one can notice that $d/dt[1/R_s] = -1/{R_s}^2(dR_s/dt)$, so that:
\begin{equation}
\frac{1}{2}\left(\frac{dR_s}{dt}\right)^2 + c_1 = GM_s\int\frac{d}{dt}\left(\frac{1}{R_s}\right),
\end{equation}
\begin{equation}
\frac{1}{2}\left(\frac{dR_s}{dt}\right)^2 + c_1 = \frac{GM_s}{R_s} + c_2,
\end{equation}
\begin{equation}
\frac{1}{2}\left(\frac{dR_s}{dt}\right)^2 = \frac{GM_s}{R_s} + c_3,
\label{F3}
\end{equation}
where $c_1$ and $c_2$ are integration constants and $c_3 = (c_2 - c_1)$. Given that $R_s$ varies in time, the volume of the sphere $V_s$ will vary as well and, as a consequence, so the matter density $\rho_m$, so that:
\begin{equation}
\rho_m = \frac{M_s}{V_s} \Rightarrow V_s = \frac{M_s}{\rho_m}.
\label{F4}
\end{equation}
Since a spherical shape is considered, Eq.\,\ref{F4} reads:
\begin{equation}
V_s = \frac{4}{3}\pi{R_s}^3 = \frac{M_s}{\rho_m},
\end{equation}
\begin{equation}
M_s = \frac{4}{3}\pi{R_s}^3\rho_m.
\label{F5}
\end{equation}
The temporal variation of $R_s$ can be inferred as a consequence of the change in the metric of space, as well as in so-called scale factor $a(t)$, so that:
\begin{equation}
R_s = a(t)r_s,
\label{F6}
\end{equation}
where $r_s$ is the comoving radius of the sphere. Essentially, the comoving radius is a definition of measuring distances in the universe that takes into account the fact that the universe is expanding over time. It helps to keep track of how far apart objects are from each other, even as the universe expands. By making use of Eq.\,\ref{F6}, the term $dR_s/dt = \dot{a}(t)r_s$. At this point, for the sake of simplicity, $a(t)$ is given solely by $a$. Then, together with Eq.\,\ref{F5}, Eq.\,\ref{F3} reads:
\begin{equation}
\frac{1}{2}(\dot{a}r_s)^2 = \frac{G}{R_s}\frac{4}{3}\pi{R_s}^3\rho_m + c_3,
\end{equation}
\begin{equation}
\frac{1}{2}{\dot{a}}^2{r_s}^2 = \frac{4\pi G}{3}{R_s}^2\rho_m + c_3.
\label{F7}
\end{equation}
Employing Eq.\,\ref{F6} into Eq.\,\ref{F7}:

\begin{equation}
\frac{1}{2}{\dot{a}}^2{r_s}^2 = \frac{4\pi G}{3}(ar_s)^2\rho_m + c_3,
\end{equation}

\begin{equation}
\frac{1}{2}{\dot{a}}^2{r_s}^2 = \frac{4\pi G}{3}{a}^2{r_s}^2\rho_m + c_3.
\label{F8}
\end{equation}

\noindent Dividing both sides of Eq.\,\ref{F8} by ${r_s}^2{a}^2/2$, it yields:

\begin{equation}
\frac{\frac{1}{2}{\dot{a}}^2{r_s}^2}{\frac{{r_s}^2{a}^2}{2}} = \frac{\frac{4\pi G}{3}{a}^2{r_s}^2\rho_m + c_3}{\frac{{r_s}^2{a}^2}{2}},
\end{equation}

\begin{equation}
\frac{{\dot{a}}^2}{{a}^2} = \frac{8\pi G}{3}\rho_m + \frac{2c_3}{{r_s}^2}\frac{1}{{a}^2},
\end{equation}

\begin{equation}
\left(\frac{\dot{a}}{a}\right)^2 = \frac{8\pi G}{3}\rho_m + \frac{2c_3}{{r_s}^2}\frac{1}{{a}^2}.
\label{F9}
\end{equation}
Equation\,\ref{F9} is the first Friedman equation in a Newtonian form. It connects the temporal behaviour of the scale factor with $\rho_m (t)$. However, the correct form of the first Friedman equation must take into account relativistic effects since it is derived from Einstein field equations in the frame of general relativity. Hence, the correct form of the first Friedman equation reads \cite{friedman}:
\begin{equation}
\vspace{0.2cm}
\boxed{\left(\frac{\dot{a}}{a}\right)^2 = \frac{8\pi G}{3c^2}\rho - \frac{\kappa c^2}{{r_s}^2}\frac{1}{{a}^2}},
\end{equation}
where $\rho$ is the energy density, $\kappa$ is the space curvature constant, and $c$ is the speed of light. Note that the relation between $\rho$ and $\rho_m$ is $\rho_m = \rho/c^2$ \cite{barbararyden}. For the derivation of the second Friedman equation, the first-law of Thermodynamics is recalled:
\begin{equation}
dS = dU + pdV_S,
\end{equation}
where $dU$ is the infinitesimal internal energy variation. Since the expansion of the universe is adiabatic $dS = 0$, so that:
\begin{equation}
dU + pdV_S = 0.
\label{F10}
\end{equation}
Assuming that $V_S(t)$ and $U(t)$, Eq.\,\ref{F10} is given by:
\begin{equation}
\dot{U} + p\dot{V_S} = 0.
\label{F10'}
\end{equation}
Recalling that $R_s(t) = ar_s$, $V_S(t)$ reads:
\begin{equation}
V_S(t) = \frac{4}{3}\pi {R_s(t)}^3 = \frac{4}{3}\pi {a}^3 {r_s}^3,
\end{equation}
thus:
\begin{equation}
\dot{V_S} = \frac{4}{3}\pi {r_s}^3 (3{a}^2\dot{a}) = V_S(t)\left(3\frac{\dot{a}}{a}\right).
\label{F11}
\end{equation}
Now, $U(t)$ can be rewritten as:
\begin{equation}
U(t) = V_S(t)\rho,
\end{equation}
then:
\begin{equation}
\dot{U} = V_S\dot{\rho} + \dot{V_S}\rho = V_S\left(\dot{\rho} + 3\frac{\dot{a}}{a}\rho\right).
\label{F12}
\end{equation}
Replacing Eqs.\,\ref{F11} and \ref{F12} into Eq.\,\ref{F10'}:
\begin{equation}
V_S\left(\dot{\rho} + 3\frac{\dot{a}}{a}\rho + 3\frac{\dot{a}}{a}p\right) = 0,
\end{equation}
\begin{equation}
\dot{\rho} + 3\frac{\dot{a}}{a}\rho + 3\frac{\dot{a}}{a}p = 0,
\end{equation}
\begin{equation}
\dot{\rho} + \frac{3\dot{a}}{a}(\rho + p) = 0.
\label{F13}
\end{equation}

Equation\,\ref{F13} is the so-called fluid equation \cite{barbararyden}, connecting the energy density and pressure with the temporal evolution of the scale factor. Now, rewriting the first Friedman equation as:
\begin{equation}
{\dot{a}}^2 = \frac{8\pi G}{3c^2}\rho{a}^2 - \frac{\kappa c^2}{{r_s}^2}.
\label{refref}
\end{equation}
Taking the time derivative in both sides of Eq.\,\ref{refref}, it reads:
\begin{equation}
2\dot{a}\ddot{a} = \frac{8\pi G}{3c^2}(\dot{\rho}{a}^2 + 2\rho a\dot{a}).
\end{equation}
Dividing both sides by $2\dot{a}a$:
\begin{equation}
\frac{\ddot{a}}{a} = \frac{4\pi G}{3c^2}\left(\dot{\rho}\frac{a}{\dot{a}} + 2\rho\right).
\label{F14}
\end{equation}
Rewriting Eq.\,\ref{F13} as:
\begin{equation}
\dot{\rho} = -\frac{3\dot{a}}{a}(\rho + p).
\label{F15}
\end{equation}
Replacing Eq.\,\ref{F15} into Eq.\,\ref{F14}:
\begin{equation}
\frac{\ddot{a}}{a} = \frac{4\pi G}{3c^2}\left\{\left[-3\cancel{\frac{\dot{a}}{a}}(\rho + p)\right]\cancel{\frac{a}{\dot{a}}} + 2\rho\right\},
\end{equation}
\begin{equation}
\frac{\ddot{a}}{a} = \frac{4\pi G}{3c^2}(-3\rho - 3p + 2\rho) = \frac{4\pi G}{3c^2}(-\rho - 3p),
\end{equation}
\begin{equation}
\boxed{\frac{\ddot{a}}{a} = -\frac{4\pi G}{3c^2}(\rho + 3p)}.
\label{F16}
\end{equation}
Equation\,\ref{F16} is the second Friedman equation, also known as the acceleration equation \cite{barbararyden}. As broadly discussed in the literature, the universe is usually described by the picture of a perfect fluid, i.e., an isotropic fluid with no viscosity \cite{weinberg}. The equation of state (EOS) of a perfect fluid is $p = \omega\rho$ \cite{weinberg,barbararyden}, where $\omega$ is the so-called EOS parameter. After the Big Bang, various eras of the universe took place. First, the radiation-dominated era and, later on, the matter-dominated era was achieved. During both radiation- and matter-dominated eras, the universe expanded in a decelerated way \cite{weinberg,barbararyden}. Currently, the universe is expanding in an accelerated way \cite{riess} because of the so-called dark-energy (DE). Essentially, DE is the vacuum energy and it is taken into account both in Einstein field equations and Friedman equations when the cosmological constant $\Lambda$ is incorporated in such equations. The term ``dark'' refers to the fact that the origin of such energy is still unknown during the writing of this Thesis. Just to mention, our universe is composed nowadays of about 69\% of dark energy, 7\% of dark-matter, and 4\% of regular matter, i.e., stars and planets, for instance \cite{barbararyden}. The relationship between $p$ and $\rho$ was different for each era, given by distinct values of $\omega$. For the radiation-dominated era $\omega = 1/3$, for the matter-dominated era $\omega \rightarrow 0$, and for the DE-dominated era $\omega = -1$. It is worth mentioning that $\omega$ is merely discussed in the literature as a numerical value, which is discussed in more details in the results section of this Thesis. Regarding the accelerated expansion, the most accepted explanation for such an acceleration is that DE is a fluid with positive $\rho$, but with a negative pressure \cite{weinberg}, namely $p = -\rho$. For the expansion to be accelerated, the condition $\ddot{a} > 0$ must be fulfilled in Eq.\,\ref{F16}, being the opposite also true for a decelerated expansion $\ddot{a} < 0$. Just to mention, since $a$ represents the metric of space, $\dot{a}$ has units of velocity while $\ddot{a}$ of acceleration \cite{barbararyden}. By assuming that the total energy density $\rho_{tot}$ of the universe is given by $\rho_{tot} = (\rho_{radiation} + \rho_{matter} + \rho_{DE})$ and that $p$ is an additive physical quantity, i.e., the total pressure $p_{tot} = (p_{radiation} + p_{matter} + p_{DE})$, Eq.\,\ref{F16} can be straightforwardly rewritten:
\begin{equation}
\frac{\ddot{a}}{a} = -\frac{4\pi G}{3c^2}(\rho_{radiation} + \rho_{matter} + \rho_{DE} + 3p_{radiation} + 3p_{matter} + 3p_{DE}).
\label{F18}
\end{equation}
Considering the perfect fluid EOS and the value of $\omega$ for the various eras:
\begin{equation}
\frac{\ddot{a}}{a} = \left(\rho_{radiation} + \rho_{matter} + \rho_{DE} + 3\frac{1}{3}\rho_{radiation} + \cancelto{0}{3\cdot0.\rho_{matter}} + 3(-1)\rho_{DE}\right),
\end{equation}
\begin{equation}
\frac{\ddot{a}}{a} = -\frac{4\pi G}{3c^2}(2\rho_{radiation} + \rho_{matter} - 2\rho_{DE}).
\label{F19}
\end{equation}
Note that as the universe expands, $a$ is increased.\,Also, based on Friedman equations, it is established in the literature that $\rho_{radiation} \propto 1/a^4$ and $\rho_{matter} \propto 1/a^3$ \cite{frieman}. Also, $\rho_{DE}$ is considered to be time-independent, which is still a topic under debate in the literature. Thus, the expansion of the universe and the increase in the scale factor makes the energy density of radiation and matter to be decreased and, in the DE-dominated era, such contributions can be considered negligible in comparison with $\rho_{DE}$, so that Eq.\,\ref{F19} reads:
\begin{equation}
\frac{\ddot{a}}{a} \simeq -\frac{4\pi G}{3c^2}(-2\rho_{DE}),
\label{F20}
\end{equation}
\begin{equation}
\frac{\ddot{a}}{a} \simeq \frac{8\pi G}{3c^2}\rho_{DE}.
\label{F21}
\end{equation}
The minus sign of $-2\rho_{DE}$, which comes from the consideration of a negative pressure associated with DE, makes $\ddot{a}/a$ to be positive in the DE-dominated assuming that $G$ is always greater than zero. This implies that DE drives the accelerated expansion of the universe. This is one of the key results emerging from the Friedman equations in modern Cosmology \cite{barbararyden}. Based on these discussions regarding the perfect fluid picture, Friedman equations, and the accelerated expansion of the universe, we have unprecedentedly connected well-established concepts of condensed matter physics, namely $\Gamma$ and $\Gamma_{eff}$, to Cosmology, which is properly discussed in Section\,\ref{gruneisencosmology}.

\clearpage\section{Results}\label{resultssection}

\subsection{The magnetic Gr\"uneisen parameter for the Brillouin paramagnet}

\begin{figure}[h!]
\centering
\includegraphics[width=0.8\textwidth]{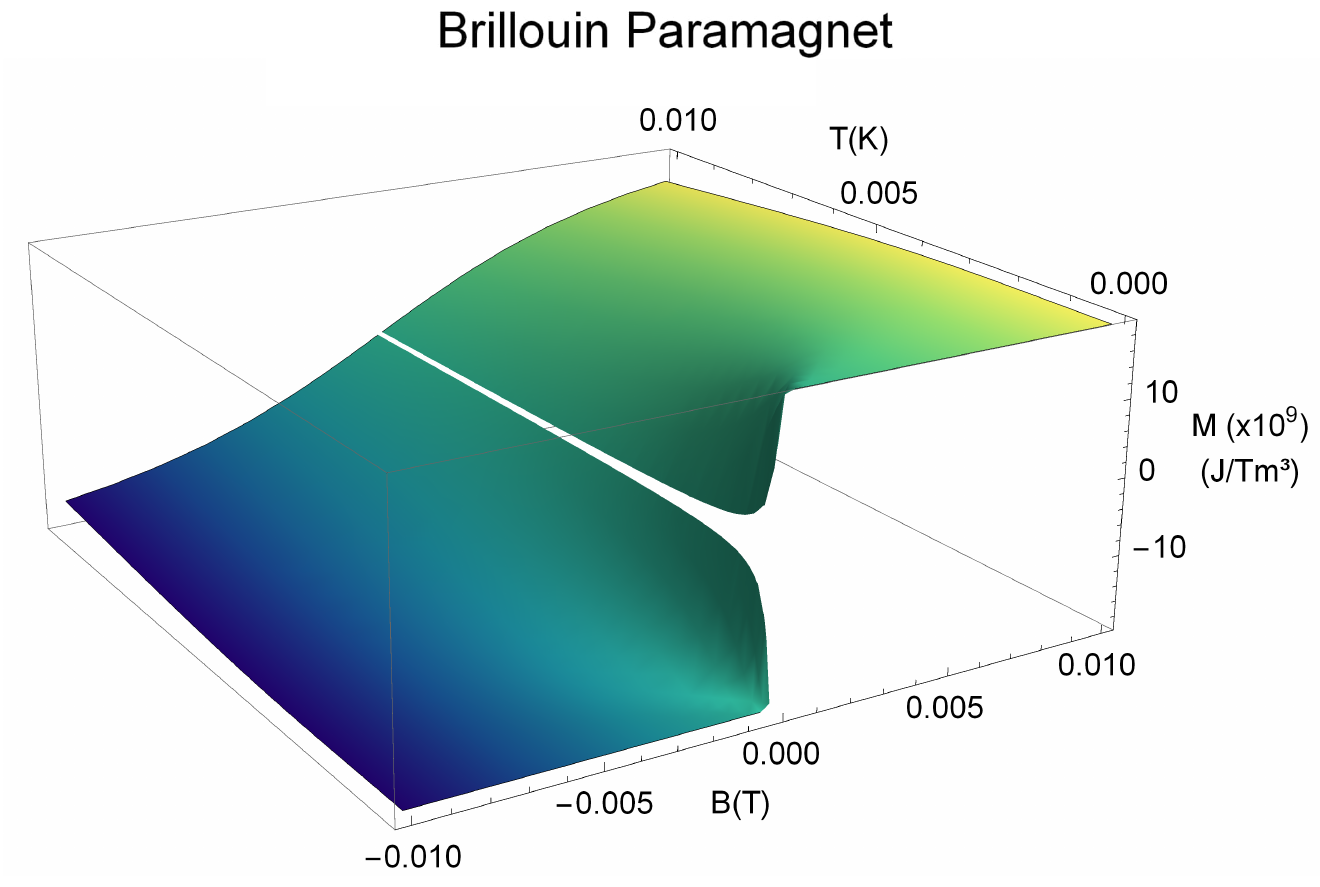}
\caption{\footnotesize Magnetization $M$ as a function of $T$ and $B$ for the Brillouin-type paramagnet. Note that in the regime of both $T$ and $B$ close to zero, a step-like behaviour of $M$ is observed. Figure extracted from Ref.\,\cite{modelsystems}.}
\label{Mbrillouin}
\end{figure}

By making use of the partition function $Z = 2\cosh{\left(\frac{\mu_B B}{k_B T}\right)}$ of the Brillouin-type paramagnet \cite{ralph}, the free energy $F$ can be computed as:
\begin{equation}
F (T,B) = -k_B T \ln{(Z)} = -k_B T\ln{\left[2\cosh{\left(\frac{\mu_B B}{k_B T}\right)}\right]}.
\end{equation}
Hence, employing $M = -(\partial F/\partial B)_T$ \cite{Stanleybook}, it reads:
\begin{equation}
M (T,B) = -\left(\frac{\partial F}{\partial B}\right)_T = -\frac{2\mu_B}{k_B T}\sinh{\left(\frac{\mu_B B}{k_B T}\right)}.
\label{magnetizationbrillouin}
\end{equation}
Employing Eq.\,\ref{magnetizationbrillouin}, the behaviour of $M$ as a function of $T$ and $B$ can be analysed. It turns out that in the regime of both $T \rightarrow 0$\,K and $B \rightarrow 0$\,T, $M$ presents a step-like behaviour, cf.\,Fig.\ref{Mbrillouin}.\newline
As a step further, the analytical calculation of $\Gamma_{mag}$ for the Brillouin-type paramagnet, namely \cite{modelsystems,zhu}:
\begin{equation}
\Gamma_{mag} = -\frac{\left(\frac{\partial M}{\partial T}\right)_B}{c_B} = -\frac{1}{T}\frac{(\partial S/\partial B)_T}{(\partial S/\partial T)_B} = \frac{1}{T}\left(\frac{\partial T}{\partial B}\right)_S,
\label{gammamagexpression}
\end{equation}
which quantifies the so-called magnetocaloric effect, i.e., the temperature change upon the adiabatic removal of $B$. For the Brillouin-type paramagnet, $\Gamma_{mag}$ was already reported in the literature by us and is temperature-independent, namely $\Gamma_{mag} = 1/B$ \cite{modelsystems,masterdissertation}, showing a divergent-like behaviour and sign-change when $B \rightarrow $ 0\,T. Such behavior of $\Gamma_{mag}$ motivated a deeper investigation of the Brillouin-type paramagnet in the frame of zero-field quantum phase transitions and the role played by the effective mutual interactions between neighboring magnetic moments in such a model. Note that, as previously discussed, the Brillouin-type paramagnet does not take into consideration the interaction between neighbouring magnetic moments and, in order to not violate the third law of Thermodynamics, the magnetic dipolar interactions must be considered. Some of the obtained theoretical results discussed in this Section were published in:
\begin{itemize}
\item Gabriel O. Gomes, Lucas Squillante, A.C. Seridonio, Andreas Ney, Roberto E. Lagos, and Mariano de Souza, Magnetic Grüneisen parameter for model systems, Physical Review B \textbf{100}, 054446 (2019).\newline
    \url{https://journals.aps.org/prb/abstract/10.1103/PhysRevB.100.054446}
\end{itemize}

\subsection{Genuine zero-field quantum phase transitions?}\label{1}

\begin{figure}[t]
\centering
\includegraphics[width=\textwidth]{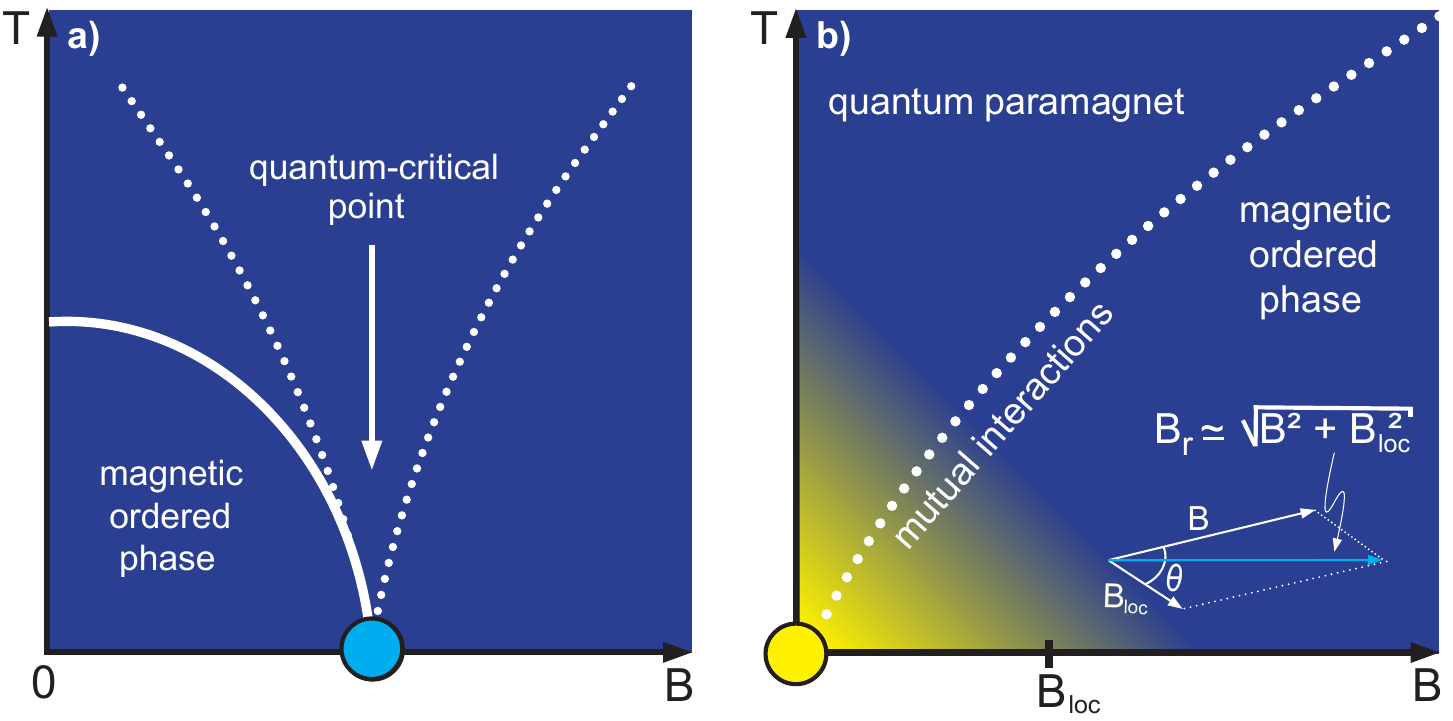}
\caption{\footnotesize Schematic representation of $T$ \emph{versus} $B$ for a magnetic-field-induced quantum phase transition. In panel a), the blue bullet represents a quantum critical point at finite $B$ where a quantum phase transition, for instance, from a magnetic ordered to a quantum disordered phase takes place. The white dotted lines represent the suppression of an energy scale upon approaching the quantum critical point \cite{zhu}, followed by the corresponding emergence of another energy scale associated with the new phase. In panel b), the hypothetical zero-field quantum critical point is depicted (yellow bullet). The yellow gradient represents the increasing role played by the mutual interactions upon decreasing temperature. It is assumed that $B_r \simeq \sqrt{B^2 + {B_{loc}}^2}$, so that the angle $\theta$ between $B$ and $B_{loc}$ is assumed to be $\theta \simeq$ 90$^\textmd{o}$. Figure extracted from Ref.\,\cite{unveiling2020}.}
\label{QPTscheme}
\end{figure}

In this Section, the focus lies on magnetic-field driven quantum phase transitions (QPT), more specifically zero-field quantum phase transitions \cite{matsumoto2011}, i.e., a quantum phase transition that takes place when both $T \rightarrow 0$\,K  and $B \rightarrow 0$\,T \cite{gegenwartgruneisen}. There are several real system reported in the literature as candidates to undergo a zero-field QPT, such as YbCo$_2$Ge$_4$ \cite{sakai2016}, $\beta$-YbAlB$_4$ \cite{matsumoto2011}, and Au-Al-Yb \cite{deguchi2012}. In this context, $\Gamma_{mag}$ plays a key role in probing a genuine magnetic-field-induced QPT, being the following criteria needed to be fulfilled for such \cite{gegenwartgruneisen}:\newline
\emph{i}) $\Gamma_{mag}$ must diverge at the critical magnetic field modulus $B_c$, i.e., the value of $B$ in which the QPT occurs; \newline
\emph{ii}) concomitant with the last condition, $\Gamma_{mag}$ must change sign upon crossing $B_c$; \newline
\emph{iii}) a typical scaling of the type $T/(B-B_c)^{\epsilon}$ must be observed, where $\epsilon$ is a scaling parameter. \newline
When such three conditions are fulfilled, a genuine magnetic-field-driven QPT is probed \cite{gegenwartgruneisen}, cf.\,Fig.\,\ref{QPTscheme}.

\begin{figure}[h!]
\centering
\includegraphics[width=0.75\textwidth]{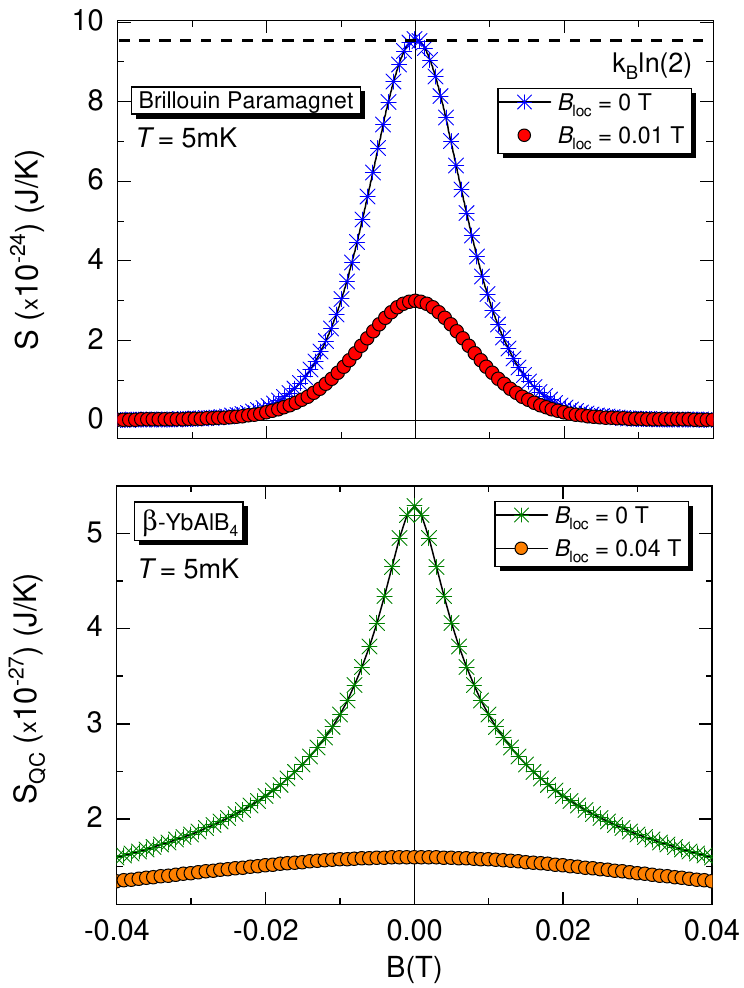}
\caption{\footnotesize Entropy $S$ \emph{versus} $B$ at a fixed $T$ = 5\,mK for the Brillouin-type paramagnet (upper panel) considering $B_{loc} = 0$\,T (blue stars) and $B_{loc}$ = 0.01\,T (red circles) and for $\beta$-YbAlB$_4$ for $B_{loc} = 0$\,T (green stars) and $B_{loc} = $0.04\,T (orange color circles). Figure extracted from Ref.\,\cite{unveiling2020}.}
\label{entropybloc}
\end{figure}

\begin{figure}[h!]
\centering
\includegraphics[width=0.75\textwidth]{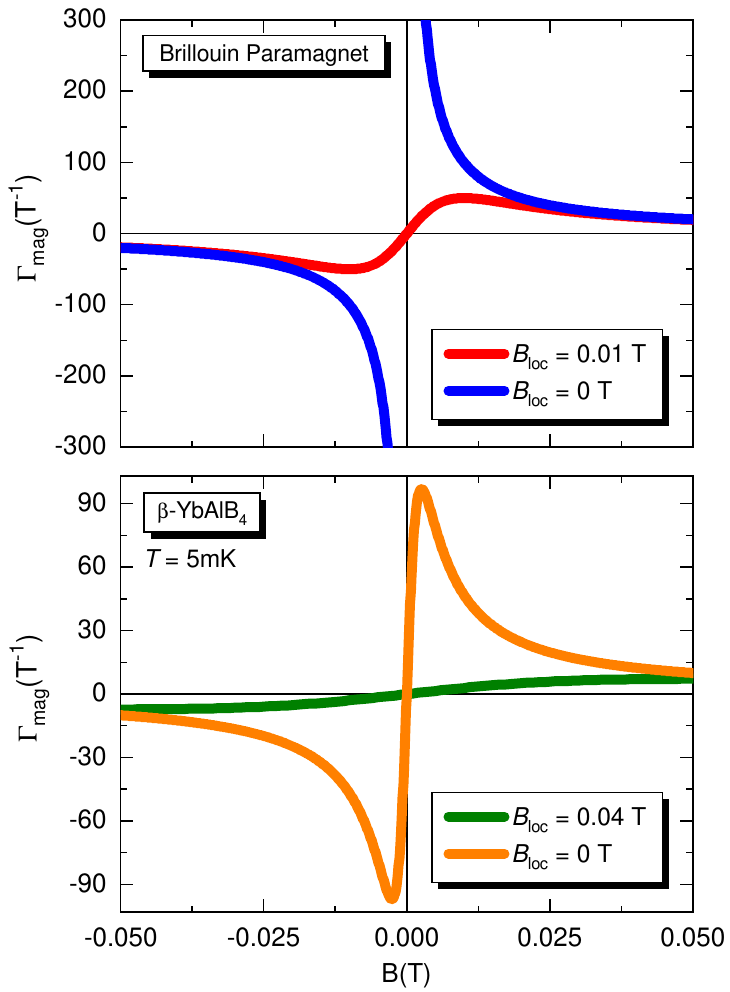}
\caption{\footnotesize Magnetic Gr\"uneisen parameter $\Gamma_{mag}$ as a function of $B$ for the Brillouin-type paramagnet (upper panel) considering $B_{loc} = 0$\,T (blue line) and $B_{loc}$ = 0.01\,T (red line) and for $\beta$-YbAlB$_4$ at $T$ = 5\,mK for $B_{loc} = 0$\,T (orange color line) and $B_{loc}$ = 0.04\,T (green line). Note that the divergence of $\Gamma_{mag}$ is suppressed at $B = 0$\,T when finite $B_{loc}$ is considered. Figure extracted from Ref.\,\cite{unveiling2020}.}
\label{gammamagbloc}
\end{figure}

\begin{figure}[h!]
\centering
\includegraphics[width=0.6\textwidth]{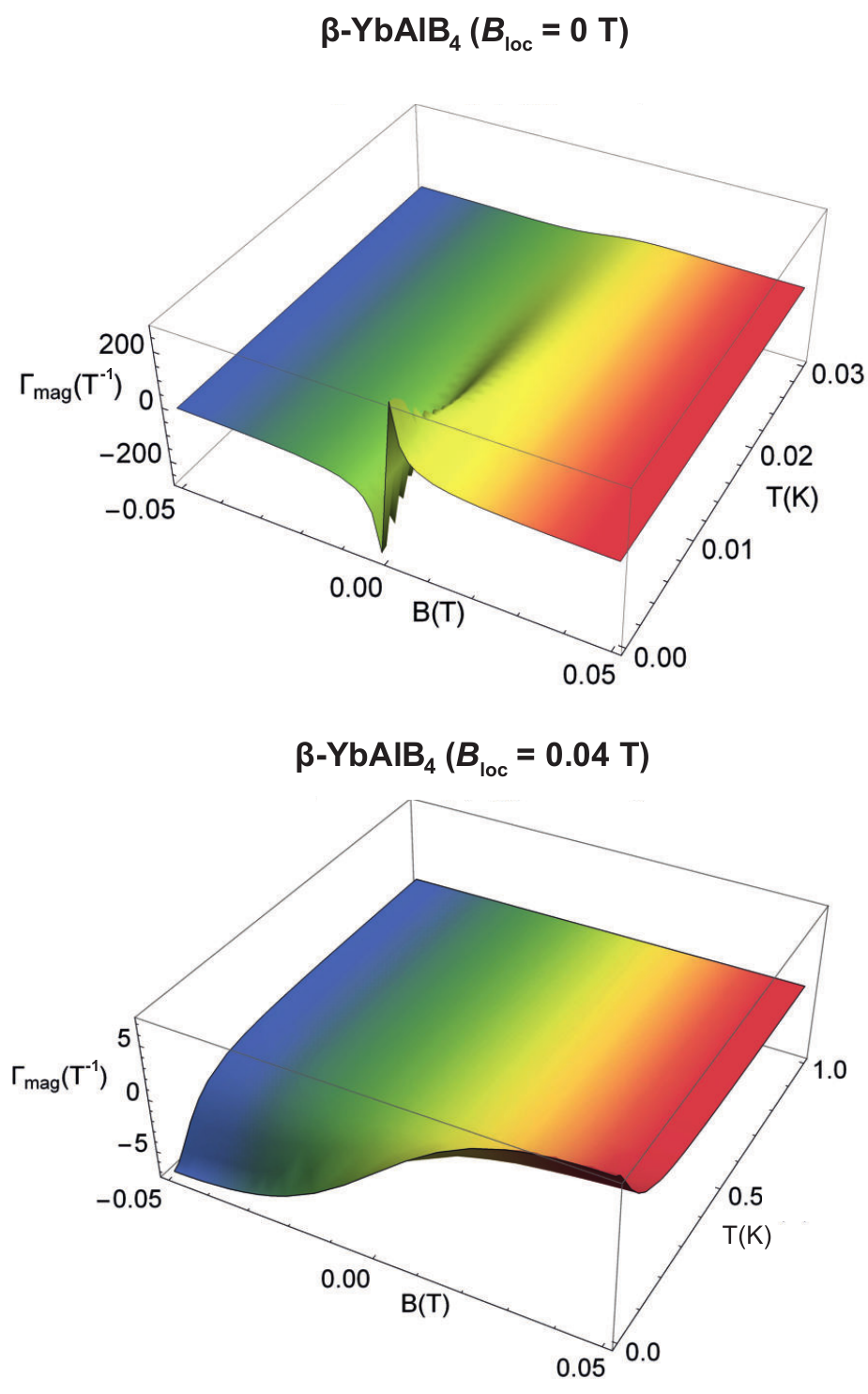}
\caption{\footnotesize Magnetic Gr\"uneisen parameter $\Gamma_{mag}$ \emph{versus} $T$ \emph{versus} $B$ for $\beta$-YbAlB$_4$ considering $B_{loc} = 0$\,T (upper panel) and $B_{loc}$ = 0.04\,T (lower panel). Figure extracted from Ref.\,\cite{unveiling2020}.}
\label{gammabetaybalb4}
\end{figure}

The divergence and sign-change of $\Gamma_{mag}$ was analysed in the frame of this Thesis when considering the intrinsic presence of $B_{loc}$ in the Brillouin-type paramagnet and also in the case of the real-system $\beta$-YbAlB$_4$ \cite{matsumoto2011}, where a divergence of $\Gamma_{mag}$ was reported for both $B$ and $T \rightarrow$ 0 \cite{matsumoto2011}. At this point, $B_{loc}$ is estimated for $\beta$-YbAlB$_4$ by making use of Eq.\,\ref{magneticenergy3} in an analogous way as for the Brillouin-type paramagnet. By making use of the effective magnetic moment of the Yb atoms $\mu \simeq 1.94\mu_B$ and a typical distance between Yb atoms $r \simeq 3.5\,\AA$ \cite{macaluso2007} $\Rightarrow$ $B_{loc} \simeq 0.04$\,T for $\beta$-YbAlB$_4$. At this point, an analysis of the impact of considering the intrinsic presence of $B_{loc}$ in both the Brillouin-type paramagnet and $\beta$-YbAlB$_4$ is carried out in terms of the behaviour of $\Gamma_{mag}$ at ultra-low temperatures for vanishing $B$. It is assumed that $B_r$ acting on the system is a sum between $B$ and $B_{loc}$ so that $B_r \simeq \sqrt{B^2 + {B_{loc}}^2}$, cf.\,Eq.\,\ref{resultantb}. Hence, $B$ is replaced by $B_r$ in the expression for entropy of the Brillouin-type paramagnetic (Eq.\,\ref{S1}) and $\Gamma_{mag}$ is computed, so that:
\begin{equation}
\boxed{\Gamma_{mag} = \frac{B}{B^2 + {B_{loc}}^2}}.
\end{equation}
Note that for vanishing $B$ $\Rightarrow$ $\Gamma_{mag} \rightarrow 0$ and its divergent-like behaviour is suppressed when $B_{loc}$ is taken into account. As a consequence, $\Gamma_{mag}$ is only enhanced for vanishing $B$ but does not diverge upon crossing $B = 0$\,T. For the real-system $\beta$-YbAlB$_4$, $\Gamma_{mag}$ can be computed via its proposed quantum critical free energy $F_{QC}$, given by \cite{matsumoto2011}:
\begin{equation}
F_{QC} (T,B) = -\frac{1}{(k_B \tilde{T})^{1/2}}{[(g\mu_B B)^2+(k_B T)^2]}^{3/4},
\label{F}
\end{equation}
where $k_B\tilde{T}$ $\approx$ 6.6\,eV $\approx$ (1.06$\times$10$^{-18}$)\,J, $\tilde{T}$ refers to a characteristic temperature and $g$ = 1.94 is the Land\'e factor \cite{yosuke}. Note that the proposed $F_{QC}$ refers to the valence-fluctuating character of 4f electrons in Yb atoms in $\beta$-YbAlB$_4$ \cite{matsumoto2011}. Using Eq.\,\ref{F}, it is straightforward to calculate the quantum critical entropy, $S_{QC}$ = $-(\partial F_{QC}/\partial T)_B$, namely:
\begin{equation}
S_{QC}(T,B) = \frac{3k_B^2 T}{2(k_B \tilde{T})^{1/2}(B^2 g^2{\mu_B}^2 + {k_B}^2 T^2)^{1/4}}.
\label{beta}
\end{equation}
An analysis of the behaviour of the entropy for both the Brillouin-type paramagnet and $\beta$-YbAlB$_4$ considering a finite $B_{loc}$ shows that the entropy accumulation at ultra-low temperatures, usually associated with the proximity of the quantum critical point, is suppressed since the presence of $B_{loc}$ brings order to the system, as can be seen in the upper and lower panels of Fig.\,\ref{entropybloc}.
Replacing $B$ by $B_r$ in Eq.\,\ref{beta} and computing $\Gamma_{mag}$ by Eq.\,\ref{gammamagexpression}, it reads:
\begin{equation}
\boxed{\Gamma_{mag} = \frac{B g^2{\mu_B}^2}{{k_B}^2T^2 + 2g^2{\mu_B}^2(B^2+B_{loc}^2)}}.
\label{gammapiers}
\end{equation}
The behaviour of $\Gamma_{mag}$ for the Brillouin-type paramagnet and for $\beta$-YbAlB$_4$ is depicted in Fig.\,\ref{gammamagbloc}, showing that a divergent-like behaviour is suppressed when $B_{loc}$ is taken into account for both cases, being only an enhancement of $\Gamma_{mag}$ observed instead of a divergent-like behaviour. For the case of $\beta$-YbAlB$_4$ without considering $B_{loc} \simeq 0.04$\,T, cf. orange color solid line in Fig.\,\ref{gammamagbloc}, $\Gamma_{mag}$ does not actually diverges for $B = 0$\,T since a fixed $T = 5$\,mK is considered.
In the case of the Brillouin-type paramagnet, $\Gamma_{mag}$ is non-temperature dependent. However, for $\beta$-YbAlB$_4$, $\Gamma_{mag}$ depends on temperature. Upon analysing the temperature dependence of $\Gamma_{mag}$ for $\beta$-YbAlB$_4$, one can notice that $\Gamma_{mag}$ presents a divergent-like behaviour followed by a sign-change when $T \rightarrow$ 0\,K and vanishing $B$, cf.\, upper panel of Fig.\,\ref{gammabetaybalb4}. Hence, upon considering finite $B_{loc}$, $\Gamma_{mag}$ is only enhanced and changes-sign at low-$T$ and vanishing field but its divergent-like behaviour is suppressed, as can be seen in the lower panel of Fig.\,\ref{gammabetaybalb4}.
Hence, it is shown that when finite $B_{loc}$ is considered in real paramagnets one of the key conditions to probe a genuine zero-field quantum phase transition is not fulfilled. Such an analysis in terms of $\Gamma_{mag}$ highlights the need of considering the intrinsic mutual interactions in real insulating paramagnets in the frame of zero-field quantum criticality. Next, a connection between $\Gamma_{mag}$ and the canonical definition of temperature is performed.

\subsection{Gr\"uneisen parameter and the canonical definition of temperature}\label{2}

\begin{figure}[h!]
\centering
\includegraphics[width=0.75\textwidth]{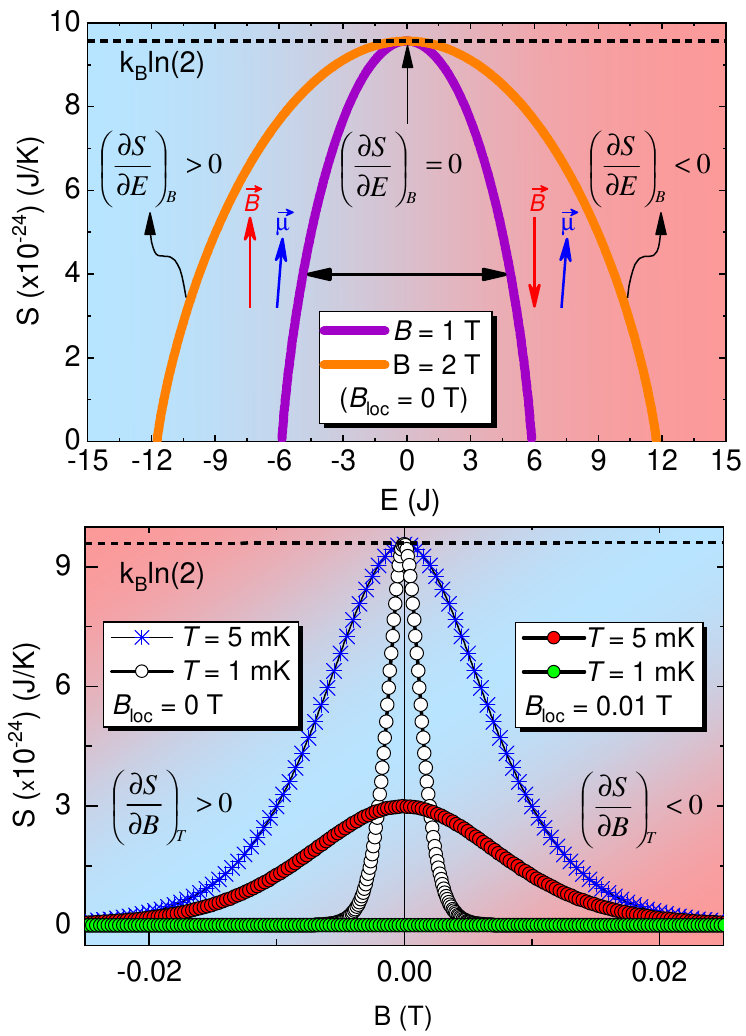}
\caption{\footnotesize Upper panel: Magnetic entropy $S$ as a function of the average magnetic energy $E$ at a fixed $B = $ 1 (purple line) and 2\,T (orange color line). The slope of $(\partial S/\partial E)_B$ indicates positive, infinite, or negative temperatures. The definition of positive and negative temperatures are associated with a parallel or an anti-parallel configuration, respectively, between $\vec{B}$ (red arrow) and the magnetic moment $\vec{\mu}$ (blue arrow). The background colors indicate that negative temperatures are hotter than positive ones. Lower panel: Magnetic entropy $S$ as a function of $B$ for $B_{loc} =$ 0\,T at $T = $ 5\,mK (blue stars) and $T = $ 1\,mK (white circles). For $B_{loc} =$ 0.01\,T the red and green circles represent $T = 5$ and 1\,mK, respectively. The background colors depict the insensitivity of $(\partial S/\partial B)_T$ regarding positive or negative temperatures. Figure extracted from Ref.\,\cite{unveiling2020}.}
\label{canonicaldefinitionoftemperature}
\end{figure}

As well-known from textbooks, the concept of temperature $T$ is universally defined as \cite{pathria,ralph}:
\begin{equation}
\frac{1}{T} = \left(\frac{\partial S}{\partial E}\right)_B,
\end{equation}
where $E$ is the average magnetic energy. The upper panel of Fig.\,\ref{canonicaldefinitionoftemperature} depicts the regimes where the slope of $(\partial S/\partial E)_B$ determines the temperature. For vanishing temperatures, $(\partial S/\partial E)_B \rightarrow \infty$, while for $(\partial S/\partial E)_B \rightarrow 0$ indicates that $T \rightarrow \infty$ \cite{ralph}. Note that for a positive (negative) slope of $(\partial S/\partial E)_B$, positive (negative) temperatures can be inferred, being positive and negative temperatures defined when a parallel or an anti-parallel alignment between $\vec{B}$ and $\vec{\mu}$ takes place \cite{pathria,ralph}, respectively. Indeed, a perfect alignment between $\vec{\mu}$ and $\vec{B}$ is prevented by Quantum Mechanics, cf.\,Appendix \ref{perfect}. The concept of negative absolute temperatures can be a little bit indigestible at first, mainly because negative temperatures are hotter than positive ones \cite{ralph}. Interestingly enough, negative temperatures were verified experimentally in the last century in the celebrated experiment of Purcell and Pound, where they used the nuclear spins of Li-7 in a LiF single-crystal that presents a long relaxation time of the order of 5 minutes under  $B \simeq 0.6376$\,T \cite{relaxationtimelitium}. Essentially, the experiment consisted in applying $\vec{B}$ on the sample, so that $\vec{B}$ and the nuclear magnetic moments are in a parallel configuration (positive temperature). Then, $\vec{B}$ was reversed in a time-scale of micro-seconds so that an anti-parallel configuration between $\vec{B}$ and the nuclear magnetic moments is achieved, making the system to be at a negative temperature.
In the regime where $E \rightarrow $ 0\,J, the value $S = k_B\ln{(2)}$ is nicely restored meaning that at very high temperatures, there is an equal probability for the spins to be aligned in a parallel or an anti-parallel configuration in respect to $\vec{B}$. Recalling that the average magnetic energy for the Brillouin-type paramagnet is given by \cite{ralph}:
\begin{equation}
E (T,B) = -N\mu_B B\tanh{\left(\frac{\mu_B B}{k_BT}\right)},
\end{equation}
it is possible to rewrite $E$ as a function of the average magnetic moment  $\langle\mu\rangle = N \mu_B \tanh{\left(\mu_B B/ k_B T\right)}$ along $\vec{B}$, as follows:
\begin{equation}
E = -\langle\mu\rangle B.
\end{equation}
From the expression for $\langle\mu\rangle$ given by \cite{ralph}:
\begin{equation}
\langle\mu\rangle = N\mu_B\tanh{\left(\frac{\mu_B B}{k_B T}\right)},
\end{equation}
it is straightforward to write $B$ as a function of $\langle\mu\rangle$:
\begin{equation}
B(T, {\langle\mu\rangle}) = \frac{k_B T}{\mu_B} \textmd{arctanh}\left(\frac{\langle\mu\rangle}{N\mu_B}\right).
\label{averagemagneticmoment}
\end{equation}
Equation \ref{averagemagneticmoment} indicates that at a certain temperature $T$, $B$ is associated with a spin configuration that corresponds to a specific average magnetic moment $\langle\mu\rangle$. Recalling that the magnetocaloric effect can be quantified by $\Gamma_{mag}$ \cite{zhu}:
\begin{equation}
\Gamma_{mag} = \frac{1}{T}\left(\frac{\partial T}{\partial B}\right)_S = \frac{1}{T}{\left(\frac{\partial B}{\partial T}\right)_S}^{-1},
\label{MCE-Eq}
\end{equation}
the temperature derivative of $B$ in Eq.\,\ref{averagemagneticmoment} can be computed and replaced it into Eq.\,\ref{MCE-Eq}, resulting:
\begin{equation}
\Gamma_{mag} = \frac{\mu_B}{Tk_B\textmd{arctanh}\left(\frac{\langle\mu\rangle}{N\mu_B}\right)}.
\end{equation}
Since $\Gamma_{mag}$ = 1/$B$ for the Brillouin-type paramagnet \cite{unveiling2020}, $T$ as a function of $B$ and $\langle\mu\rangle$ is given by:
\begin{equation}
\boxed{T (B,\langle\mu\rangle) = \frac{B \mu_B}{k_B \textmd{arctanh}\left(\frac{\langle\mu\rangle}{N\mu_B}\right)}}.
\label{connection}
\end{equation}
Equation \ref{connection} connects in an unprecedented way the Purcell and Pound experiment \cite{purcellandpound} and $\Gamma_{mag}$ itself. When $\vec{B}$ and $\langle\mu\rangle$ are $\parallel$, positive temperatures are inferred. However, when $\vec{B}$ and $\langle\mu\rangle$ are anti-$\parallel$, then negative temperatures can be associated. Remarkably, the definition of temperature is encoded in $\Gamma_{mag}$ and vice-versa.

\subsection{Adiabatic magnetization in a paramagnetic insulating system} \label{3}

\begin{figure}[h!]
\centering
\includegraphics[width=0.8\textwidth]{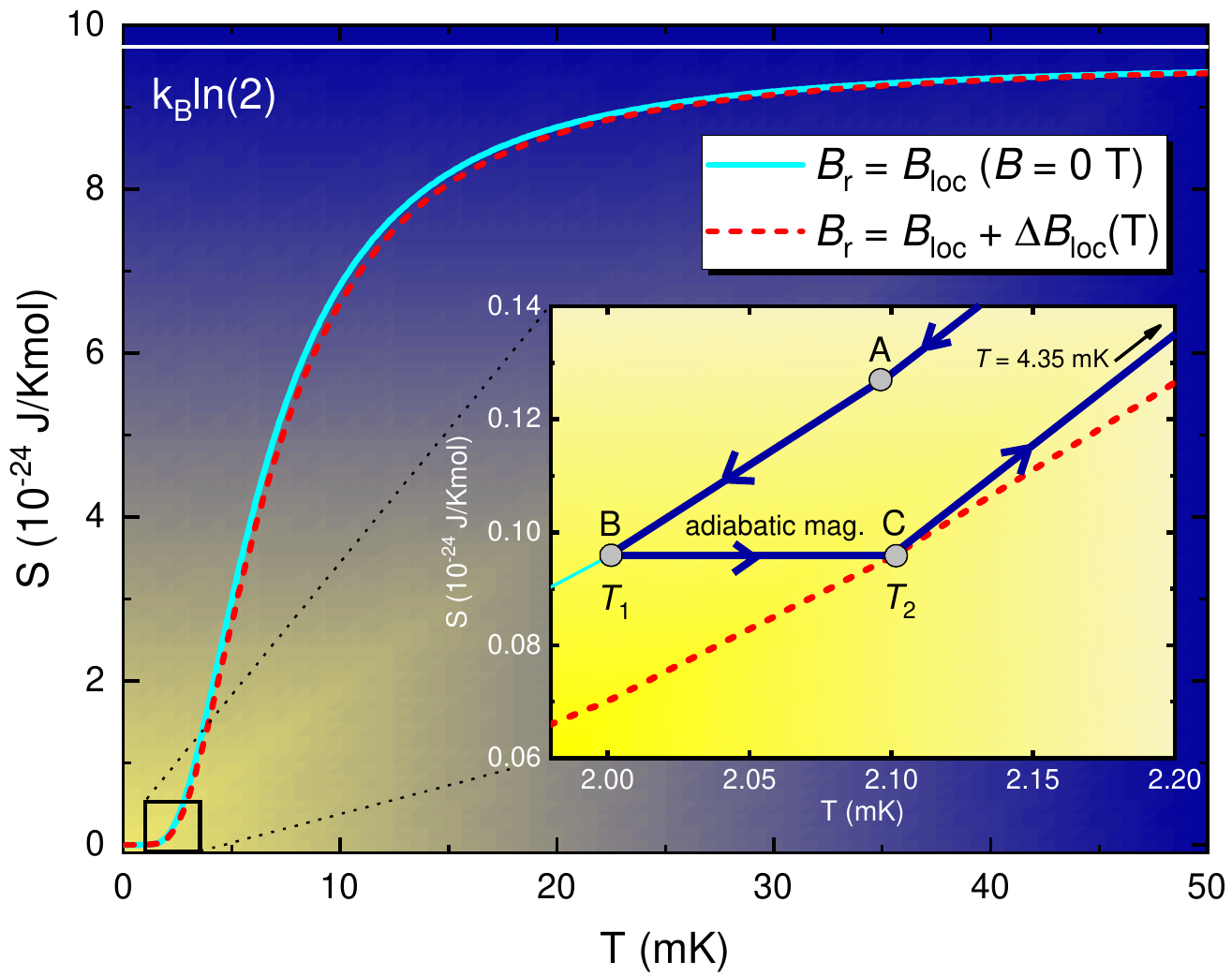}
\caption{\footnotesize Magnetic entropy $S$ as a function of temperature $T$ for the Brillouin-type paramagnet considering $B_r = B_{loc}$ (cyan solid line) and $B_r = (B_{loc} + \Delta B_{loc})$ (red dashed line). The inset shows a zoom in the low-$T$ regime describing the proposed steps for performing the adiabatic magnetization. First, the system should be cooled to a temperature where the magnetic energy associated with $B_{loc}$ is comparable to or higher than the thermal one, which is represented by going from point A to B. Then, at an initial temperature $T_1$, the temperature is adiabatically increased from $T_1$ to $T_2$ (path B to C) and, to keep the entropy constant, the mutual interactions must rearrange itself to increase the magnetic energy associated with $B_{loc}$ to increase it by an increment $\Delta B_{loc}$, giving rise to the proposed adiabatic magnetization. The cycle can be restarted upon increasing the temperature non-adiabatically and then cooling down the system again to point A. Figure extracted from Ref.\,\cite{unveiling2020}.}
\label{adiabaticmagnetization}
\end{figure}

The so-called adiabatic demagnetization is well-known from textbooks \cite{ralph}. In short, it refers to the cooling of a paramagnetic system upon adiabatically demagnetizing it. The basic concept behind this powerful cooling technique lies in the adiabaticity of the process, i.e., when $B$ is adiabatically removed, the thermal energy must be reduced to maintain the entropy constant and, as a consequence, the system cools down. But what if an opposite situation sets in? What if the temperature is varied adiabatically instead of $B$? Although counter-intuitive, it is indeed possible to vary the temperature adiabatically employing the application of stress \cite{Ikeda2019}, which is discussed in more details in the next Section. Regarding a paramagnetic insulating system without the influence of any $B$, $B_{loc}$ still remains finite due to the intrinsic mutual interactions between neighboring spins, cf.\,previously discussed. Upon decreasing temperature until the magnetic energy associated with $B_{loc}$ is comparable to the thermal energy, usually below $\approx$ 6.7\,mK, the mutual interactions become more and more relevant. In this temperature regime, the process of adiabatic magnetization can be performed, cf.\,shown step-by-step in Fig.\,\ref{adiabaticmagnetization}.

As can be seen in Fig.\,\ref{adiabaticmagnetization}, the crucial point in performing the adiabatic magnetization refers to an adiabatic temperature increase that makes the neighboring spins rearrange themselves to increase $B_{loc}$ in order to compensate the adiabatic increase of the thermal energy. Upon increasing the temperature adiabatically from an initial temperature $T_1$ to a final temperature $T_2$, the ratio between magnetic and thermal energies can be written as \cite{unveiling2020}:
\begin{equation}
\frac{\mu_B B_{loc}}{k_B T_1} = \frac{\mu_B \sqrt{{B_{loc}}^2 + {\Delta B_{loc}}^2}}{k_B T_2},
\label{blocincrease}
\end{equation}
where $B_r$ after the adiabatic temperature increase is given by $B_r = \sqrt{{B_{loc}}^2 + {\Delta B_{loc}}^2}$. From Eq.\,\ref{blocincrease}, the local magnetic field increment $\Delta B_{loc}$ is given by \cite{unveiling2020}:
\begin{equation}
\Delta B_{loc} = B_{loc}\sqrt{\frac{{T_2}^2}{{T_1}^2} - 1}.
\end{equation}
Upon assuming $T_1 = 2$\,mK, $T_2 = 2.1$\,mK, and $B_{loc} = 0.01$\,T:
\begin{equation}
\Delta B_{loc} = 0.01\sqrt{\frac{2.1^2}{2^2}-1} \simeq 0.0032 \textmd{T} \simeq 3.2\textmd{mT}.
\label{deltabloc}
\end{equation}
Amazingly, such adiabatic magnetization can be carried out by solely manipulating the mutual interaction upon increasing the temperature adiabatically, i.e., without applying $B$ to the paramagnetic system. As a step further, two experimental setups to perform and detect the adiabatic magnetization were proposed by us in the literature, which is discussed in the next Section.
The obtained theoretical results discussed in Sections\,\ref{1}, \ref{2}, and \ref{3} were published in:
\begin{itemize}
\item Lucas Squillante, Isys F. Mello, Gabriel O. Gomes, A.C. Seridonio, R.E. Lagos-Monaco, H. Eugene Stanley, Mariano de Souza, Unveiling the physics of the mutual interactions in paramagnets, Scientific Reports \textbf{10}, 7981 (2020).\newline
    \url{https://www.nature.com/articles/s41598-020-64632-x}
\end{itemize}

\subsection{Elastocaloric-induced effect adiabatic magnetization} \label{4}

\begin{figure}[h!]
\centering
\includegraphics[width=0.8\textwidth]{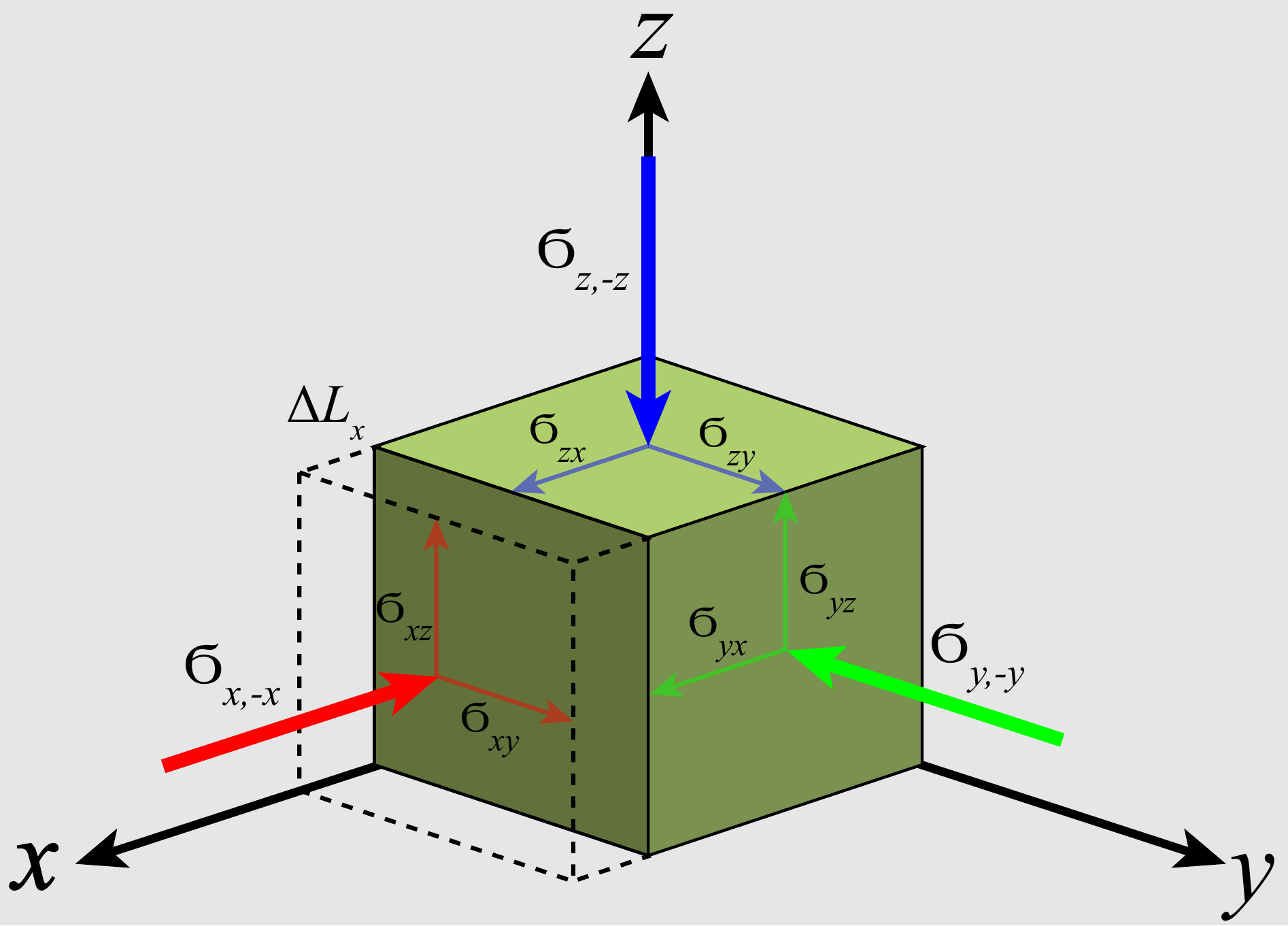}
\caption{\footnotesize Schematic representation of both the normal ($\sigma_{x,-x}$, $\sigma_{y,-y}$, and $\sigma_{z,-z}$) and shear ($\sigma_{xy}$, $\sigma_{xz}$, $\sigma_{yx}$, $\sigma_{yz}$, $\sigma_{zx}$, and $\sigma_{zy}$) stress components in a solid. The dashed lines indicate the compression of the solid only in the $x$ direction by the application of a uniaxial compressive stress component $\sigma_{x,-x}$, which in turn decreases the length $L$ of the solid by a factor $\Delta L_x$. Figure extracted from Ref.\,\cite{elastocaloric}.}
\label{stresscomponents}
\end{figure}

\begin{figure}[h!]
\centering
\includegraphics[width=\textwidth]{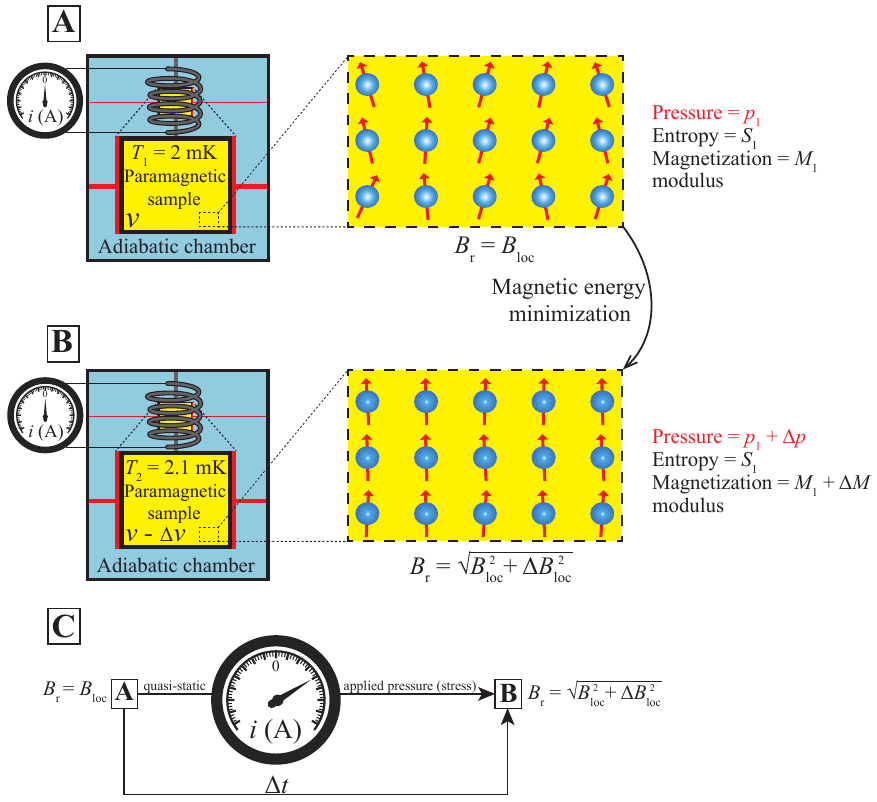}
\caption{\footnotesize Schematic representation of the experimental setup for attaining adiabatic magnetization. In panel A, the paramagnetic insulating sample is inserted into an adiabatic chamber inside a coil connected to an ampere meter and it is supported by two lateral pistons. Essentially, an adiabatic chamber is a thermally isolated compartment designed to prevent heat exchange with the outside environment. The sample is at an initial temperature $T_1 = 2$\,mK, pressure $p_1$, entropy $S_1$, magnetization $M_1$, and $B_r = B_{loc}$. In panel B, the two pistons compress adiabatically the sample so its volume is reduced by a factor of $\Delta v$. Hence, the final temperature is $T_2 = 2.1$\,mK, so that the entropy remains constant but the magnetization is increased in order to minimize the magnetic energy of the system so that $B_r = \sqrt{{B_{loc}^2 + {\Delta B_{loc}}^2}}$. In C, it is depicted that, upon going from A to B in a time interval $\Delta t$, the increase in $B_r$ generates a magnetic flux variation in time, which in turn leads to an electromotive force in the coil generating thus an electric current that is measured by the ampere meter. Based on the measured electrical current, $\Delta B_{loc}$ can be determined. Figure extracted from Ref.\,\cite{elastocaloric}.}
\label{adiabaticchamber}
\end{figure}

\begin{figure}[t]
\centering
\includegraphics[width=\textwidth]{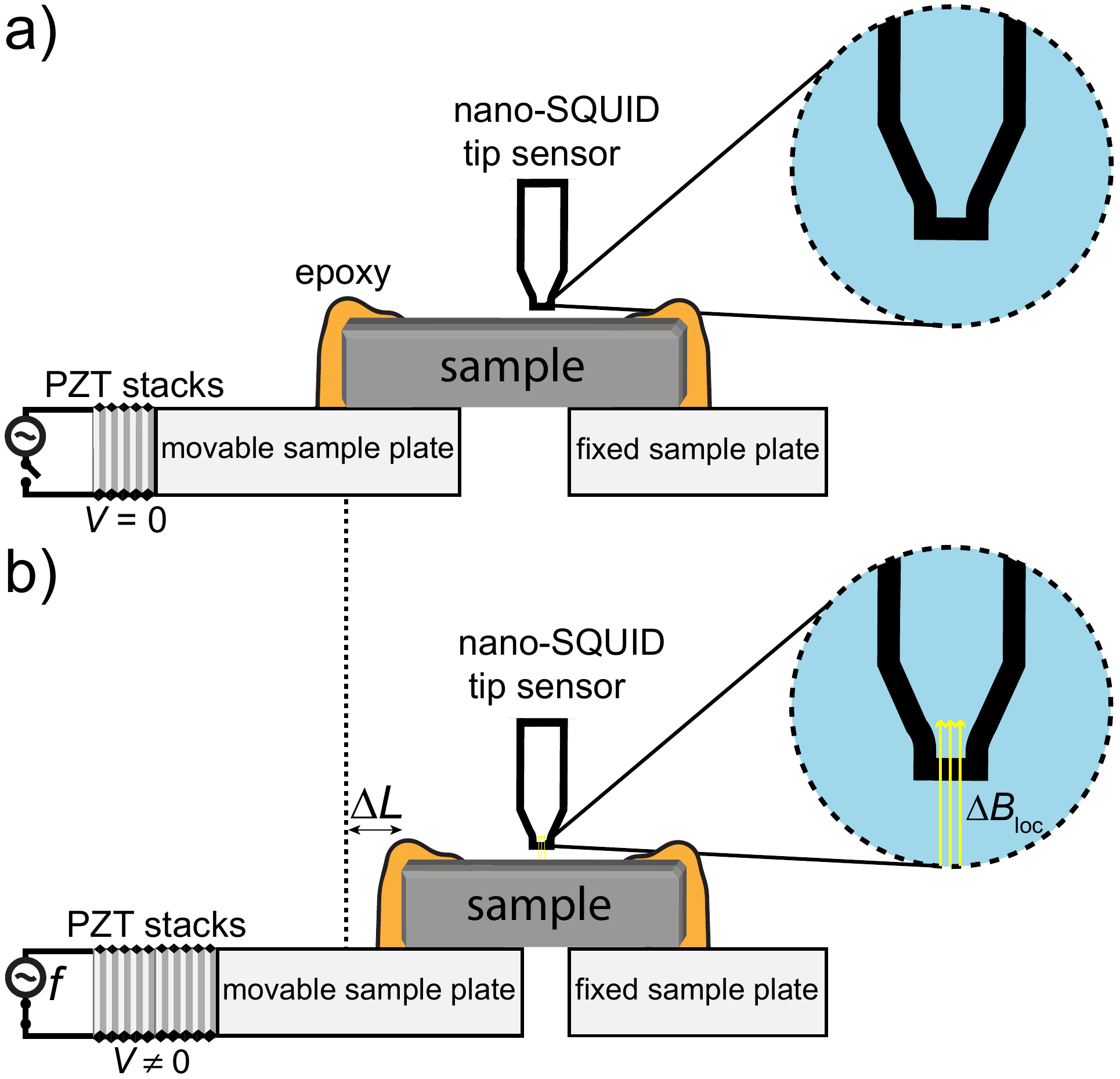}
\caption{\footnotesize Schematic representation of the proposed experimental setup to carry out and detect the elastocaloric-induced adiabatic magnetization due to the mutual interactions. a) The sample is attached with epoxy to a movable plate, which is connected to a piezoelectric device \cite{Ikeda2019, hicks2014, gannon2015}. b) When A.C. voltage $V$ is applied on the piezoelectric stacks, the sample is continuously uniaxially compressed and decompressed by a factor $\Delta L$. Due to the adiabatic compression, $B_{loc}$ is increased by a factor $\Delta B_{loc}$, which is detected by a nano-SQUID tip sensor \cite{zeldov}. Figure extracted from Ref.\,\cite{elastocaloric}.}
\label{pzt}
\end{figure}

As previously mentioned, adiabatic temperature changes can be seem, at first glance, as counter-intuitive. However, this is totally possible due to an adiabatic compression \cite{ralph}. At this point, the first law of Thermodynamics is recalled, which reads \cite{Stanleybook}:
\begin{equation}
dQ = dU + dW,
\label{firstlaw}
\end{equation}
where $dQ$ is the infinitesimal heat variation and $dW$ is the infinitesimal work performed on the system regarding its compression. As a consequence of an adiabatic compression, the temperature will change and so the internal energy, so that \cite{Stanleybook}:
\begin{equation}
dU = \left(\frac{\partial U}{\partial T}\right)_v dT = c_v dT.
\label{internalenergyfinal}
\end{equation}
The infinitesimal work performed on the system due to compression, is then given by \cite{Stanleybook}:
\begin{equation}
dW = -pdv.
\label{workdone}
\end{equation}
The pressure under adiabatic conditions can be expressed in terms of the variation of $U$ in response to a change of $v$, so that $p = -(\partial U/\partial v)_S$ \cite{Stanleybook}. Thus, Eq.\,\ref{workdone} reads:
\begin{equation}
dW = - pdv = \left(\frac{\partial U}{\partial v}\right)_S dv = \left(\frac{\partial U}{\partial S}\right)_v \left(\frac{\partial S}{\partial v}\right)_T dv.
\label{workmaxwell}
\end{equation}
Note that the term $(\partial U/ \partial S)_v = T$, following the canonical temperature definition \cite{ralph}. Employing the Maxwell relation $(\partial S/\partial v)_T = (\partial p/ \partial T)_v$ \cite{Stanleybook}, Eq.\,\ref{workmaxwell} now reads:
\begin{equation}
dW = T\left(\frac{\partial p}{ \partial T}\right)_v dv.
\label{workfinal}
\end{equation}
Upon putting Eqs.\,\ref{internalenergyfinal} and \ref{workfinal} into Eq.\,\ref{firstlaw}, it becomes:
\begin{equation}
dQ = c_v dT + T\left(\frac{\partial p}{ \partial T}\right)_v dv.
\label{dQ}
\end{equation}
Since $dQ = TdS$ \cite{ralph,Stanleybook} and that in an adiabatic process $dS = 0$, hence $dQ = 0$ as well, so that:
\begin{equation}
TdS = c_v dT + T\left(\frac{\partial p}{ \partial T}\right)_v dv
\end{equation}
\begin{equation}
0 = c_v dT + T\left(\frac{\partial p}{ \partial T}\right)_v dv
\end{equation}
\begin{equation}
c_v dT = -T\left(\frac{\partial p}{ \partial T}\right)_v dv.
\label{cvdt}
\end{equation}\newline
Employing the Thermodynamic relation \cite{Stanleybook}:
\begin{equation}
\left(\frac{\partial p}{\partial T}\right)_v \left(\frac{\partial T}{\partial v}\right)_p \left(\frac{\partial v}{\partial p}\right)_T = -1,
\end{equation}
it can be rewritten as:
\begin{equation}
\left(\frac{\partial p}{\partial T}\right)_v = -\frac{1}{\left(\frac{\partial T}{\partial v}\right)_p \left(\frac{\partial v}{\partial p}\right)_T}.
\label{maxwellrelation2}
\end{equation}
Note that the isothermal compressibility is given by $\kappa_T = -1/v (\partial v/\partial p)_T$ \cite{ralph}, so that the term in Eq.\,\ref{maxwellrelation2} $\left(\frac{\partial v}{\partial p}\right)_T = -\kappa_T v$. Also, following the definition of the volumetric thermal expansion $\beta = 1/v(\partial v/\partial T)_p$, the term in Eq.\,\ref{maxwellrelation2} $\left(\frac{\partial T}{\partial v}\right)_p = 1/\beta v$. Employing these substitutions in Eq.\,\ref{maxwellrelation2}, it reads:
\begin{equation}
\left(\frac{\partial p}{\partial T}\right)_v = -\frac{1}{\frac{1}{\beta v}[-\kappa_T v]} = \frac{\beta}{\kappa_T}.
\label{substitutions}
\end{equation}
Then, replacing Eq.\,\ref{substitutions} into Eq.\,\ref{cvdt}:
\begin{equation}
c_v dT = -\frac{T\beta}{\kappa_T} dv.
\label{cvdt2}
\end{equation}
Equation\,\ref{cvdt2} can then be rewritten as:
\begin{equation}
\boxed{dT = -\frac{T\beta}{c_v \kappa_T} dv}.
\label{key}
\end{equation}
Equation\,\ref{key} is key in analysing the temperature variation in response to an adiabatic compression of a system. Considering that $T$, $\beta$, $c_v$, and $\kappa_T$ are positive quantities, the sign of the adiabatic temperature variation is ruled solely by $dv$. On one hand, if $dv < 0$, the temperature of the system is adiabatically increased ($dT > 0$) in response to the decrease in its volume due to an adiabatic compression. On the other hand, if $dv > 0$, the temperature of the system is adiabatically decreased ($dT < 0$) as a consequence of an increase in the volume of the system due to an adiabatic decompression. The very same analysis can be performed now in terms of $p$ instead of $v$. Considering that the entropy of the system can be varied due to $p$ or $T$, so that:
\begin{equation}
dS(T, p) = \left(\frac{\partial S}{\partial T}\right)_p dT + \left(\frac{\partial S}{\partial p}\right)_T dp.
\label{entropyvariation}
\end{equation}
Considering the definition of the heat capacity at constant pressure $c_p = T(\partial S/\partial T)_p$ \cite{ralph}, the term in Eq.\,\ref{entropyvariation} $(\partial S/\partial T)_p = c_p/T$ and the Maxwell relation $(\partial S/ \partial p)_T = -(\partial v/\partial T)_p$ \cite{Stanleybook}, Eq.\,\ref{entropyvariation} can now be rewritten as:
\begin{equation}
dS(T, p) = \frac{c_p}{T} dT - \left(\frac{\partial v}{\partial T}\right)_p dp.
\end{equation}
Since an adiabatic process is considered, i.e., $dS(T, p) = 0$:
\begin{equation}
\frac{c_p}{T}dT = \left(\frac{\partial v}{\partial T}\right)_p dp.
\end{equation}
Recalling that $\beta = 1/v(\partial v/\partial T)_p$, then $(\partial v/\partial T)_p = \beta v$ so that:
\begin{equation}
\boxed{dT = \frac{\beta v T}{c_p} dp.}
\label{adiabaticpressure}
\end{equation}
Equation\,\ref{adiabaticpressure} is similar to Eq.\,\ref{key}, since for an adiabatic pressurization the final $p$ is higher than the initial one, i.e., $dp > 0$, the volume of the system is decreased and the temperature of the system is adiabatically increased ($dT > 0$), while for an adiabatic decompression the final $p$ is lower than the initial one ($dp < 0$), the volume of the system is increased and thus the temperature is adiabatically decreased. The effect of varying the temperature upon an adiabatic pressure variation is called the barocaloric effect, which is quantified by the singular portion of $\Gamma_{eff}$, the so-called Gr\"uneisen ratio $\Gamma = \beta/c_p = 1/Tv(\partial T/\partial p)_S$ \cite{zhu, EJP2016}. Indeed, $\Gamma$ is naturally encoded in Eq.\,\ref{adiabaticpressure}, which can be rewritten as:
\begin{equation}
\left(\frac{\partial T}{\partial p}\right)_S = \frac{\beta v T}{c_p},
\end{equation}
so that:
\begin{equation}
\frac{1}{Tv}\left(\frac{\partial T}{\partial p}\right)_S = \frac{\beta}{c_p} = \Gamma.
\label{ratio}
\end{equation}
Experimentally, an adiabatic temperature change can be carried out due to the application of stress $\sigma$ (Fig.\,\ref{stresscomponents}), which refers to the compression or decompression of a particular direction.
All the 9 stress components can be represented by a 3$\times$3 matrix, given by \cite{mechanicalbehavior}:
\begin{equation}
\sigma_{ij} =
\begin{bmatrix}
\sigma_{11} & \sigma_{12} & \sigma_{13} \\
\sigma_{21} & \sigma_{22} & \sigma_{23} \\
\sigma_{31} & \sigma_{32} & \sigma_{33}
\end{bmatrix},
\end{equation}
where $\sigma_{ij}$ represents all the stress components in Voigt's abbreviated notation \cite{mechanicalbehavior, Barron1980}, being the index $i$ the normal plane on which the stress component is acting and $j$ the direction of the applied stress. Hence, analogously to Eq.\,\ref{entropyvariation}, it can now be written as:
\begin{equation}
dS(T, \sigma) = \left(\frac{\partial S}{\partial T}\right)_\sigma dT + \left(\frac{\partial S}{\partial \sigma_{ij}}\right)_{T,\sigma} d\sigma_{ij}.
\label{entropyvariation2}
\end{equation}
Note that the second term on the right side of Eq.\,\ref{entropyvariation2}, represents the entropy variation due to the application of only one stress component while the others are kept constant, being also $T$ kept constant. In our analysis it is considered that only one of the normal components is varied, i.e., a uniaxial application of stress (Fig.\,\ref{stresscomponents}), being this the reason why such an entropy variation is given at a fixed $T$ and $\sigma$. An analogous decomposition of $\sigma$ can be made in terms of the thermal expansion, so that $\alpha_{ij} = (\partial S/\partial \sigma_{ij})_{T,\sigma}$ \cite{Ikeda2019}. Note that, usually, stress has units of pressure [Pa], however, in this particular case, the notation of $\sigma_{ij}$ is given in terms of stress components per unit of volume, so that $\sigma_{ij}$ = [J] \cite{mechanicalbehavior}. Also, considering that the heat capacity at constant stress $c_{\sigma}/T = (\partial S/ \partial T)_{\sigma}$, Eq.\,\ref{entropyvariation2} can be rewritten as:
\begin{equation}
dS = \frac{c_{\sigma}}{T}dT + \alpha_{ij}d\sigma_{ij} = 0
\end{equation}
\begin{equation}
\frac{c_{\sigma}}{T}dT = -\alpha_{ij}d\sigma_{ij},
\end{equation}
\begin{equation}
dT = -\frac{T}{c_{\sigma}}\alpha_{ij}d\sigma_{ij}.
\label{dTsigma}
\end{equation}
It becomes clear that in Eq.\,\ref{dTsigma}, $dT$ is the infinitesimal adiabatic temperature change, from a starting temperature $T$, due to the application of an adiabatic stress component $\sigma_{ij}$. Conventionally, $\sigma_{ij} < 0$ means that an uniaxial compression is carried out, while $\sigma_{ij} > 0$ is associated with an uniaxial decompression \cite{mechanicalbehavior}. Hence, considering that $c_{\sigma}$, $T$, and $\alpha_{ij}$ are positive, for $d\sigma_{ij} < 0$ it implies in $dT > 0$, while for $d\sigma_{ij} > 0$, $dT < 0$. In other words, upon adiabatically compressing (decompressing) the system along a particular axis, its length is reduced (increased) and, as a consequence, its temperature is adiabatically increased (reduced) to keep the system's entropy constant. This is the so-called elastocaloric-effect, which is similar to the barocaloric effect, previously discussed. Analogously to the fact that $\Gamma$ (Eq.\,\ref{ratio}) quantifies the barocaloric effect, Eq.\,\ref{dTsigma} can be rewritten as \cite{elastocaloric}:
\begin{equation}
\left(\frac{\partial T}{\partial \sigma_{ij}}\right)_S = \frac{T\alpha_{ij}}{c_{\sigma}}
\end{equation}
\begin{equation}
\boxed{\frac{1}{T}\left(\frac{\partial T}{\partial \sigma_{ij}}\right)_S = \frac{\alpha_{ij}}{c_{\sigma}} = \Gamma_{ec}}.
\label{elasticgruneisen}
\end{equation}\newline
Equation\,\ref{elasticgruneisen} accounts for the definition of the elastic Gr\"uneisen parameter $\Gamma_{ec}$, which quantifies the elastocaloric effect and was proposed in the literature in the frame of this Ph.D. Thesis, cf.\,Ref.\,\cite{elastocaloric}. Just to mention, after the proposal of $\Gamma_{ec}$ it was employed by researchers of the Max Planck Institute - Dresden to investigate the phase diagram of the unconventional superconductor Sr$_2$RuO$_4$ \cite{mackenzie}. Now, the proposed adiabatic magnetization of a paramagnetic insulating system due to an adiabatic temperature increase in terms of $\sigma_{ij}$ and the magnetization $M$ of the system are analysed, so that it is possible to write:
\begin{equation}
dS(T, M, \sigma_{ij}) = \left(\frac{\partial S}{\partial T}\right)_{M,\sigma} dT + \left(\frac{\partial S}{\partial M}\right)_{T,\sigma} dM + \left(\frac{\partial S}{\partial \sigma_{ij}}\right)_{T,M} d\sigma_{ij} = 0,
\end{equation}
so that:
\begin{equation}
-\left(\frac{\partial S}{\partial M}\right)_{T,\sigma} dM = \left(\frac{\partial S}{\partial T}\right)_{M,\sigma} dT + \left(\frac{\partial S}{\partial \sigma_{ij}}\right)_{T,M} d\sigma_{ij}.
\end{equation}
Employing the well-known Maxwell relation $(\partial S/\partial M)_T = -(\partial B/\partial T)_S$ \cite{Stanleybook}, it is possible to write that:
\begin{equation}
\left(\frac{\partial B}{\partial T}\right)_S dM = \left(\frac{\partial S}{\partial T}\right)_{M,\sigma} dT + \left(\frac{\partial S}{\partial \sigma_{ij}}\right)_{T,M} d\sigma_{ij}.
\label{mainelastocaloric}
\end{equation}
Equation\,\ref{mainelastocaloric} is key in analysing the proposed adiabatic magnetization upon an adiabatic uniaxial compression. Since no $B$ is applied to the system, it becomes evident that $B$ in Eq.\,\ref{mainelastocaloric} can be recognized as $B_{loc}$. Essentially, Eq.\,\ref{mainelastocaloric} tell us that, in response to an adiabatic temperature variation due to the application of uniaxial stress in the system, there should be an increase in the magnetization of the system by a factor $dM$ to keep the entropy constant, which in turn is originated by a variation of $B_{loc}$ by a factor $\Delta B_{loc}$ due to the temperature increase of the system, cf.\,previously discussed. At this point, it is tempting to make an analogy between positive and negative temperatures in the celebrated Purcell and Pound's experiment \cite{purcellandpound} in terms of either an adiabatic expansion or contraction under application of stress. However, this is not in agreement with the well-established conditions for attaining negative temperatures \cite{Ramsey1956}, as discussed in Appendix\,\ref{stressnegativetemperatures}. Just to mention, the proposed adiabatic magnetization resembles the celebrated Pomeranchuk effect \cite{pomeranchuk}, which predicts that $^3$He might solidify upon heating. In the case of the adiabatic magnetization, the system is magnetized upon adiabatically heating it. Next, two proposed experimental setups are discussed to carry out the adiabatic magnetization. In the first case, the paramagnetic insulating specimen is inserted into an adiabatic chamber inside a coil with two pistons holding it, cf.\,Fig.\,\ref{adiabaticchamber}.
As depicted in Fig.\,\ref{adiabaticchamber}, the sample is compressed adiabatically by the pistons decreasing its volume by a factor $\Delta v$. Hence, since it is an adiabatic compression, its temperature rises and the sample is adiabatically magnetized. The modulus of $\vec{B}_r$ is increased from $B_r = B_{loc}$ to $B_r = \sqrt{{B_{loc}^2 + {\Delta B_{loc}}^2}}$. The change in $B_r$ in time generates a magnetic flux variation in time, which in turn gives rise to an electromotive force $\xi = -(d\Phi/dt)$ \cite{ash123} opposing to such a change in flux $\Phi$, so that an electric current flows through the coil and it can be detected by the ampere meter. Based on the electrical current measured, $\Delta B_{loc}$ can be determined. At this point, it is plausible to ask ``How to apply uniaxial stress to a sample?'', and more importantly, ``How to actually attain the adiabatic character that is crucial for the experimental realization of the elastocaloric effect?''. The answer to both of these questions lies in the piezoelectric device proposed in Ref.\,\cite{Ikeda2019}, cf.\,depicted in Fig.\,\ref{pzt}.
The proposed experimental setup for attaining adiabatic temperature changes is depicted in Fig.\,\ref{pzt} \cite{Ikeda2019,elastocaloric}. The paramagnetic insulating sample is attached with epoxy to a movable plate, which is connected to a piezoelectric (PZT) device. As soon as A.C. voltage with a chosen frequency $f$ is applied to the PZT, the movable plate starts to oscillate back and forwards with the same frequency $f$ of the A.C. voltage, continuously compressing and decompressing the sample. Upon tuning $f$ to achieve such an oscillation in a time scale higher than the one associated with thermal relaxation, the adiabatic character is attained \cite{Ikeda2019}. In other words, the compression/decompression of the sample occurs faster than it can exchange heat with its surroundings. In this tiny time window of adiabaticity of about 30\,ms \cite{Ikeda2019}, the temperature of the paramagnetic sample is continuously increased/decreased, following Eq.\,\ref{dTsigma}. Hence, considering the case of its adiabatic compression, its temperature increases and the sample is adiabatically magnetized, cf.\,previously discussed. To detect such $\Delta B_{loc}$, we proposed that a nano-SQUID on a tip \cite{zeldov} can be employed, which is broadly used to detect magnetic fields of a few mT \cite{zeldov}. Just for the sake of completeness, the magnetic flux $\Phi$ through a SQUID with surface area $S'$ is given by \cite{ash123}:
\begin{equation}
\Phi = \int_{S'}\vec{B}\cdot d\vec{S'} = n\left(\frac{h}{2e}\right),
\end{equation}
where $n$ is an integer number, $h$ is Planck's constant, $d\vec{S'}$ a normal vector in respect to the SQUID's junction inner area $S'$, and $e$ the fundamental electron charge. The factor $(h/2e)$ is the quantum of $\Phi$ and it dictates the SQUID sensitivity to small values of $\Phi$. Although the nano-SQUID on a tip \cite{zeldov} has its resolution compromised by its small $S'$, its spatial resolution is outstanding making it very sensitive to small magnetic moments \cite{zeldov}, which in turn can detect $\Delta B_{loc} \simeq 3.2$\,mT (Eq.\,\ref{deltabloc}). It is worth mentioning that the parameters involving the detection of $\Delta B_{loc}$ employing the nano-SQUID on a tip \cite{zeldov} shall depend on the physical properties of the investigated system, the temperature in which the experiment is carried out, and, of course, the characteristics of the nano-SQUID on a tip itself.

Based on the proposed adiabatic magnetization due to an adiabatic temperature increase of a paramagnetic insulating specimen, it is natural to extend the concept of an adiabatic temperature increase to investigate other interacting systems \cite{elastocaloric}, such as Bose-Einstein condensates in magnetic insulators \cite{giamarchi} and spin-ice systems \cite{bramwell}. In such cases, we propose that the many-body interactions between magnetic moments are rearranged as a whole when the temperature is adiabatically increased similarly to the one proposed for the paramagnetic insulating system. Hence, this technique for adiabatically increasing temperature can be employed to explore the many-body character of the magnetic interactions in other systems. For instance, in the case of the Bose-Einstein condensate in an insulator, the Hamiltonian of the system incorporates a Zeeman term $-g\mu_B B_z S^z$, cf.\,Eq.\,1 of Ref.\,\cite{giamarchi}, that governs the density of triplons in the system, being $B_z$ the modulus of the applied external magnetic field in the $z$ direction, and $S^z$ the spin projection in the $z$ direction. Considering that the resulting magnetic field $B_r$ in the system is solely given by the effective local magnetic field generated by the mutual interactions between adjacent magnetic moments, an adiabatic temperature increase makes the magnetic moments to rearrange to compensate for the adiabatic thermal energy increase. As a consequence, the Zeeman term of the Hamiltonian is varied and so is the density of triplons. This is only one particular example regarding how the adiabatic temperature increase can be employed to investigate the many-body character of the magnetic interactions in various systems \cite{elastocaloric}.

\subsection{Gr\"uneisen parameter and second-order phase transitions} \label{5}

\begin{figure}[h!]
\centering
\includegraphics[width=\textwidth]{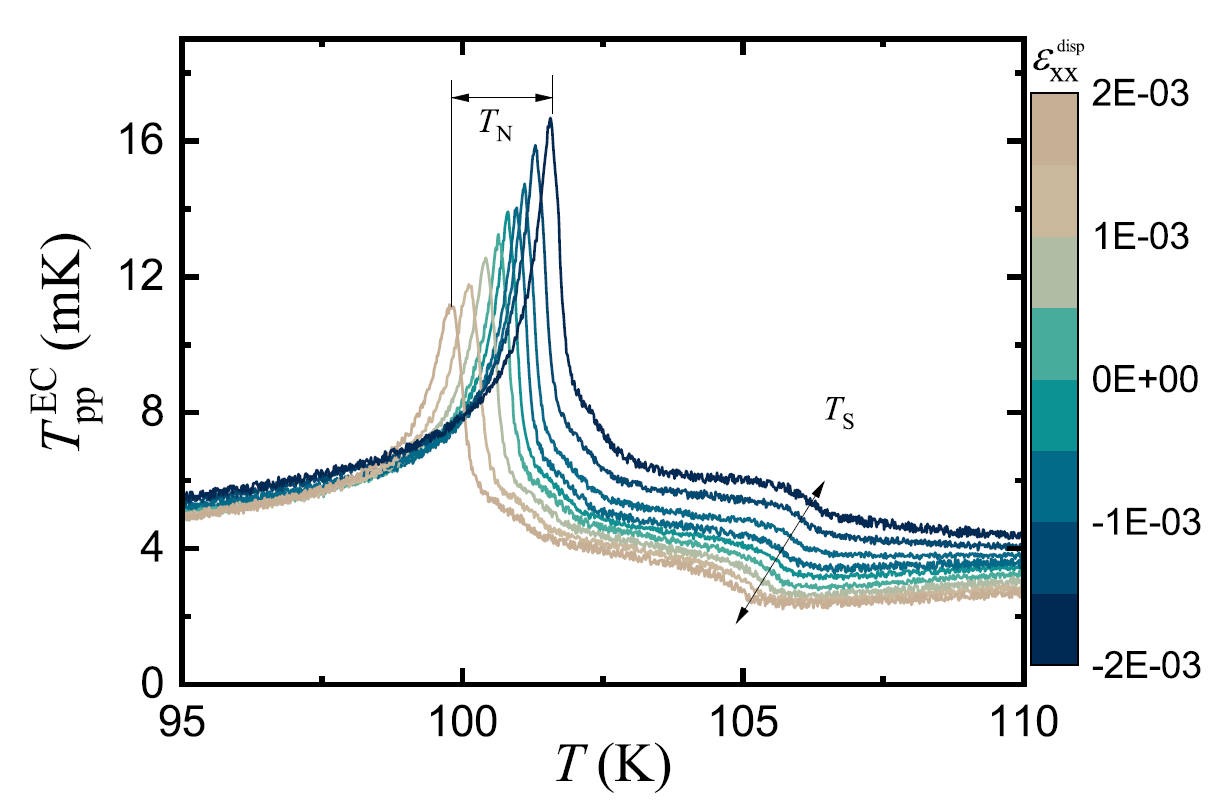}
\caption{\footnotesize Peak to peak amplitude of the elastocaloric temperature oscillation $T_{pp}^{EC}$ \emph{versus} temperature $T$ in which the elascoaloric effect was performed for Ba(Fe$_{0.979}$Co$_{0.021}$)$_2$As$_2$ \cite{Ikeda2019}. The uniaxial strain in the $x$ direction $\varepsilon_{xx}^{disp}$ is also indicated. It is worth mentioning that $\varepsilon_{xx}^{disp} > 0$ means that the specimen length is higher than the original one (decompressed), while $\varepsilon_{xx}^{disp} < 0$ implies that its length was reduced from its original size (compressed). Following discussions in the main text, note that when the sample is compressed $\varepsilon_{xx}^{disp} < 0$, both the antiferromagnetic transition temperature $T_N$ and the temperature in which a nematic/structural phase transition $T_S$ takes place are increased, while for $\varepsilon_{xx}^{disp} > 0$, $T_N$ and $T_S$ decreases. Figure extracted from Ref.\,\cite{Ikeda2019}.}
\label{ironbased}
\end{figure}

As discussed in the last Sections, $\Gamma$ quantifies caloric effects and also the entropy change upon tuning the system close to the critical point. In a step further, we have proposed that $\Gamma$ can also quantify the critical temperature $T_c$ shift in the vicinity of a second-order phase transition upon varying a tuning parameter, which can be pressure, electric or magnetic field, for instance, in an analogous way as the canonical Ehrenfest relation, which is given by \cite{ehrenfest}:
\begin{equation}
\frac{dT_c}{dp} = Tv\left(\frac{\Delta \alpha}{\Delta c_p}\right),
\label{ehrenfest}
\end{equation}
where $\Delta \alpha$ and $\Delta c_p$ refer to the jumps in the thermal expansion and the heat capacity at constant pressure due to a second-order phase transition. Equation\,\ref{ehrenfest} quantifies the shift in $T_c$ upon applying pressure in terms of $\Delta \alpha$ and $\Delta c_p$. For the case of an adiabatic stress application, the entropy variation reads \cite{Ikeda2019}:
\begin{equation}
dS = -\frac{c_{\sigma}^{cr}}{T}\left(\frac{d T_c}{d\epsilon}\right)_T d\epsilon + \frac{c_{\sigma}}{T}dT = 0,
\label{secondorder}
\end{equation}
where $c_{\sigma}^{cr}$ is the critical contribution to the heat capacity at constant stress, $\epsilon = (L - L_0)/L_0$ refers to the strain, which is connected to $\sigma$ through the Young modulus $YM = \sigma/\epsilon$ \cite{mechanicalbehavior}, $L$ and $L_0$ refer to the final and initial sample lengths, respectively. The term of Eq.\,\ref{secondorder} $\frac{c_{\sigma}}{T}dT$ accounts for the entropy variation due to the elastocaloric effect, while $-\frac{c_{\sigma}^{cr}}{T}\left(\frac{d T_c}{d\epsilon}\right)_T$ refers to shifting $T_c$ and thus compensating the contribution to $S$ from the elastocaloric effect in order to keep the global entropy constant. Upon adiabatically compressing the sample, for instance, its temperature is increased contributing to the increase of the system's entropy and thus the term $\frac{c_{\sigma}}{T}dT$ contributes to the enhancement of $S$. To keep the entropy constant, the term on Eq.\,\ref{secondorder} $-\frac{c_{\sigma}^{cr}}{T}\left(\frac{d T_c}{d\epsilon}\right)_T$ must be negative, i.e., $(dT_c/d\epsilon)_T$ must be positive. In other words, upon adiabatically compressing the sample, its $T_c$ is increased. The opposite is also true, when the sample is adiabatically decompressed, its temperature decreases and the term in Eq.\,\ref{secondorder} $\frac{c_{\sigma}}{T}dT$ contributes to decrease $S$. Hence, in order to keep $S$ constant, the term in Eq.\,\ref{secondorder} $-\frac{c_{\sigma}^{cr}}{T}\left(\frac{d T_c}{d\epsilon}\right)_T$ must be positive, contributing thus to compensate the decrease in $S$, so that the total $S$ variation is zero. For such a term to be positive, $(dT_c/d\epsilon)_T$ must be negative, i.e., its $T_c$ is decreased. In summary, upon adiabatically compressing (decompressing) the system, its $T_c$ is increased (decreased). An experimental verification of this analysis is reported in Ref.\,\cite{Ikeda2019} for the iron-based superconductor Ba(Fe$_{0.979}$Co$_{0.021}$)$_2$As$_2$, cf.\,Fig.\,\ref{ironbased}.
Rewriting Eq.\,\ref{secondorder}:
\begin{equation}
\frac{c_{\sigma}^{cr}}{T}\left(\frac{d T_c}{d\epsilon}\right)_T d\epsilon = \frac{c_{\sigma}}{T}dT
\end{equation}
\begin{equation}
\left(\frac{dT}{d\epsilon}\right)_S = \frac{c_{\sigma}^{cr}}{c_{\sigma}}\left(\frac{dT_c}{d\epsilon}\right)_T,
\label{cri}
\vspace{0.2cm}
\end{equation}
which connects the elastocaloric effect, now written in terms of $\epsilon$ instead of $\sigma$, with the $T_c$ shift. Employing the definition of the proposed $\Gamma_{ec}$ (Eq.\,\ref{elasticgruneisen}) and multiplying both sides of Eq.\,\ref{cri} by $1/T$, it now reads:
\begin{equation}
\boxed{\frac{1}{T}\left(\frac{dT}{d\epsilon}\right)_S = \frac{1}{T}\frac{c_{\sigma}^{cr}}{c_{\sigma}}\left(\frac{dT_c}{d\epsilon}\right)_T = \Gamma_{ec}^{cr}},
\label{shifttc}
\end{equation}
where $\Gamma_{ec}^{cr}$ quantifies the shift in $T_c$ due to the elastocaloric effect in terms of the critical contribution of the heat capacity at constant pressure. Equation\,\ref{shifttc} demonstrates that, when $(dT/d\epsilon)_S > 0$, i.e., the temperature of the system is adiabatically increased as a result of a strain (adiabatic compression), $(dT_c/d\epsilon)_T$ is also positive, which means that $T_c$ is shifted to higher temperatures. But when $(dT/d\epsilon)_S < 0$, it means that the temperature of the system is adiabatically decreased due to an adiabatic stretching of the sample, which implies on a shift in $T_c$ to lower temperatures [$(dT_c/d\epsilon)_T$ < 0]. Another important finding of this analysis is that close to a second-order phase transition, the elastocaloric effect is pronounced due to the entropy variation associated with the phase transition close to $T_c$ \cite{elastocaloric}. Note that close to $T_c$, it is possible to write an analogous relation as Eq.\,\ref{dTsigma}, so that \cite{ehrenfest, mori}:
\begin{equation}
dT_c = -T_c\frac{\Delta \alpha_{ij}}{\Delta c_{\sigma}}d\sigma_{ij}
\end{equation}

\begin{equation}
\left(\frac{dT_c}{d\sigma_{ij}}\right)_S = -T_c\frac{\Delta \alpha_{ij}}{\Delta c_{\sigma}},
\label{ehrenfest2}
\end{equation}
which resembles the Ehrenfest relation, cf.\,Eq.\,\ref{ehrenfest} \cite{ehrenfest}. More specifically, Eq.\,\ref{ehrenfest2} can be rewritten as \cite{elastocaloric}:
\begin{equation}
\boxed{\Gamma_{ec}^{cr} = \frac{1}{T_c}\left(\frac{dT_c}{d\sigma_{ij}}\right)_S = -\frac{\Delta \alpha_{ij}}{\Delta c_{\sigma}}},
\label{ehrenfest3}
\end{equation}
which quantifies the $T_c$ shift in terms of an adiabatic stress application in analogy with the canonical Ehrenfest relation \cite{ehrenfest}. Note that $\Delta \alpha_{ij}$ and $\Delta c_{\sigma}$ represent the difference between the ordinary and critical contributions of both $\alpha_{ij}$ and $c_{\sigma}$ close to $T_c$. It is worth mentioning that Eq.\,\ref{ehrenfest3} was already reported in Refs.\,\cite{ehrenfest,mori}, but the connection to $\Gamma_{ec}$ was still lacking. Furthermore, Eq.\,\ref{ehrenfest3} can be extended to the magnetic and electric cases, i.e., to the magnetocaloric and electrocaloric effects \cite{modelsystems,unveiling2020,mrb}, so that \cite{elastocaloric}:
\begin{equation}
\boxed{\Gamma_{mag}^{cr} = \frac{1}{T_c}\left(\frac{dT_c}{dB}\right)_S = -\frac{\Delta \alpha_{B}}{\Delta c_{B}}},
\end{equation}
and
\begin{equation}
\boxed{\Gamma_{ece}^{cr} = \frac{1}{T_c}\left(\frac{dT_c}{dE}\right)_S = -\frac{\Delta \alpha_{E}}{\Delta c_{E}}},
\end{equation}\newline
where $\Gamma_{mag}^{cr}$ and $\Gamma_{ece}^{cr}$ accounts for the critical contributions of the magnetic and electric Gr\"uneisen parameters, which quantifies the magnetocaloric and electrocaloric effects, respectively, and $E$ is the modulus of the electric field. Hence, not only $\Gamma$ quantifies caloric effects, but also it quantifies the $T_c$ shift of a second-order phase transition upon adiabatically varying the control parameter, which can be $\sigma_{ij}$, $B$ or $E$, for instance. Based on such discussions, a generalization of $\Gamma$ in terms of the tuning parameter can be made, which is discussed in the next Section.

The obtained theoretical results discussed in Sections\,\ref{4} and \ref{5} were published in:
\begin{itemize}
\item Lucas Squillante, Isys F. Mello, Gabriel O. Gomes, A.C. Seridonio, Mariano de Souza, Elastocaloric-effect-induced adiabatic magnetization in paramagnetic salts due to the mutual interactions, Scientific Reports \textbf{11}, 9431 (2021).\newline
    \url{https://www.nature.com/articles/s41598-021-88778-4}
\end{itemize}

\subsection{Generalization of the Gr\"uneisen parameter}\label{generalization}

In Section\,\ref{theoreticalbackground}, it has became notorious that Maxwell relations are connected to the definition of $\Gamma$. In this Section, it is demonstrated that in fact $\Gamma$ is naturally derived by making simple mathematical manipulations of the Maxwell relations associated with adiabatic processes. Although quite simple, to the best of our knowledge, this particular point was not properly discussed in the literature until now. Based on the well-known Maxwell relation \cite{Stanleybook}:
\begin{equation}
-\left(\frac{\partial S}{\partial p}\right)_T = \left(\frac{\partial v}{\partial T}\right)_p.
\label{10}
\end{equation}
The term $(\partial S/\partial p)_T$ can be rewritten following Eq.\,\ref{generalization2}, so that:
\begin{equation}
\left(\frac{\partial S}{\partial p}\right)_T = -\left(\frac{\partial T}{\partial p}\right)_S \left(\frac{\partial S}{\partial T}\right)_p.
\label{11}
\end{equation}
Replacing Eq.\,\ref{11} into Eq.\,\ref{10}, it reads:
\begin{equation}
\left(\frac{\partial T}{\partial p}\right)_S \left(\frac{\partial S}{\partial T}\right)_p = \left(\frac{\partial v}{\partial T}\right)_p,
\end{equation}
\begin{equation}
\left(\frac{\partial T}{\partial p}\right)_S = \frac{\left(\frac{\partial v}{\partial T}\right)_p}{\left(\frac{\partial S}{\partial T}\right)_p}.
\label{12}
\end{equation}\newline
Considering $c_p = T/N(\partial S/\partial T)_p$ and $\alpha_p = 1/v(\partial v/\partial T)_p$, Eq.\,\ref{12} becomes:\newline
\begin{equation}
\left(\frac{\partial T}{\partial p}\right)_S = \frac{\alpha_p v}{\frac{c_p N}{T}} = \frac{\alpha_p v T}{c_p N} = \frac{\alpha_p T v_m}{c_p}.
\label{13}
\end{equation}
Dividing both sides of Eq.\,\ref{13} by $1/T$, it is given by:\newline
\begin{equation}
\boxed{\frac{1}{v_mT}\left(\frac{\partial T}{\partial p}\right)_S = \frac{\alpha_p}{c_p} = \Gamma},
\end{equation}\newline
which is the definition of $\Gamma$, following Eq.\,\ref{gamma3}. As a consequence of the thermodynamic relations, it can be seen that the definition of $\Gamma$ is naturally encoded in the Maxwell relation of Eq.\,\ref{10}. A similar analysis can be made for $\Gamma_{mag}$ employing the Maxwell relation \cite{Stanleybook}:
\begin{equation}
\left(\frac{\partial S}{\partial B}\right)_T = \left(\frac{\partial M}{\partial T}\right)_B.
\label{14}
\end{equation}
In the case of thermodynamic relations for magnetic systems, it is widely discussed the following analogy \cite{Stanleybook}:
\begin{equation}
v \rightarrow -M,
\label{15}
\end{equation}
\begin{equation}
p \rightarrow B.
\end{equation}
The minus sign of the analogy between $v$ and $M$ in Eq.\,\ref{15} is because $M$ increases (decreases) due to the increase (decrease) of $B$, while $v$ increases (decreases) due to the decrease (increase) of $p$ \cite{Stanleybook}. In other words, $M$ and $B$ are directly proportional while $p$ and $v$ are inversely proportional. Therefore, employing such an analogy, Eq.\,\ref{generalization2} can be rewritten for a magnetic system by \cite{Stanleybook}:
\begin{equation}
\left(\frac{\partial T}{\partial B}\right)_S \left(\frac{\partial S}{\partial T}\right)_B \left(\frac{\partial B}{\partial S}\right)_T = -1,
\end{equation}
\begin{equation}
\left(\frac{\partial S}{\partial B}\right)_T = -\left(\frac{\partial T}{\partial B}\right)_S \left(\frac{\partial S}{\partial T}\right)_B.
\label{16}
\vspace{0.2cm}
\end{equation}
Replacing Eq.\,\ref{16} into Eq.\,\ref{14}:
\begin{equation}
-\left(\frac{\partial T}{\partial B}\right)_S \left(\frac{\partial S}{\partial T}\right)_B = \left(\frac{\partial M}{\partial T}\right)_B,
\end{equation}
\begin{equation}
\left(\frac{\partial T}{\partial B}\right)_S  = -\frac{\left(\frac{\partial M}{\partial T}\right)_B}{\left(\frac{\partial S}{\partial T}\right)_B}.
\label{17}
\end{equation}
Consdering that $c_B = T(\partial S/\partial T)_B$ \cite{Stanleybook} in Eq.\,\ref{17}, it reads:
\begin{equation}
\left(\frac{\partial T}{\partial B}\right)_S = -\frac{\left(\frac{\partial M}{\partial T}\right)_B}{\frac{c_B}{T}} = -T\frac{\left(\frac{\partial M}{\partial T}\right)_B}{c_B}.
\label{18}
\end{equation}
Multiplying both sides of Eq.\,\ref{18} by $1/T$:
\begin{equation}
\boxed{\frac{1}{T}\left(\frac{\partial T}{\partial B}\right)_S = -\frac{\left(\frac{\partial M}{\partial T}\right)_B}{c_B} = \Gamma_{mag}}.
\label{19}
\end{equation}
Equation\,\ref{19} is the definition of $\Gamma_{mag}$, following Eqs.\,\ref{mgp} and \ref{mgp2}.

Furthermore, a similar analysis can be made for the electric case where the tuning parameter is the electric field $E$. Given the Maxwell relation \cite{Stanleybook,reese}:
\begin{equation}
\left(\frac{\partial S}{\partial E}\right)_T = \left(\frac{\partial P}{\partial T}\right)_E,
\label{20}
\end{equation}
where $P$ is the electric polarization, the term $(\partial S/\partial B)_T$ can be rewritten following the relation \cite{Stanleybook}:\newline
\begin{equation}
\left(\frac{\partial S}{\partial T}\right)_E \left(\frac{\partial T}{\partial E}\right)_S \left(\frac{\partial E}{\partial S}\right)_T = -1 \Rightarrow \left(\frac{\partial S}{\partial E}\right)_T = -\left(\frac{\partial S}{\partial T}\right)_E \left(\frac{\partial T}{\partial E}\right)_S.
\label{21}
\end{equation}\newline
Replacing Eq.\,\ref{21} into Eq.\,\ref{20}:
\begin{equation}
-\left(\frac{\partial S}{\partial T}\right)_E \left(\frac{\partial T}{\partial E}\right)_S = \left(\frac{\partial P}{\partial T}\right)_E \Rightarrow \left(\frac{\partial T}{\partial E}\right)_S = -\frac{\left(\frac{\partial P}{\partial T}\right)_E}{\left(\frac{\partial S}{\partial T}\right)_E}.
\label{22}
\end{equation}
Considering that the heat capacity at constant $E$ is given by $c_E = T(\partial S/\partial T)_E$ \cite{Stanleybook}, Eq.\,\ref{22} now reads:
\begin{equation}
\left(\frac{\partial T}{\partial E}\right)_S = -\frac{\left(\frac{\partial P}{\partial T}\right)_E}{\frac{c_E}{T}} = -T\frac{\left(\frac{\partial P}{\partial T}\right)_E}{c_E}.
\label{23}
\end{equation}
Multiplying both sides of Eq.\,\ref{23} by $1/T$:
\begin{equation}
\boxed{\frac{1}{T}\left(\frac{\partial T}{\partial E}\right)_S = -\frac{\left(\frac{\partial P}{\partial T}\right)_E}{c_E} = \Gamma_E},
\end{equation}
which is the definition of the electric Gr\"uneisen parameter $\Gamma_E$ proposed for the first time in the literature in Ref.\,\cite{jap}. At this point, it is worth mentioning that, in order to avoid confusion, note that lower case $p$ refers to pressure, while the capital letter $P$ is electric polarization. Last, but not least, the polar Gr\"uneisen parameter $\Gamma_P$, proposed for the first time in the literature in Ref.\,\cite{mrb} can be derived from the Maxwell relation \cite{Stanleybook,reese}:
\begin{equation}
\left(\frac{\partial S}{\partial P}\right)_T = -\left(\frac{\partial E}{\partial T}\right)_P
\label{24}
\end{equation}
Analogously to the previously discussed cases, the term $(\partial S/\partial P)_T$ can be rewritten employing the thermodynamic relation \cite{Stanleybook}:
\begin{equation}
\left(\frac{\partial S}{\partial T}\right)_P \left(\frac{\partial T}{\partial P}\right)_S \left(\frac{\partial P}{\partial S}\right)_T = -1 \Rightarrow \left(\frac{\partial S}{\partial P}\right)_T = -\left(\frac{\partial S}{\partial T}\right)_P \left(\frac{\partial T}{\partial P}\right)_S.
\label{25}
\vspace{0.2cm}
\end{equation}
Replacing Eq.\,\ref{25} into Eq.\,\ref{24}:
\begin{equation}
-\left(\frac{\partial S}{\partial T}\right)_P \left(\frac{\partial T}{\partial P}\right)_S = -\left(\frac{\partial E}{\partial T}\right)_P,
\end{equation}
\begin{equation}
\left(\frac{\partial T}{\partial P}\right)_S = \frac{\left(\frac{\partial E}{\partial T}\right)_P}{\left(\frac{\partial S}{\partial T}\right)_P}.
\label{26}
\end{equation}
Assuming that the heat capacity at constant $P$ is $c_P = T(\partial S/\partial T)_P$ \cite{Stanleybook}, Eq.\,\ref{26} reads:
\begin{equation}
\left(\frac{\partial T}{\partial P}\right)_S = \frac{\left(\frac{\partial E}{\partial T}\right)_P}{\frac{c_P}{T}} = T\frac{\left(\frac{\partial E}{\partial T}\right)_P}{c_P}.
\label{27}
\end{equation}
Multiplying both sides of Eq.\,\ref{27} by $1/T$, it reads:
\begin{equation}
\boxed{\frac{1}{T}\left(\frac{\partial T}{\partial P}\right)_S = \frac{\left(\frac{\partial E}{\partial T}\right)_P}{c_P} = \Gamma_P},
\end{equation}
which is the definition of $\Gamma_P$ \cite{mrb}.

It becomes evident that the various forms of the Gr\"uneisen parameter, namely magnetic, electric, polar, and the Gr\"uneisen ratio, are naturally embedded in the widely employed Maxwell relations. Table\,\,\ref{tablegruneisen} summarizes the result discussed in this regard.
\begin{table}[!h]
\centering
\begin{tabular}{|c|c|}
\hline\hline
Maxwell-Relation & Gr\"uneisen Parameter  \\ \hline\hline
$\left(\frac{\partial S}{\partial p}\right)_T = -\left(\frac{\partial v}{\partial T}\right)_p$ & $\Gamma = -\frac{1}{Tv_m}\frac{\left(\frac{\partial S}{\partial p}\right)_T}{\left(\frac{\partial S}{\partial T}\right)_p} = \frac{1}{T v_m}\left(\frac{\partial T}{\partial p}\right)_S$     (\textbf{BC})               \\ \hline
$\left(\frac{\partial S}{\partial B}\right)_T = \left(\frac{\partial M}{\partial T}\right)_B$ & $\Gamma_{mag} = -\frac{1}{T}\frac{\left(\frac{\partial S}{\partial B}\right)_T}{\left(\frac{\partial S}{\partial T}\right)_B} = \frac{1}{T}\left(\frac{\partial T}{\partial B}\right)_S$ \hspace{0.15cm}(\textbf{MC})                     \\ \hline
$\left(\frac{\partial S}{\partial E}\right)_T = \left(\frac{\partial P}{\partial T}\right)_E$ & $\Gamma_E = -\frac{1}{T}\frac{\left(\frac{\partial S}{\partial E}\right)_T}{\left(\frac{\partial S}{\partial T}\right)_E} = \frac{1}{T}\left(\frac{\partial T}{\partial E}\right)_S$         \hspace{0.3cm} (\textbf{EC})           \\ \hline
$\left(\frac{\partial S}{\partial P}\right)_T = -\left(\frac{\partial E}{\partial T}\right)_P$ & $\Gamma_P = -\frac{1}{T}\frac{\left(\frac{\partial S}{\partial P}\right)_T}{\left(\frac{\partial S}{\partial T}\right)_P} = \frac{1}{T}\left(\frac{\partial T}{\partial P}\right)_S$              \hspace{0.45cm}(\textbf{PC})        \\ \hline
\end{tabular}
\caption{\footnotesize The Gr\"uneisen ratio $\Gamma$, magnetic $\Gamma_{mag}$, electric $\Gamma_E$, and polar $\Gamma_P$ Gr\"uneisen parameters derived from the corresponding Maxwell relations, which quantify the barocaloric (BC), magnetocaloric (MC), electrocaloric (EC), and polarcaloric (PC) effects, respectively. Table extracted from Ref.\,\cite{mrb}.}
\label{tablegruneisen}
\end{table}

Hence, a generalized form of the Gr\"uneisen ratio $\Gamma_g$ is proposed in terms of an adiabatic change of an arbitrary tuning parameter $g$, so that \cite{mrb}:
\begin{equation}
\boxed{\Gamma_g = -\frac{1}{T}\frac{\left(\frac{\partial S}{\partial g}\right)_T}{\left(\frac{\partial S}{\partial T}\right)_g} = \frac{1}{T}\left(\frac{\partial T}{\partial g}\right)_S}.
\end{equation}
Therefore, the magnitude of any caloric effect can be quantified by $\Gamma_g$ depending if $g$ is $p$, $E$, $\sigma$ or $B$ \cite{mrb}. In the next Section, the maximization of caloric effects in terms of entropy arguments and the Gr\"uneisen parameter is discussed.

\subsection{Maximization of caloric effects}\label{maximization}

\begin{figure}[h!]
\centering
\includegraphics[width=\textwidth]{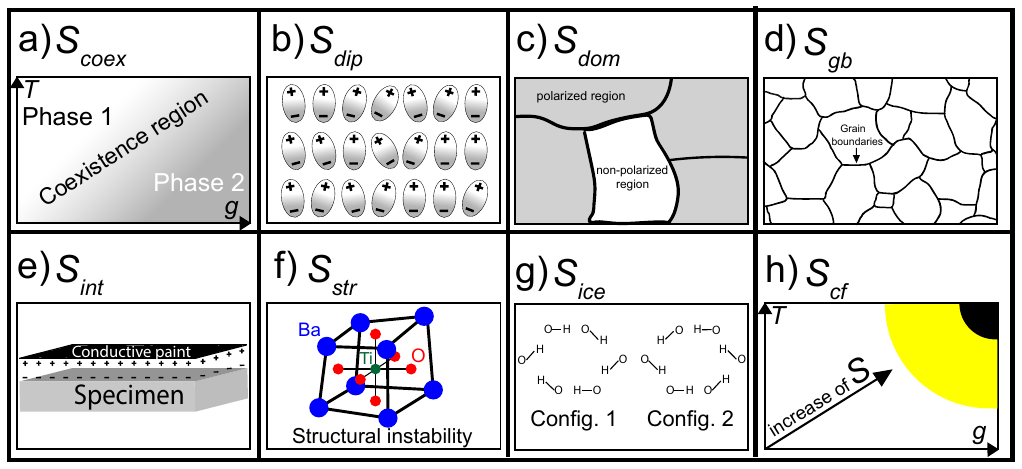}
\caption{\footnotesize  Schematic representation of several contributions to the total entropy: a) coexistence region between two arbitrary phases 1 and 2, b) distinct dipolar configurations in a ferroelectric, c) presence of domain walls in a relaxor-like ferroelectric, d) presence of grain boundaries in poly-crystalline or powdered materials, e) interface polarization due to the Maxwell-Wagner effect \cite{maxwell}, f) structural instabilities, as it occurs, for instance, in BaTiO$_3$ BCC unit cell \cite{trautmann}, g) residual entropy regarding two distinct molecular configurations \cite{pauling}, h) critical fluctuations associated with the proximity to the critical point. Figure extracted from Ref.\,\cite{mrb}.}
\label{entropycontributions}
\end{figure}

The quantification of the effectiveness of cooling devices that employ the concept of decreasing temperature due to an adiabatic change of a control parameter is key for the optimization of caloric effects. As a matter of fact, the traditional refrigerators that are vastly used in our daily lives incorporate an adiabatic compression of the refrigerant gas in order to cool its interior \cite{callen}. However, other types of cooling devices that are more efficient and eco-friendly has been widely investigated, such as devices that employ the concept of the electrocaloric effect to cool down, which refers to the temperature decrease due to an adiabatic removal of $E$ \cite{mrb}. It is worth mentioning that the search for more sustainable and eco-friendly cooling devices is in line with some of the 17 goals of the United Nations, namely: 7) industry, innovation, and infrastructure, 11) sustainable cities and communities, 12) responsible consumption and production, and 13) climate action \cite{onu}. In order to quantify the temperature change $\Delta T$ in caloric devices, it is usually employed the so-called caloric strength $CS = \Delta T/ \Delta g$, where $\Delta g$ is the adiabatic change of the tuning parameter and $\Delta T$ the corresponding adiabatic temperature change. Essentially, $CS$ quantifies the amount of $\Delta g$ needed to produce an adiabatic $\Delta T$. As previously discussed, $\Gamma$ is not solely employed to investigate phase transitions and critical phenomena, but it also quantifies caloric effects \cite{mrb}, cf.\,Table\,\ref{tablegruneisen}. However, when quantifying caloric effects in terms of the Gr\"uneisen parameters, namely $\Gamma$, $\Gamma_{mag}$, and $\Gamma_E$, for instance, additional physical information can be extracted in terms of the entropy variation of the system upon tuning $g$.
\begin{figure}[t]
\centering
\includegraphics[width=0.7\textwidth]{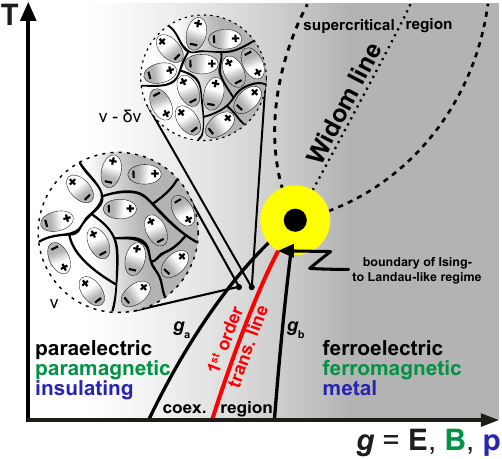}
\caption{\footnotesize Schematic representation of a generic phase diagram temperature $T$ \emph{versus} tuning parameter $g$, which can be electric field $E$, magnetic field $B$, or pressure $p$, for instance. Above the critical point, there is a supercritical region and its corresponding Widow line. Below the critical point, black bullet, there is a coexistence region between a paraelectric and ferroelectric, paramagnetic and ferromagnetic, or insulating and metal phases. The coexistence region is delimited by the spinodal lines $g_a$ and $g_b$. Within the phases coexistence region, there is a first-order transition line (red color). The yellow region around the critical point represents the boundary of the transition from Ising- to Landau-like regime \cite{garst2012}. The case of electric dipoles separated by domain walls in a ferroelectric phase is depicted in an initial volume $v$ and, upon applying pressure, its volume is decreased by $\delta v$ and the entropy is increased. Figure extracted from Ref.\,\cite{mrb}.}
\label{phasediagram}
\end{figure}
Hence, it becomes natural to investigate the optimization of caloric effects in terms of entropy arguments based on the various definitions of the Gr\"uneisen parameter, cf.\,Table\,\ref{tablegruneisen}. The idea is that the more contributions to $S$ are associated with a particular system, more expressive $\Delta T$ is going to be when a caloric effect is performed, i.e., when a tuning parameter is adiabatically varied. Interestingly, as widely discussed in the literature, there is an intrinsic accumulation of $S$ close to both classical \cite{Stanleybook} and quantum critical points \cite{qimiao2011,rost2009} due to the presence of competing phases in their vicinities. Hence, the high $S$ accumulation in the proximity to a critical point alone is key to enhance the performance of caloric effects, cf.\,Fig.\,\ref{phasediagram}. Also, the phases coexistence region in the vicinities of the first-order transition line close to the critical point plays a crucial role on optimizing caloric effects since the mixture of phases contributes to the enhancement of $S$. However, there are other key ingredients in terms of material design that can be incorporated to a system in order to maximize $\Delta T$ due to the presence of several additional contributions to $S$, cf.\,Fig.\,\ref{entropycontributions}.
Hence, a possible generalized expression for $S$ regarding a candidate system that can attain a giant caloric effect is given by \cite{mrb}:
\begin{equation}
S_{total} = (S_{coex} + S_{dip} + S_{dom} + S_{gb} + S_{int} + S_{str} + S_{ice} + S_{cf}),
\end{equation}
where $S_{total}$ is the total entropy, $S_{coex}$, $S_{dip}$, $S_{dom}$, $S_{gb}$, $S_{int}$, $S_{str}$, $S_{ice}$, and $S_{cf}$ are the contributions to $S_{total}$ regarding, respectively, the phases coexistence region near the critical point, different dipolar configurations, the presence of domain walls, the presence of grain boundaries in poly-crystalline or powdered materials, interface phenomena, such as Maxwell-Wagner polarization \cite{maxwell}, structural instabilities \cite{trautmann}, residual entropy in a solid \cite{pauling}, and the critical fluctuations associated with the proximity to the critical point. The various contributions to $S_{total}$ are schematically depicted in Fig.\,\ref{entropycontributions}. Such a proposal of the maximization of caloric effects in terms of entropy arguments enables the interpretation of the recently reported large electrocaloric effect at room-temperature in a multilayered capacitor \cite{nair}. Also, based on this proposal, a material can be designed to incorporate as much contributions to $S$ as it can to attain giant caloric effects. Yet, a connection between quantum paraelectricity and enhanced caloric effects was proposed, which is discussed in the next Section.

\subsection{Electric Gr\"uneisen parameter and quantum paraelectricity}\label{quantumparaelectric}

It is well-known from textbooks that the dielectric constant $\epsilon$ is enhanced upon approaching the Curie temperature and vanishes below it indicating a phase transition from a paraelectric to a ferroelectric phase \cite{maxwell}. However, some systems present a quite unusual behaviour in $\epsilon$ below the Curie temperature, such as SrTiO$_3$, where $\epsilon$ enhances close to the Curie temperature, but does not vanish. Instead, $\epsilon$ stabilizes in a plateau below it \cite{burkard}. This behaviour is associated with the presence of a quantum critical point in the system, the so-called quantum paraelectric behaviour \cite{rowley}.
\begin{figure}[h!]
\centering
\includegraphics[width=0.9\textwidth]{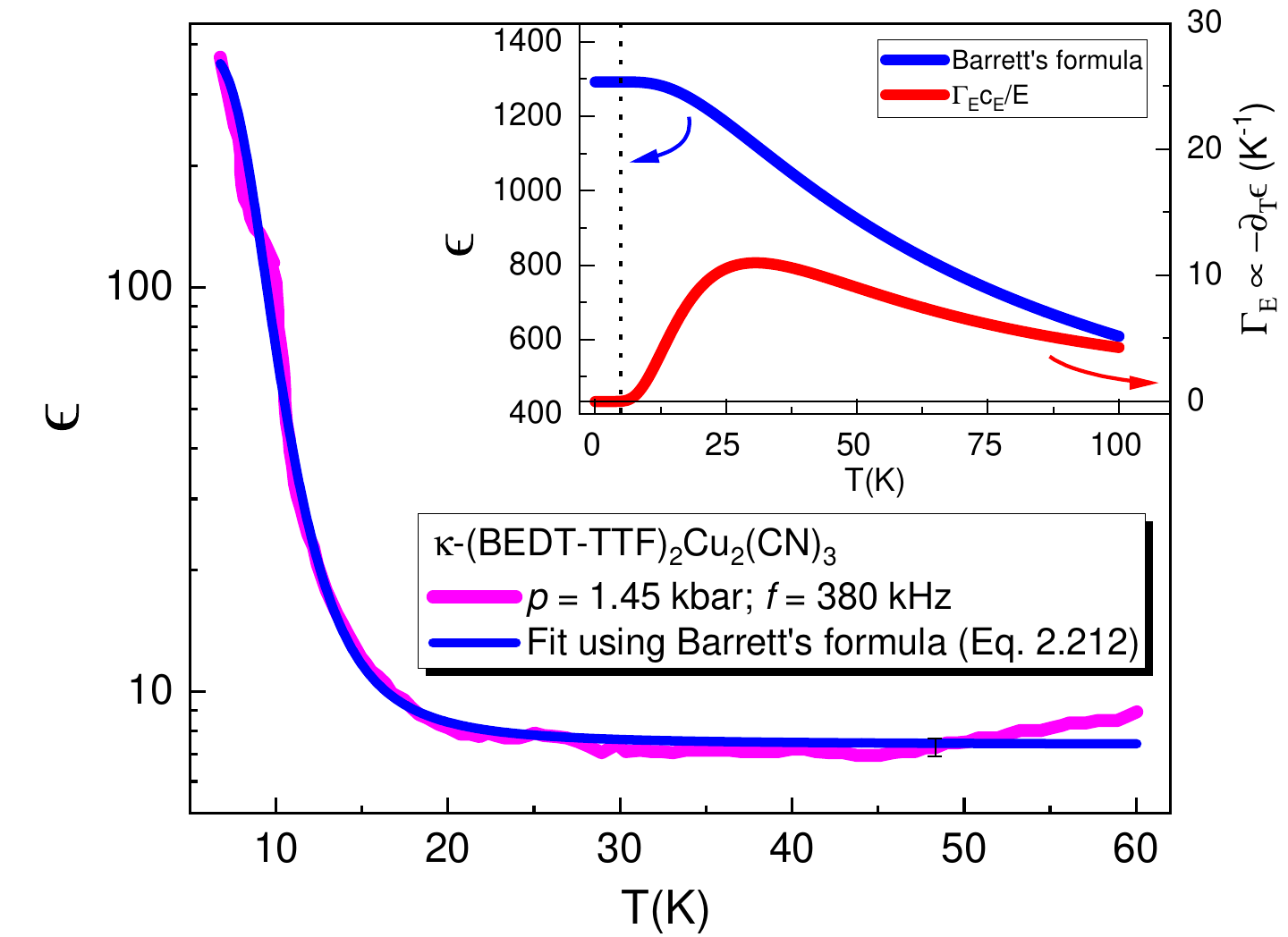}
\caption{\footnotesize Main panel: experimental dielectric constant $\epsilon$ results as a function of temperature $T$ for the spin-liquid candidate $\kappa$-(BEDT-TTF)$_2$Cu$_2$(CN)$_3$ (pink line) under a pressure of $p$ = 1.45\,kbar and frequency $f$ = 380\,kHz. Data extracted from Ref.\,\cite{dressel2021}. The blue solid line represents a fitting of the experimental data employing Barrett's formula \cite{barrett} (Eq.\,\ref{barrettformula}). The error bar represents a 5\% error in tracing the experimental data. Inset: dielectric constant $\epsilon$ \emph{versus} temperature employing the Barrett's formula (blue line) with parameters $A = 0$, $m$ = (8.4$\times$10$^4$)\,K (CGS units), $T_1$ = 60\,K, and $T_0$ = $-$35\,K \cite{barrett}. The temperature derivative of electric Gr\"uneisen parameter $\Gamma_E$ multiplied by the ratio between the heat capacity at constant electric field $c_E$ and the electric field $E$ (red line) showing that close to the $\epsilon$ plateau, the electrocaloric effect is enhanced. Figure extracted from Ref.\,\cite{mrb}.}
\label{barrett}
\end{figure}
In systems presenting such a quantum paraelectric behaviour, it was proposed that $\epsilon$ has the following $T$-dependence \cite{barrett}:
\begin{equation}
\epsilon = A + \frac{m}{\left(\frac{T_1}{2}\right)\coth{\left(\frac{T_1}{2T}\right)}-T_0},
\label{barrettformula}
\end{equation}
where $A$ is a non-universal constant, $T_0$ is a parameter related either to the ferroelectric ($T_0 >$ 0) or antiferroelectric ($T_0 <$ 0) effective dipolar interaction, $T_1$ is the temperature below which quantum fluctuations dominate, and $m = n_{ed}\mu_{ed}^{2}/k_B$, where $n_{ed}$ is the electric-dipole density, and $\mu_{ed}$ refers to the local electric-dipole moment. The typical behaviour of $\epsilon$ given by Barrett's formula \cite{barrett} is depicted in the inset of Fig.\,\ref{barrett} (blue line). Furthermore, $T_1$ can be defined as the balance between thermal and the vibrational energies associated with a harmonic oscillator in an electric field by the form $T_1 = h\nu/k_B$ \cite{barrett}, where $\nu$ is the oscillation frequency. Essentially, in this analysis the response of the electric-dipoles to an external $E$ is treated in the harmonic approximation. The quantum critical regime is set when the energy associated with such quantum fluctuations overcomes the thermal energy. In general terms, the fingerprint of a quantum paraelectric phase is a plateau in $\epsilon$, i.e., the quantum fluctuations affect the ferroelectric phase and instead of presenting a vanishing $\epsilon$ for $T \rightarrow $ 0, as it occurs in classical ferroelectric-type transitions, $\epsilon$ remains constant. It is worth mentioning that such a plateau will be observed only when the ferroelectric transition temperature is close to $T_1$, otherwise the quantum critical fluctuations will not affect $\epsilon$ \cite{barrett}. Note that the plateau in $\epsilon$ occurs because the so-called soft modes associated with the instability of the ferroelectric transition do not vanish at the Curie temperature. Instead, the soft modes survive below the Curie temperature and, as a consequence, the quantum fluctuations stabilizes the ferroelectric transition and ``locks'' $\epsilon$ in a finite value instead of presenting a vanishing $\epsilon$ below the Curie temperature as it is usually observed for ferroelectrics \cite{rubio2021}. As proposed in Ref.\,\cite{jap}, $\Gamma_E$ is given by:
\begin{equation}
\Gamma_E = -\frac{\left(\frac{\partial P}{\partial T}\right)_E}{c_E}.
\end{equation}
Assuming a linear system, i.e., a system which its electric polarization increases linearly with $E$ so that $P = \epsilon E$, $\Gamma_E$ can be rewritten as \cite{jap}:
\begin{equation}
\Gamma_E = -\frac{\left(\frac{\partial \epsilon E}{\partial T}\right)_E}{c_E} = -E\frac{\left(\frac{\partial \epsilon}{\partial T}\right)_E}{c_E}.
\label{gammae}
\end{equation}
Employing the temperature derivative of $\epsilon$ following Barrett's formula in Eq.\,\ref{barrettformula}, $\Gamma_E$ reads \cite{mrb}:
\begin{equation}
\Gamma_E = \frac{\varepsilon_0 E}{c_E}\frac{m {T_1}^2}{T^2 \left[T_1 \cosh
   \left(\frac{T_1}{2 T}\right)-2
   T_0 \sinh \left(\frac{T_1}{2
   T}\right)\right]^2},
   \label{Barret}
\end{equation}
where $\epsilon_0$ is the vacuum permittivity. Upon analysing the behaviour of $\Gamma_E$ close to the plateau of $\epsilon$, cf.\,inset of Fig.\,\ref{barrett} (red line), it is clear that $\Gamma_E$ is maximized due to the presence of the plateau on $\epsilon$. This indicates that systems presenting such a quantum paralectric behaviour, there is an intrinsic enhanced electrocaloric effect in the regime where quantum fluctuations starts to play a key role since $\Gamma_E$, which quantifies the electrocaloric effect, is enhanced close to the plateau of $\epsilon$, cf.\,inset of Fig.\,\ref{barrett}. Thus, not only $\Gamma_E$ can be employed, together with $\epsilon$ measurements, to detect and probe quantum paraelectricity, but also the analysis in terms of $\Gamma_E$ predicts a giant electrocaloric effect right at the onset of quantum paraelectricity.

The theoretical results discussed in Sections\,\ref{generalization}, \ref{maximization} and \ref{quantumparaelectric} were published in:
\begin{itemize}
\item Lucas Squillante, Isys F. Mello, Gabriel O. Gomes, A.C. Seridonio, Mariano de Souza, Giant caloric effects close to \emph{any} critical end point, Materials Research Bulletin \textbf{142}, 111413 (2021).\newline
    \url{https://www.sciencedirect.com/science/article/abs/pii/S0025540821002105}
\end{itemize}

\subsection{Gr\"uneisen parameter and the expansion of the universe}\label{gruneisencosmology}

The accelerated expansion of the universe was originally probed via type Ia Supernova measurements in 1998 \cite{refs}. Although intensive investigations, the physical mechanism causing the accelerated expansion of the universe is still a matter of debate in the literature \cite{Kamionkowski2019}. The Planck collaboration has provided precise measurements of cosmological parameters in recent years \cite{Planck2020}. The so-called $\Lambda$-Cold Dark Matter (CDM) model is one of the most prominent models in the literature to account for the expansion of the universe \cite{Bonometto2006,Wu2007,Grande2007,Neupane2008,CDM,Weinberg1972}. Essentially, it considers that the universe is composed by dark energy (DE), dark matter (DM), and ordinary matter. Nowadays, the universe is assumed to be dominated by vacuum energy $\Lambda$, i.e., dark energy, with a constant energy density $\rho$ \cite{Peebles1984}. Alternative models using the approach of a dark fluid have also been proposed to unify DE and DM \cite{Chavanis}. Various EOS have been suggested, such as the Chaplygin one \cite{Kamenshchik2001}.
\begin{figure}[h!]
\centering
\includegraphics[width=\textwidth]{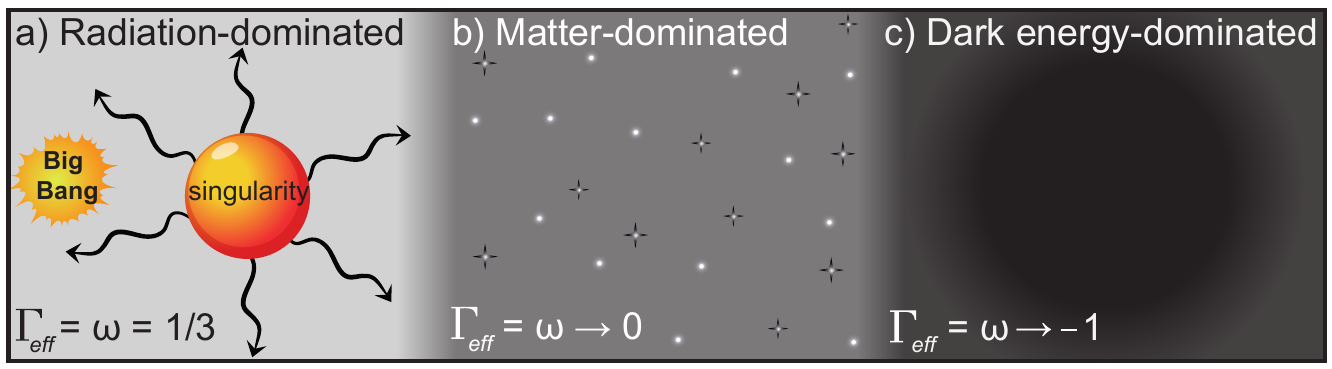}
\caption{\footnotesize  Schematic representation of the various eras of the universe, namely a) radiation-, b) matter-, and c) DE-dominated eras. The values of $\omega$, namely $\Gamma_{eff}$, for each particular era is also shown. Figure extracted from Ref.\,\cite{meetseinstein}.}
\label{cosmology}
\end{figure}
However, the Chaplygin EOS conflicts with experimental observations \cite{Zhu2004}, being already surpassed. Essentially, the available experimental data regarding the dimensionless coordinate distances and the X-ray gas mass fractions do not follow the EOS for the Chaplygin gas within a 99\% confidence level \cite{Kamenshchik2001}. In this context, the discrepancy between cosmological expansion models and the expansion mechanisms throughout different eras still remains. Hence, there is a need to come out with new ideas and approaches in order to bring new physical elements to such a discussion in modern Cosmology \cite{review}. In this context, the unprecedented connection between well-established concepts of condensed matter Physics and Cosmology in terms of $\Gamma_{eff}$ and $\Gamma$ can shed light into the discrepancies regarding the description of an expanding universe. As discussed in Section\,\ref{friedmanequations}, the universe is modeled as a perfect fluid, where the radiation, matter, and DE dominated eras are characterized by $\omega = 1/3, 0,$ and $-1$, respectively \cite{Weinberg1972}. It has been postulated that the late stage of the universe is predominantly composed of vacuum energy ($\omega \rightarrow -1$), or DE, with $\rho \propto a(t)^{-3(1 + \omega)}$, and $-1 \le \omega <0$ for various eras, where $a$ is the scale factor \cite{Huterer2008}. Remarkably, the EOS for a perfect fluid $p = \omega\rho$ is similar to the well-known Mie-Gr\"uneisen EOS, i.e. $p_a = \Gamma_{eff}E_a/v$, where $p_a$ and $E_a$ are the contributions of atomic vibrations to $p$ and energy, respectively, and is widely used in the field of Condensed Matter Physics for solids and fluids \cite{Gruneisen1926,Shalom1986,Chernyshev2011}. In the case of fluids, $\Gamma_{eff}$ links $p$ and the kinetic/thermal energy of the particles, while in solids, $\Gamma_{eff}$ measures the ratio of pressure to energy when $v$ changes with a tuning parameter, such as temperature or stress. Hence, $\Gamma_{eff}$ can be straightforwardly recognized as $\omega$ \cite{meetseinstein}, providing an alternative interpretation to $\omega$ besides being comprised to a numerical value. This is one of the key results of this Section to be discussed into more details in the following. Now, considering Friedmann equations \cite{friedman}, the relationship between $\rho$, $p$, and specific entropy $\sigma_S$ can be expressed as \cite{Weinberg1972,phdthesis}:
\begin{equation}
nTd\sigma_S = d\rho - \frac{p + \rho}{n}dn,
\label{gibbslaw}
\end{equation}
where $n = N/v$ is the number of particles $N$ per $v$. Considering that $\rho = \rho(n, T)$, we write:
\begin{equation}
d\rho = \left(\frac{\partial\rho}{\partial T}\right)_n dT + \left(\frac{\partial\rho}{\partial n}\right)_T dn.
\label{gibbslaw2}
\end{equation}
Rewriting Eq.\,\ref{gibbslaw}:
\begin{equation}
d\sigma_S = \frac{1}{nT}\left(d\rho - \frac{\rho + p}{n}dn\right),
\end{equation}
and employing Eq.\,\ref{gibbslaw2}, we have:
\begin{equation}
d\sigma_S = \frac{1}{nT}\left[\left(\frac{\partial\rho}{\partial T}\right)_n dT + \left(\frac{\partial\rho}{\partial n}\right)_T dn - \left(\frac{\rho + p}{n}\right)dn\right].
\label{exactderivative}
\end{equation}
Since entropy is a state function, i.e., $d\sigma_S$ must be exact. Considering a general differential function $dx = Jdy + Hdz$, if $dx$ is exact it means that $(\partial J/\partial z)_y = (\partial H/\partial y)_z$ \cite{ralph}. Therefore, from Eq.\,\ref{exactderivative} we have:
\begin{equation}
\frac{\partial}{\partial T}\left\{\frac{1}{nT}\left[\left(\frac{\partial\rho}{\partial n}\right)_T - \left(\frac{\rho+p}{n}\right)\right]\right\}_n = \frac{\partial}{\partial n}\left[\frac{1}{nT}\left(\frac{\partial\rho}{\partial T}\right)_n\right]_T,
\end{equation}
\begin{equation}
\left\{\frac{\partial}{\partial T}\left[\frac{1}{nT}\left(\frac{\partial\rho}{\partial n}\right)_T - \frac{1}{nT}\left(\frac{\rho+p}{n}\right)\right]\right\}_n = -\frac{1}{n^2 T}\left(\frac{\partial\rho}{\partial T}\right)_n,
\end{equation}
\begin{equation}
-\frac{1}{nT^2}\left(\frac{\partial\rho}{\partial n}\right)_T - \overbrace{\frac{\partial}{\partial T}\left[\frac{1}{nT}\left(\frac{\rho + p}{n}\right)\right]}^\Upsilon = -\frac{1}{n^2 T}\left(\frac{\partial\rho}{\partial T}\right)_n.
\label{s7}
\end{equation}\newline
Since $\rho = \rho(T, n)$ and $p = p(T, n)$, the term $\Upsilon$ is given by:
\begin{equation}
\Upsilon = \frac{\partial}{\partial T}\left[\frac{1}{nT}\left(\frac{\rho + p}{n}\right)\right] = \frac{1}{n^2}\left[\frac{\partial}{\partial T}\left(\frac{\rho + p}{T}\right)\right] = \frac{1}{n^2}\left[\frac{\partial}{\partial T}\left(\frac{\rho}{T} + \frac{p}{T}\right)\right],
\end{equation}
\begin{equation}
\Upsilon = \frac{1}{n^2}\left[\frac{\partial}{\partial T}\left(\frac{\rho}{T}\right) + \frac{\partial}{\partial T}\left(\frac{p}{T}\right)\right] = \frac{1}{n^2}\left[\frac{T\left(\frac{\partial\rho}{\partial T}\right)_n - \rho}{T^2} + \frac{T\left(\frac{\partial p}{\partial T}\right)_n - p}{T^2}\right],
\end{equation}
which in turn can be rewritten as:
\begin{equation}
\Upsilon = \frac{1}{n^2T^2}\left[T\left(\frac{\partial \rho}{\partial T}\right)_n - \rho + T\left(\frac{\partial p}{\partial T}\right)_n - p\right].
\label{upsilon}
\end{equation}
Replacing $\Upsilon$ from Eq.\,\ref{upsilon} into Eq.\,\ref{s7}, we have:
\begin{equation}
-\frac{1}{nT^2}\left(\frac{\partial \rho}{\partial n}\right)_T - \frac{1}{n^2T}\left(\frac{\partial \rho}{\partial T}\right)_n + \frac{\rho}{n^2T^2} - \frac{1}{n^2T}\left(\frac{\partial p}{\partial T}\right)_n + \frac{p}{n^2T^2} = -\frac{1}{n^2T}\left(\frac{\partial \rho}{\partial T}\right)_n,
\end{equation}
\begin{equation}
\frac{1}{n^2T}\left(\frac{\partial p}{\partial T}\right)_n = \frac{\rho}{n^2T^2} + \frac{p}{n^2T^2} - \frac{1}{nT^2}\left(\frac{\partial \rho}{\partial n}\right)_T.
\label{s12}
\end{equation}\newline
Multiplying both sides of Eq.\,\ref{s12} by $T^2n^2$ we achieve the key relation: \newline
\begin{equation}
T\left(\frac{\partial p}{\partial T}\right)_n = \rho + p - n\left(\frac{\partial \rho}{\partial n}\right)_T.
\label{keyrelation}
\end{equation}\newline
Upon analysing the temporal behaviour of $\rho$, the following relation can be written:
\begin{equation}
\dot{\rho}= \left(\frac{\partial \rho}{\partial n}\right)_T\dot{n} + \left(\frac{\partial \rho}{\partial T}\right)_n\dot{T}.
\label{temporal}
\end{equation}
At this point, we recall the so-called equations of motion for an imperfect fluid in equilibrium \cite{phdthesis}:
\begin{equation}
u_\mu T_{;\nu}^{\mu\nu} = \dot{\rho} + (\rho + p)\Theta + u_\mu\Delta T_{;\nu}^{\mu\nu} = 0,
\label{motion1}
\end{equation}
\begin{equation}
N_{;\mu}^{\mu} = \dot{n} + n\Theta + \Delta N_{;\mu}^{\mu} - \Psi = 0,
\label{motion2}
\end{equation}\newline
where $N_{;\mu}^{\mu}$ the particle flow tensor, $\Psi$ represents the particle source ($\Psi > 0$) or sink ($\Psi < 0$), and $\Theta$ is the fluid expansion rate. Note that the semicolon notation index is employed to represent the so-called covariant derivative. The upper index represents the 4 space coordinates. For instance, the notation of the particle flow tensor $N_{;\mu}^{\mu}$ represents the covariant derivative of the particle flow in the 4 space coordinates, namely, one temporal and 3 spatial ones \cite{barbararyden}. Thus, Eq.\,\ref{temporal} can be rewritten as:
\begin{equation}
\left(\frac{\partial \rho}{\partial T}\right)_n \dot{T} = \dot{\rho} - \left(\frac{\partial\rho}{\partial n}\right)_T\dot{n}.
\label{mot}
\end{equation}
Solving Eqs.\,\ref{motion1} and \ref{motion2} for $\dot{\rho}$ and $\dot{n}$, we have:
\begin{equation}
\dot{\rho} = -(\rho + p)\Theta - u_{\mu}\Delta T_{;\nu}^{\mu\nu},
\label{motion3}
\end{equation}
\begin{equation}
\dot{n} = -n\Theta - \Delta N_{;\mu}^{\mu} + \Psi.
\label{motion4}
\end{equation}\newline
Replacing $\dot{\rho}$ and $\dot{n}$ from Eqs.\,\ref{motion3} and \ref{motion4} into Eq.\,\ref{mot}, it reads:
\begin{equation}
\left(\frac{\partial \rho}{\partial T}\right)_n \dot{T} = -(\rho + p)\Theta - u_{\mu}\Delta T_{;\nu}^{\mu\nu} - \left(\frac{\partial \rho}{\partial n}\right)_T [\Psi - n\Theta - \Delta N_{;\nu}^{\nu}],
\end{equation}
\begin{equation}
\left(\frac{\partial \rho}{\partial T}\right)_n \dot{T} = \Theta\left[n\left(\frac{\partial \rho}{\partial n}\right)_T - (\rho + p)\right] - \left(\frac{\partial \rho}{\partial n}\right)_T[\Psi - \Delta N_{;\mu}^{\mu}] - u_{\mu}\Delta T_{;\nu}^{\mu\nu}.
\label{ss}
\end{equation}\newline
Rewriting Eq.\,\ref{keyrelation} as:
\begin{equation}
\left(\frac{\partial p}{\partial T}\right)_n = \frac{\rho + p - n\left(\frac{\partial \rho}{\partial n}\right)_T}{T} \Rightarrow \left(\frac{\partial \rho}{\partial n}\right)_T = \frac{-T\left(\frac{\partial p}{\partial T}\right)_n + \rho + p}{n},
\end{equation}
and inserting it into Eq.\,\ref{ss}, it reads:
\begin{equation}
\left(\frac{\partial \rho}{\partial T}\right)_n \dot{T} = \Theta\left[-T\left(\frac{\partial p}{\partial T}\right)_n\right] - \left(\frac{\partial \rho}{\partial n}\right)_T[\Psi - \Delta N_{;\mu}^{\mu}] - u_{\mu}\Delta T_{;\nu}^{\mu\nu}.
\label{sss}
\end{equation}\newline
Multiplying both sides of Eq.\,\ref{sss} by $(\partial \rho/\partial T)^{-1}_n$, we have:
\begin{equation}
\dot{T} = -T\left(\frac{\partial p}{\partial \rho}\right)_n\Theta - \left(\frac{\partial \rho}{\partial T}\right)^{-1}_n\left\{\left(\frac{\partial \rho}{\partial n}\right)_T[\Psi - \Delta N_{;\mu}^{\mu}] - u_{\mu}\Delta T_{;\nu}^{\mu\nu}\right\}.
\label{aaa}
\end{equation}\newline
Multiplying both sides of Eq.\,\ref{aaa} by $1/T$, it reads:\newline
\begin{equation}
\frac{\dot{T}}{T} = -\left(\frac{\partial p}{\partial \rho}\right)_n\Theta -\frac{1}{T}\left(\frac{\partial \rho}{\partial T}\right)^{-1}_n\left\{\left(\frac{\partial \rho}{\partial n}\right)_T[\Psi - \Delta N_{;\mu}^{\mu}] - u_{\mu}\Delta T_{;\nu}^{\mu\nu}\right\}
\label{generallaw}
\end{equation}\newline
Equation\,\ref{generallaw} is the general law of the temperature evolution of an imperfect fluid. For the case of a perfect fluid, there are no dissipation fluxes, so that $\Psi$, $\Delta T_{;\nu}^{\mu\nu}$, and $\Delta N_{;\mu}^{\mu}$ are zero and $\Theta = -\dot{n}/n$. Hence, Eq.\,\ref{generallaw} reads \cite{phdthesis}: \newline
\begin{equation}
\boxed{\frac{\dot{T}}{T} = \left(\frac{\partial p}{\partial \rho}\right)_n\frac{\dot{n}}{n}.}
\label{keyeq}
\end{equation}\newline
Equation\,\ref{keyeq} is the so-called temperature evolution law for the perfect fluid in the adiabatic regime. Considering the EOS of a perfect fluid, the term $(\partial p/\partial \rho)_n$ is recognized as $\omega$. Since $\rho = E/v$, we have:
\begin{equation}
\omega = \left(\frac{\partial p}{\partial \rho}\right)_n = \left(\frac{\partial p}{\partial [E/v]}\right)_n = v\left(\frac{\partial p}{\partial E}\right)_n.
\label{omega}
\end{equation}
The definition of $\omega$ in Eq.\,\ref{omega} is exactly the same as the definition of the effective Gr\"uneisen parameter $\Gamma_{eff}$, cf.\,Eq.\,\ref{dpdegru}. Also, as previously discussed, we have identified $\omega$ as $\Gamma_{eff}$ in the frame of a perfect fluid picture employing the Mie-Gr\"uneisen EOS. Now, Eq.\,\ref{keyeq} can be rewritten as: \newline
\begin{equation}
\frac{dT}{T} = \left(\frac{\partial p}{\partial \rho}\right)_n\frac{dn}{n}.
\label{keyeq2}
\end{equation}\newline
Integrating both sides of Eq.\,\ref{keyeq2} and considering that $(\partial p/\partial \rho)_n = \omega = \Gamma_{eff}$, we have:
\begin{equation}
\ln{(T)} + C_1 = \Gamma_{eff}[\ln{(n)} + C_2],
\end{equation}
\begin{equation}
\ln{(T)} + C_1 = \ln{(n^{\Gamma_{eff}}) + C_3},
\end{equation}
\begin{equation}
\ln{(T)} = \ln{(n^{\Gamma_{eff}})} + C_4,
\end{equation}
\begin{equation}
\ln{\left(\frac{T}{n^{\Gamma_{eff}}}\right)} = C_4,
\end{equation}
\begin{equation}
\frac{T}{n^{\Gamma_{eff}}} = C_5,
\end{equation}
where $C_1$ and $C_2$ are integration constants and $C_3$, $C_4$, and $C_5$ are arbitrarily constants. Upon recalling that $n \propto 1/v$ \cite{phdthesis}, we have:
\begin{equation}
\frac{T}{\left(\frac{1}{v}\right)^{\Gamma_{eff}}} = C_5,
\end{equation}
which in turn can be rewritten as:
\begin{equation}
\boxed{Tv^{\Gamma_{eff}} = constant}.
\label{tvgamma}
\end{equation}
Equation\,\ref{tvgamma} relates $T$ and $v$ during an adiabatic expansion of the universe in the context of a perfect fluid analysis. The temperature of the universe is consistently reduced as it expands during the radiation- and matter-dominated eras, which can be interpreted as a barocaloric effect \cite{Mariano2021}. The latter can be quantified using $\Gamma$ (Eq.\,\ref{gamma3}), which is the singular contribution to $\Gamma_{eff}$, as discussed in Section\,\ref{deductionofgamma}. In condensed matter Physics, the barocaloric effect is initiated by applying pressure, followed by its adiabatic removal, resulting in cooling. In the case of the universe, the singularity before the Big Bang could be considered as the first step in performing the barocaloric effect. However, due to the inherent uncertainty in estimating precisely $T$ and $p$ of the expanding universe during each era \cite{Melendres2021}, quantifying this (baro)caloric effect in terms of $\Gamma$ is challenging. During the radiation- and matter-dominated eras, $\Gamma_{eff} > 0$, causing the temperature of the universe to decrease in response to its expansion, cf.\,Eq.\,\ref{tvgamma}. However, during the DE-dominated era, $T$ increases during adiabatic expansion, with $\Gamma_{eff} < 0$, as described by Eq.\,\ref{tvgamma}. This, in turn, can be interpreted as an inverse barocaloric effect \cite{Majumdar2011}, which is associated to the portion of DE in the universe, which in turn can affect the temperature evolution of the universe in late stages \cite{Komatsu2020}. At this point, an analysis in terms of the volume- and temperature-dependence of $S$ across various eras is performed. Based on Eq.\,\ref{gammaeffective} and the fact that $\mathcal{B}_T = 1/{\kappa_T}$, $\Gamma_{eff} = v\alpha_p/\kappa_Tc_v$ \cite{EJP2016}. By employing the thermodynamic relation $\alpha_p = \kappa_T(\partial S/\partial v)_T$, it can be written:
\begin{equation}
v\left(\frac{\partial S}{\partial v}\right)_T = \Gamma_{eff} c_v.
\end{equation}
Considering that $\Gamma_{eff} = \omega$ and $c_v = T(\partial S/\partial T)_v$, it yields:
\begin{equation}
v\left(\frac{\partial S}{\partial v}\right)_T = \omega T \left(\frac{\partial S}{\partial T}\right)_v.
\end{equation}
For the various eras:
\begin{align}
v\left(\frac{\partial S}{\partial v}\right)_T &= \frac{T}{3} \left(\frac{\partial S}{\partial T}\right)_v, &&\text{(radiation-dominated era)} \label{rad} \\
\left(\frac{\partial S}{\partial v}\right)_T &\rightarrow 0, &&\text{(matter-dominated era)} \label{mat} \\
v\left(-\frac{\partial S}{\partial v}\right)_T &= T \left(\frac{\partial S}{\partial T}\right)_v. &&\text{(DE-dominated era)} \label{de}
\end{align}\newline
Equation\,\ref{rad} highlights that during the radiation-dominated era, the isothermal entropy change caused by the universe expansion was more pronounced than the one associated with its isochoric temperature change by a factor of 3. During the matter-dominated era, $S$ remains approximately constant, cf.\,Eq.\,\ref{mat}. By considering $(\partial S/\partial v)_T = p/T$ \cite{Lavenda2015}, Eq.\,\ref{de} can only hold true if negative pressures are assumed. Therefore, the concept of negative pressure associated with dark energy is supported by the analysis of Eq.\,\ref{de}. According with Landau \cite{Landau1980}, a negative value of $(\partial S/\partial v)_T$ implies a metastable state. In this case, the volume portion associated with DE, spontaneously contracts and, as a result, $S$ increases due to the potential formation of cavities \cite{Lavenda2015,Landau1980}. This contraction of the volume portion associated with dark energy can explain the accelerated expansion during the dark energy-dominated era, as proposed in Ref.\,\cite{Melendres2021}. The contraction of such a volume portion related to DE can be considered as the driving force associated with the accelerated expansion during the DE-dominated era, given that the so-called Hubble flow increases due to this contraction \cite{Melendres2021}. Some authors suggest that the concept of negative temperatures could be associated with the DE-dominated era \cite{bloch}, referred to as the phantom regime \cite{Weinberg1972}. However, this assumption is questionable since the concept of negative temperatures require thermodynamic equilibrium, which is not the case of an expanding universe with a time-dependent $T$. This is discussed in details in Ref.\,\cite{ramsey} and is in contrast to the canonical definition of temperature based on the magnetic energy of a paramagnetic system \cite{unveiling}, cf.\,discussed in the end of Appendix\,\ref{stressnegativetemperatures} of this Thesis. At this point, an analysis assuming a closed spherical universe approach \cite{silk,Melendres2021} to describe its speed of expansion in connection with $\Gamma_{eff}$ is provided. Recalling the first law of Thermodynamics \cite{ralph}:
\begin{equation}
dQ = dU + dW.
\end{equation}
Given that $dQ = TdS$ and $dW = +pdv$, we have \cite{Melendres2021}:
\begin{equation}
TdS = dU + pdv,
\end{equation}
which results in $dU = -pdv$ for an adiabatic process \cite{Melendres2021}. The Hubble parameter is defined as $H = \dot{a}/a$ \cite{Knox2022}, which represents the normalized rate of expansion. Analogously, one can determine $H$ as the normalized rate of the universe's volume change over time \cite{Melendres2021}:
\begin{equation}
H = \frac{\dot{v}}{v} = \frac{\frac{dv}{dt}}{\frac{4}{3}\pi r^3},
\label{hubble}
\end{equation}
where $r$ is the radius of the assumed spherical universe. Equation\,\ref{hubble} can be rearranged as \cite{Melendres2021}:
\begin{equation}
H = \frac{3}{4\pi r^3} \frac{dv}{dU}\frac{dU}{dt},
\end{equation}
which, in turn, gives us:
\begin{equation}
\frac{dU}{dt} = \frac{4}{3}\pi r^3 H \frac{dU}{dv}.
\label{dedt}
\end{equation}\newline
Following discussions in Ref.\,\cite{Busca2013}, during the radiation-dominated era the universe expanded at a decelerated rate, i.e., $dU/dt < 0$. The speed of expansion reached a minimum at the matter-dominated era, i.e., $dU/dt \rightarrow 0$, and then started to increase as the universe entered in the DE-dominated era, i.e., $dU/dt > 0$, indicating that the universe is expanding in an accelerated way. Given that $dU = -pdv$ and that the volume variation of the universe is positive due to its expansion, a negative $p$ implies that the internal energy of the universe is increasing in the DE-dominated era \cite{Melendres2021}. This allows to establish a connection between $\Gamma_{eff}$, the universe's expansion speed, and the temporal evolution of $U$. Given that $dU/dv = -p$ for an adiabatic expansion and $p = \omega\rho = \Gamma_{eff}\rho$ for a perfect fluid, $dU/dv = -\Gamma_{eff}\rho$. Hence, Eq.\,\ref{dedt} can be rewritten as:
\begin{equation}
\Gamma_{eff} = \frac{3}{4\pi r^3 H \rho}\left(-\frac{dU}{dt}\right).
\label{lambdacdm}
\end{equation}
Assuming that the physical quantities $r$, $H$, and $\rho$ are always positive, the sign of $\Gamma_{eff}$ is determined solely by $-dU/dt$. Based on this, when $dU/dt < 0$, $\Gamma_{eff} > 0$; when $dU/dt \rightarrow 0$, $\Gamma_{eff} \rightarrow 0$; and when $dU/dt > 0$, $\Gamma_{eff} < 0$. The sign-change of $\Gamma_{eff}$ during the transition from decelerated to an accelerated expansion is reminiscent of a phase transition near a critical end point in condensed matter physics \cite{Mottiscool}. Using Noether's theorem \cite{Noether}, the change in the time dependence of $U$ during this transition can be interpreted as a symmetry breaking, which is analogous to a canonical phase transition \cite{Landau1980}. It is worth mentioning that it is still under debate whether $\omega$ has indeed converged to $-$1 in the DE-dominated era or if it will vary over time as the universe continues to expand \cite{Planck2020}. Under the light of the proposed interpretation of $\omega$ as $\Gamma_{eff}$, it is tempting to conclude that $\omega$, i.e., $\Gamma_{eff}$, will become time-dependent as the thermodynamic quantities embedded in $\Gamma_{eff}$, such as $\kappa_T$ and $c_v$, change in response to the continuous temperature change and expansion of the universe. Since $\omega$, namely $\Gamma_{eff}$, is linked to $\Lambda$ through $p = \omega\rho_{\Lambda} = -\rho_{\Lambda} = -c^2\Lambda/8\pi G$ \cite{Weinberg1972}, where $c$ is the speed of light and $\rho_{\Lambda}$ is the vacuum energy density, a time-dependent $\omega$ would imply a time-dependent $\Lambda(t)$ and/or $G(t)$, as suggested in recent works \cite{pimentel,tutusaus,shukla}. To resolve this issue, accurate observations of $\rho_{\Lambda}$ over time are crucial \cite{frieman}. The analogy of the sign-change in $\Gamma_{eff}$ between the matter- and DE-dominated eras and the proposal of a temporal behaviour of $\Lambda$ under the light of $\Gamma_{eff}$ are another key results in this Thesis in the frame of the application of $\Gamma_{eff}$ to Cosmology. Last but not least, the connection between $\Gamma_{eff}$ and Einstein field equations is made in the following. The renowned Einstein field equations, as reported in Ref.\,\cite{Einstein1916}, link the curvature of spacetime to the distribution of energy and momentum in the universe. This set of equations is given by \ref{EFE} \cite{Einstein1916}:
\begin{equation}
R_{\mu\nu} - \frac{1}{2}g_{\mu\nu}R + \Lambda g_{\mu\nu} = 8\pi G T_{\mu\nu},
\label{EFE}
\end{equation}
where $R_{\mu\nu}$ represents the Ricci curvature tensor, $g_{\mu\nu}$ is the metric tensor, and $R$ stands for the curvature scalar. Upon considering a perfect fluid, $T_{\mu\nu}$ is related to $p$ and $\rho$ through the relation $T_{\mu\nu} = (p+\rho)u^{\mu}u^{\nu}+pg_{\mu\nu}$ \cite{Weinberg1972,Dalarsson2015,Shang2005}, where $u$ is the four-velocity vector field. Using the EOS of a perfect fluid, $T_{\mu\nu}$ can be expressed as a function of $\omega$ as $T_{\mu\nu} = (1+\omega)\rho u^{\mu}u^{\nu}+\rho\omega g_{\mu\nu}$ \cite{Weinberg1972}. Hence, Einstein field equations can be rewritten in terms of $\Gamma_{eff}$ so that:
\begin{equation}
R_{\mu\nu} - \frac{1}{2}g_{\mu\nu}R + \Lambda g_{\mu\nu} = 8\pi G[(1+\Gamma_{eff})\rho u^{\mu}u^{\nu}+\rho\Gamma_{eff} g_{\mu\nu}].
\label{gammaeinstein}
\end{equation}
It is then evident that Einstein field equations implicitly incorporate $\Gamma_{eff}$ through $T_{\mu\nu}$. It is worth mentioning that this is the very first time in the literature that it is reported about the connection between Einstein field equations and $\Gamma_{eff}$. Note that the relationship between $T_{\mu\nu}$ and $\Gamma_{eff}$ comes from the fact that $T_{\mu\nu}$ incorporates all pressure components, shear stresses, and energy density associated with the gravitational field. The connection between $\Gamma_{eff}$ and $T_{\mu\nu}$ is particularly relevant in the context of imperfect fluids, which have been suggested as a way to explain observational models that deviate from the $\Lambda$-CDM one by incorporating additional anisotropic energy fluxes and stress tensor components \cite{Algoner2019}. In such a case, $T_{\mu\nu}$ embodies a combination of perfect and imperfect fluid contributions \cite{Algoner2019}. The imperfect fluid model is consistent with recent experimental observations of an anisotropic universe expansion resulting from the formation of galaxy clusters of varying sizes during the early stages of the universe \cite{Lovisari2020}. Such an anisotropic expansion has been verified through measurements of redshift space distortions \cite{Huff2022}. Furthermore, it has been reported in the literature that the quantum vacuum in the DE-dominated era is highly inhomogeneous and anisotropic, with $\nabla a$ and $\nabla^2a$ being expressive \cite{unruh}. Inspired by the work reported in Ref.\,\cite{Mottiscool} on the metal-insulator coexistence region of the Mott transition, similar arguments to those reported in Ref.\,\cite{Mottiscool} were employed to describe the anisotropic expansion of the universe. In Ref.\,\cite{Mottiscool}, using the Avramov/Casalini's model, different relaxation times for metallic and insulating bubbles were assumed in the coexistence region of the Mott transition. As reported in Refs.\,\cite{Barto,prl2010}, anisotropic effects in thermal expansion are observed upon crossing the first-order transition line within the coexistence region. An analogy to the universe by considering different rotation periods for different galaxy clusters is straightforward, where the galaxies clusters play the role of metallic/insulating bubbles. As the universe transitions from radiation-, to matter-, and DE-dominated eras, $\Gamma_{eff}$ changes sign, a behaviour analogous to that observed in the pressure-induced Mott metal-to-insulator transition near a second-order critical end point \cite{Mottiscool}. Therefore, an anisotropic expansion of the universe can be inferred based on the proposal of critical points related to the universe expansion was discussed in Ref.\,\cite{Melendres2021}. The anisotropic expansion effects can be described through the components of the stress tensor, which, in turn, can be associated with the elastic Gr\"uneisen parameter, $\Gamma_{ec} = 1/T(\partial T/\partial \sigma_{ij})_S$ \cite{Mariano2021} (Eq.\,\ref{elasticgruneisen}). Since $\Gamma_{eff}$ is related to $\Gamma_{ec}$, the anisotropic expansion effects can be explored through $\Gamma_{eff}$ embodied in the energy-momentum tensor in Einstein field equations. If DE continuously drives the universe towards an endless accelerated expansion, one of its possible fates is the so-called ``big rip'' \cite{Weinberg2003,katie,skibba}. Such a connection between $\Gamma_{eff}$, $\Gamma_{ec}$, and Einstein field equations offer an alternative perspective to explore such a scenario, which incorporates all stress tensor components associated with an anisotropic expansion in $T_{\mu\nu}$ and, as a consequence, in Einstein field equations. The is another key result of this Section, which provides a new avenue in the field of Cosmology for interpreting the universe's expansion and addressing the limitations of the current standard cosmological scenario, as highlighted in Ref.\,\cite{review}.

The theoretical results discussed in this Section were published in:
\begin{itemize}
\item Lucas Squillante, G.O. Gomes, I.F. Mello, G. Nogueira, A.C. Seridonio, R.E. Lagos-Monaco, M. de Souza, Exploring the expansion of the universe using the Gr\"uneisen parameter, Results in Physics \textbf{57}, 107344 (2024). \newline
    \url{https://www.sciencedirect.com/science/article/pii/S2211379724000263}
\end{itemize}
\begin{savequote}[8cm]
``When you want something in life you just gotta reach out and grab it.''
  \qauthor{--- Christopher McCandless, Into The Wild.}
\end{savequote}

\chapter{\label{ch:3}Experimental investigations on the possible multiferroic character of the Fabre salts}

\minitoc


\section{Strongly correlated electronic phenomena in the (TMTTF)$_2$X salts}

\begin{figure}[!htb]
\centering
\includegraphics[width=0.95\textwidth]{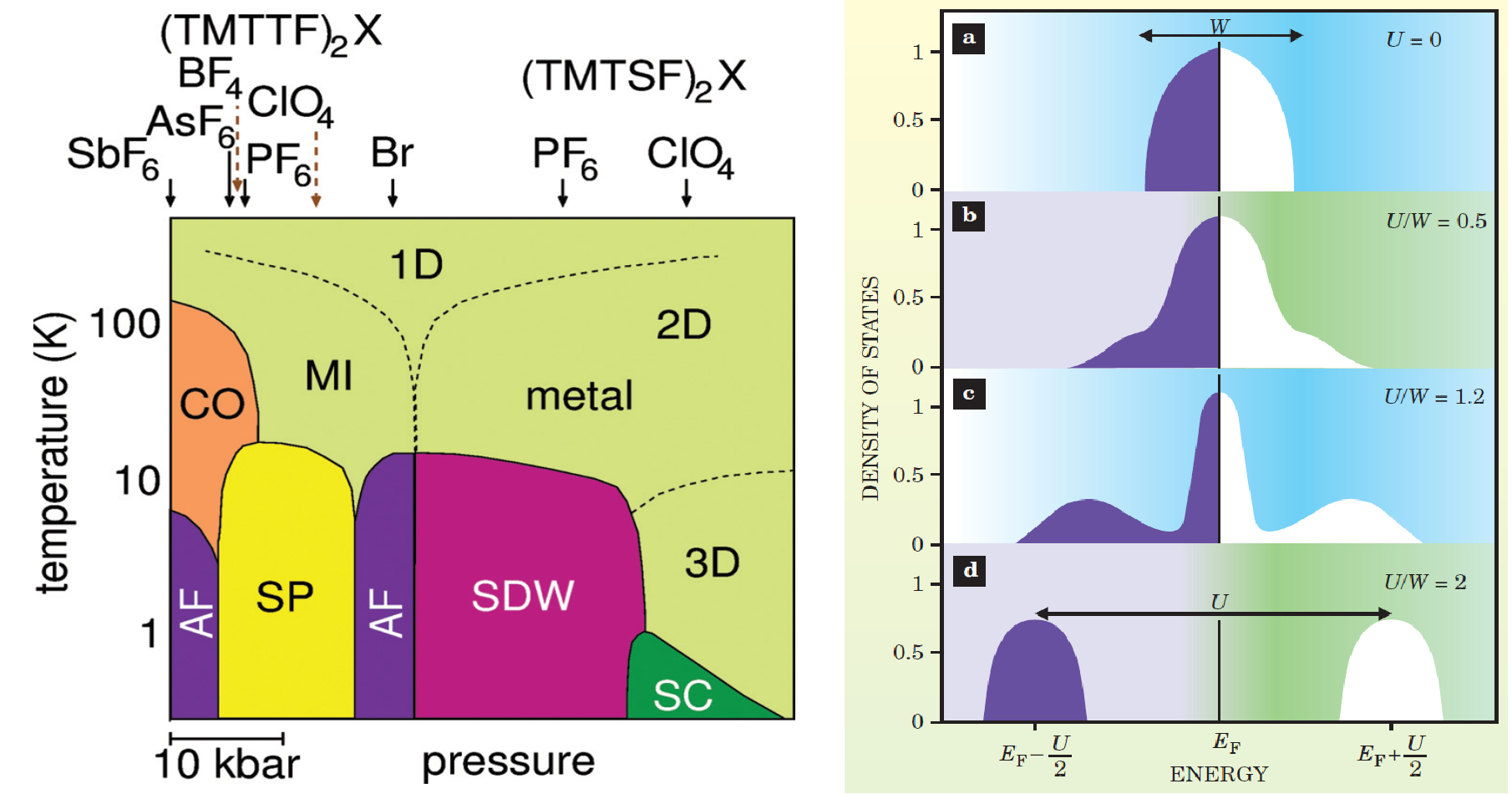}
\caption{\footnotesize Left panel: Schematic temperature \emph{versus} pressure phase diagram for the Fabre-Bechgaard salts. The position of the different compounds under ambient pressure is indicated above by the black arrows, such difference is directly connected with the counter-anion chemical pressure. The phases shown are Mott insulator (MI), charge-ordering (CO), antiferromagnetic (AF), spin-Peierls (SP), spin-density wave (SDW) and superconductivity (SC). The dashed lines represent cross-overs from a 1, 2, and 3D system \cite{jacko}. Right panel: Density of states (DOS) as a function of the energy for several values of on-site Coulomb interaction $U$ between electrons. Without any Coulomb repulsion taken into account ($U$ = 0), the DOS has a half elliptical shape (blue = filled, white = empty) and the Fermi level $E_F$ is located in the middle. In the case where the electron interactions are strong enough, the bands are split and thus the Mott metal-insulator transition occurs \cite{physicstoday}.}
\label{jacko}
\end{figure}

\begin{figure}[!htb]
\centering
\includegraphics[width=0.9\textwidth]{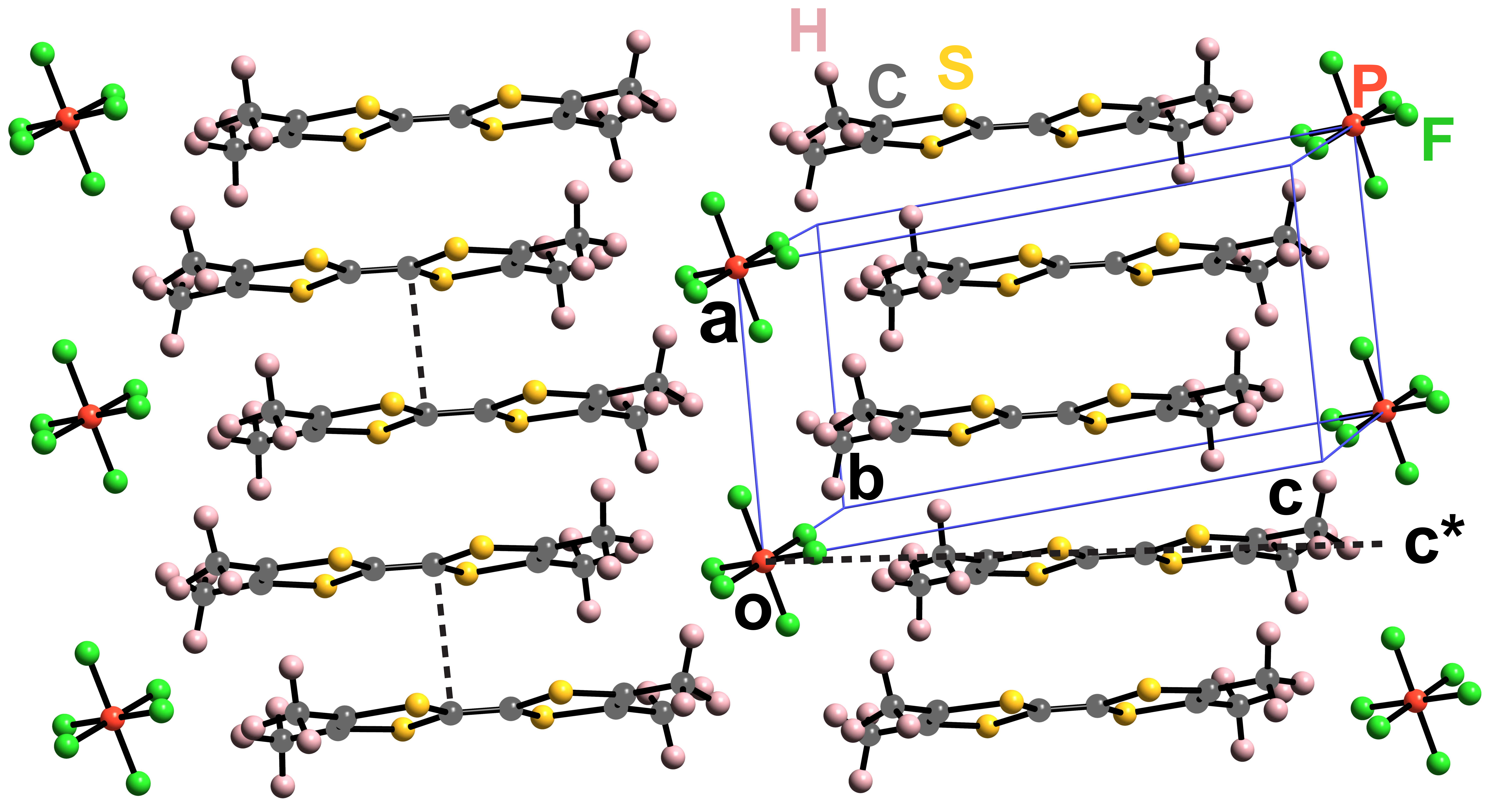}
\caption{\footnotesize Molecular structure of the (TMTTF)$_2$$X$ \cite{prbrioclaro}, where the counter anions $X$ are shown and the dotted lines between two TMTTF molecules represent the dimers. The triclinic unit cell is outlined by the blue lines and \emph{a}, \emph{b}, \emph{c} and $c^*$ (dashed horizontal line) represent the crystallographic axes.}
\label{tmttf}
\end{figure}

\begin{figure}[!htb]
  \centering
  \includegraphics[width=0.8\textwidth]{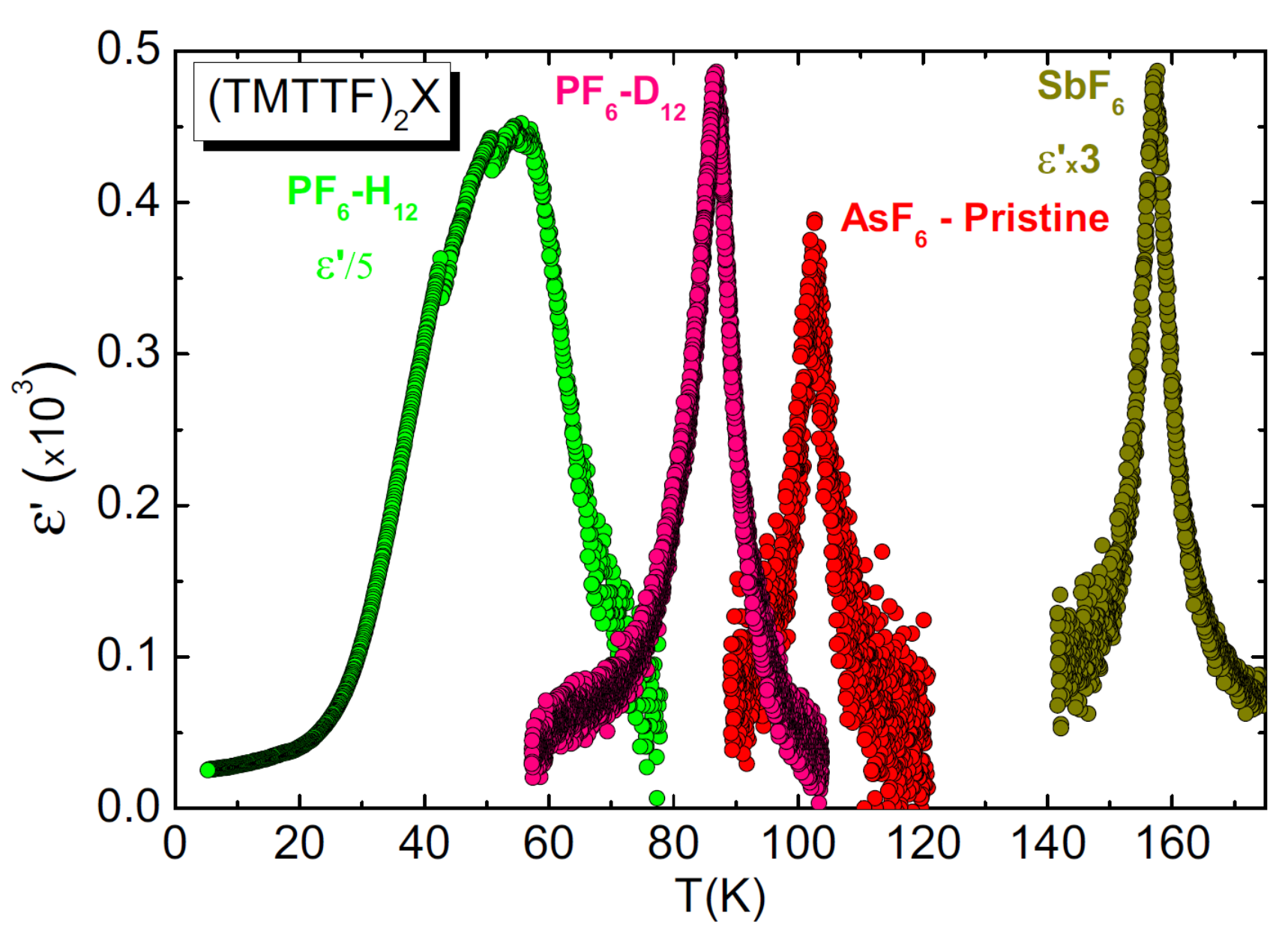}
  \caption{\footnotesize Quasi-static (fixed frequency $f$ = 1\,kHz) dielectric constant as a function of temperature measured in the $c^*$-axis for the fully-hydrogenated (PF$_6$-H$_{12}$) and 97.5\% deuterated (PF$_6$-D$_{12}$) (TMTTF)$_2$PF$_6$ systems (green and pink circles, respectively), pristine (TMTTF)$_2$AsF$_6$ (red circles), and (TMTTF)$_2$SbF$_6$ (dark yellow circles) systems upon cooling down the system \cite{prbrioclaro}. All the experimental results in Ref.\,\cite{prbrioclaro} were obtained in the Solid State Physics Laboratory at the S\~ao Paulo State University in Rio Claro, SP - Brazil.}\label{molecules22}
\end{figure}

\begin{figure}[!htb]
\centering
\includegraphics[width=\textwidth]{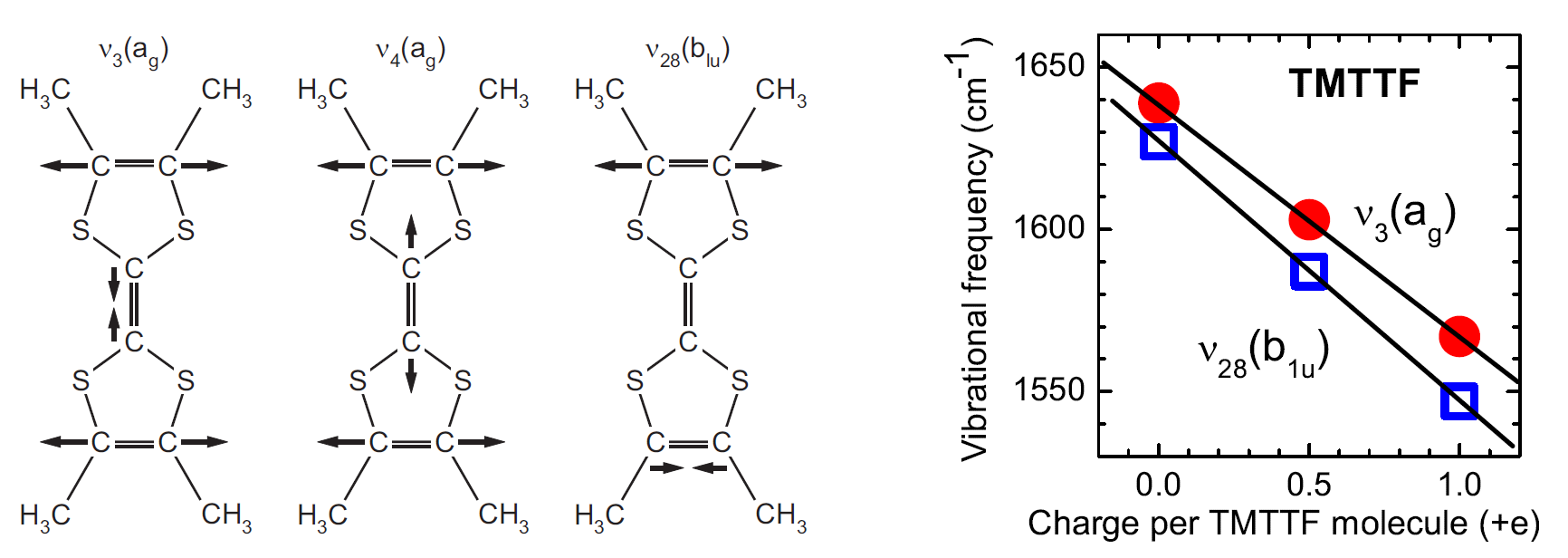}
\caption{\footnotesize  Schematic representation of the symmetric $a_g$ modes $\nu_3$ and $\nu_4$ and the asymmetric $b_{1u}$ mode $\nu_{28}$ of the TMTTF molecule (left panel). Vibrational frequency as a function of charge per TMTTF molecule (right panel) for the vibrational modes $\nu_3$($a_g$) and $\nu_{28}$($b_{1u}$) \cite{menegetthi}.}
\label{ramanmolecules}
\end{figure}

\begin{figure}[!htb]
\centering
\includegraphics[width=\textwidth]{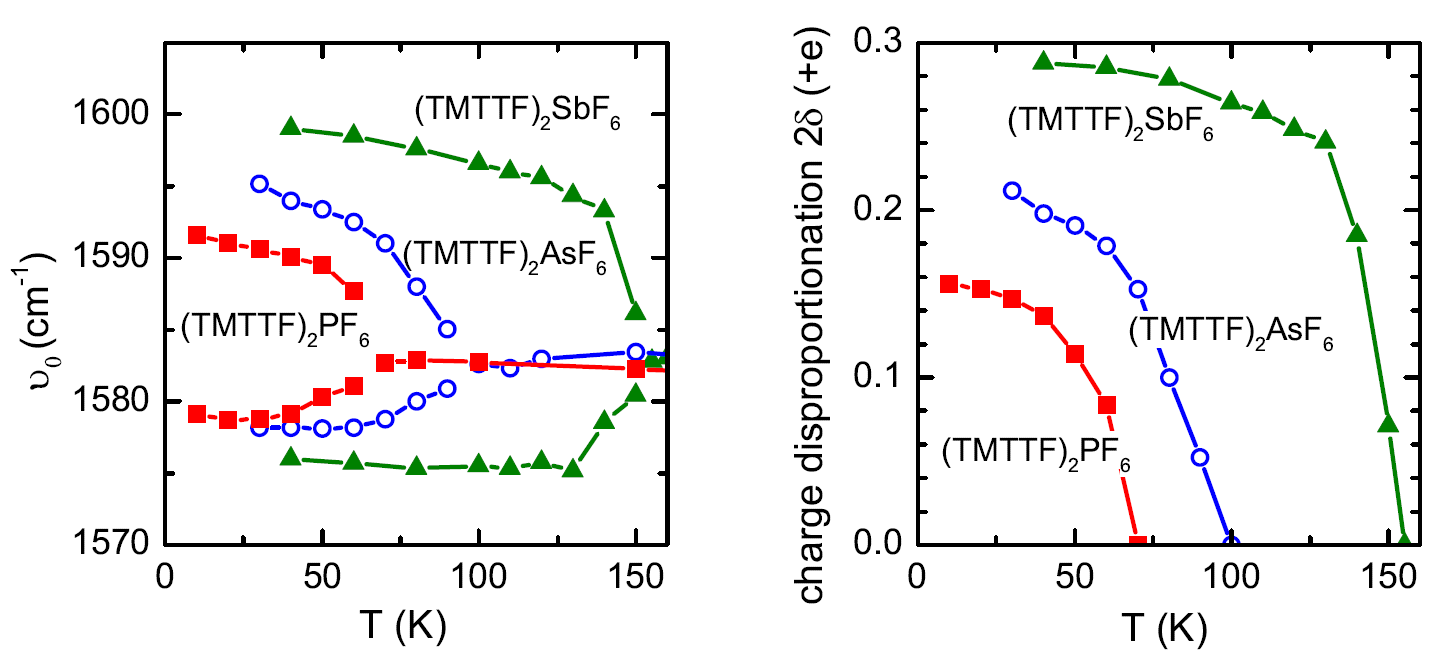}
\caption{\footnotesize Left panel: resonance frequency $\nu_0$ associated with the $\nu_{28}$ vibrational mode as a function of temperature $T$ for the fully-hydrogenated (TMTTF)$_2$X (X = PF$_6$, AsF$_6$, and SbF$_6$) systems, where a splitting of the resonance frequency is shown at $T_{co}$ for each salt, since below this temperature a charge disproportionation takes place and the resonance frequency of the charge-rich ($\rho_0$ + $\delta$) and -poor ($\rho_0$ - $\delta$) sites differs. Right panel: charge disproportionation 2$\delta$ as a function of the temperature $T$. Above $T_{co}$ there is no charge disproportionation and below it the charge disproportionation is finite \cite{dresselraman}.}
\label{splitting}
\end{figure}

One of the most investigated phenomena at low temperatures is superconductivity, where the electrical resistance $R_{el}$ of a determined system goes to zero at a critical temperature $T_c$. In other words, there is a current flow without energy dissipation by the well-known Joule effect ($P_d = R_{el}\cdot i^2$), where $P_d$ is the dissipated power, and $i$ is the electrical current. After several years since Kamerlingh Onnes first observed superconductivity in a mercury (Hg) sample with $T_c$ $\sim$ 4\,K \cite{Heike}, Meissner and Ochsenfeld \cite{Meissner} demonstrated experimentally what would be named afterward as the Meissner-Ochsenfeld effect, where a magnetic field $\vec{B}$ is expelled from a superconductor and a magnetic levitation occurs. Such an effect is crucial in defining superconductivity as a new state of matter. This phenomenon is so counter-intuitive that more than half-century passed since the famous BCS theory \cite{bcs} was developed, which proposed an electron-electron coupling mediated by the lattice phonons in the so-called Cooper pairs\,[7].
\begin{figure}[t]
  \centering
    \includegraphics[width=\textwidth]{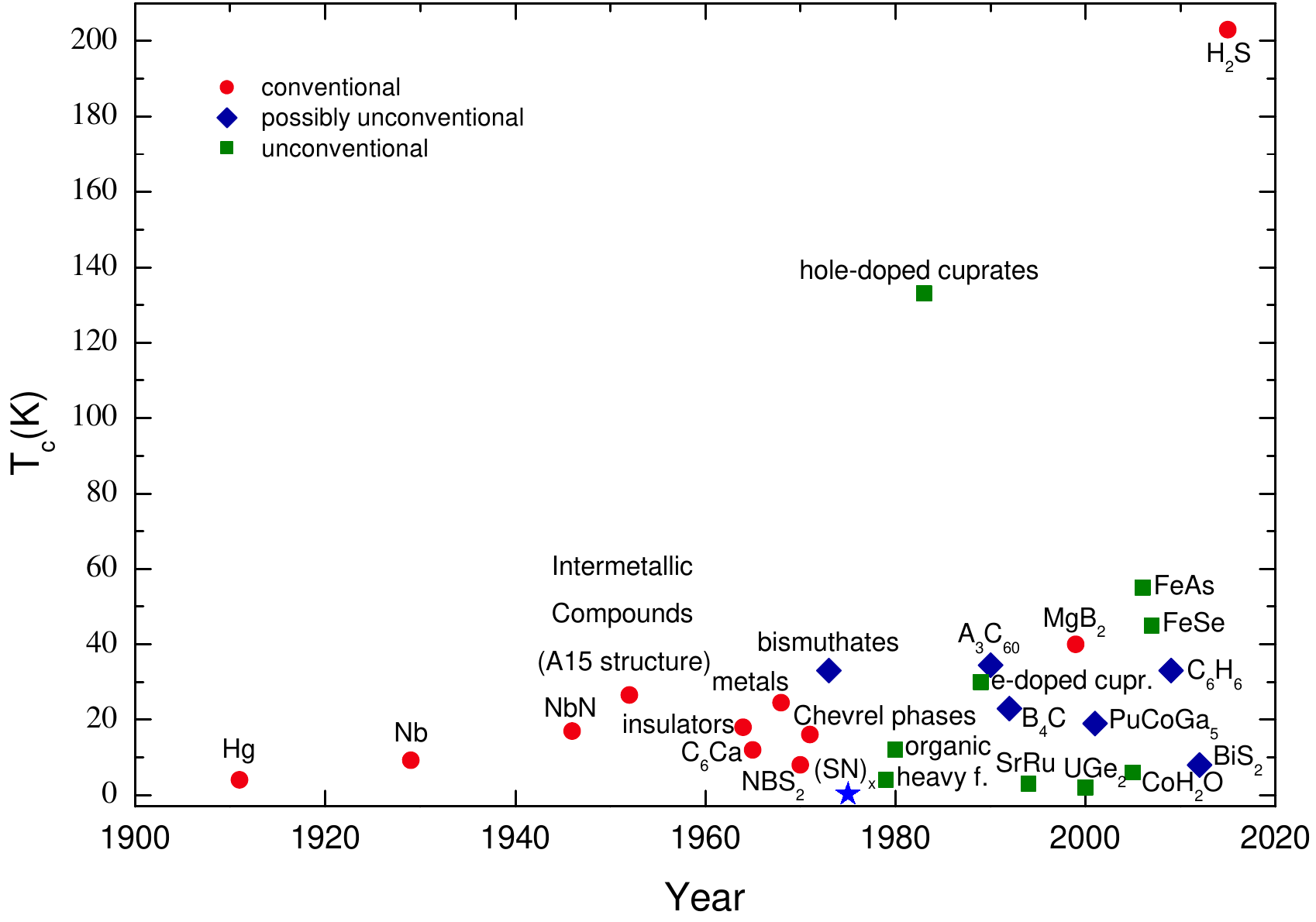}
  \caption{\footnotesize Evolution of the critical temperature $T_c$ over the years of conventional, possibly unconventional, and unconventional superconductors. Figure adapted from Ref.\,\cite{tcxyears}.}\label{tcxyears}
\end{figure}
There is a technological interest in synthesizing new superconducting materials with higher critical temperatures \cite{naturewire}. This is because of the use of superconducting materials in helping connect grids as better safety switches than conventional inductors \cite{refinductors}, a base to develop quantum computers \cite{naturequantumcomp} and even in the development of Maglev trains technology \cite{maglev}. Most metallic elements were discovered to superconduct below 10\,K and even a group of metallic alloys have shown superconductivity \cite{buckel,Blundell,annett}. However, until the 80's critical temperatures above 30\,K had not been observed, as shown in Fig.\,\ref{tcxyears}. In 1986, Bednorz and M\"uller observed for the first time superconductivity in the so-called cuprates (alloys containing copper and oxygen) above the 40\,K range \cite{Bednorz}. Also, superconductivity was discovered in the system YBa$_2$Cu$_3$O$_{7-x}$ (YBCO) with a critical temperature of 90\,K \cite{natureybco}. In the last decade, superconductivity was discovered in iron-based structures, the so-called iron pnictides or iron-based superconductors \cite{IBSC_Dysco}. Such a discovery was unexpected for two main reasons: it was believed that the magnetic ions would destroy superconductivity and the fact that the non-superconducting phases of this family are metallic, totally different from the cuprate families which are Mott insulators in the underdoped regime \cite{paglione,norman}. It is worth mentioning that other systems of interest also superconduct, such as the heavy fermion compound CeCu$_2$Si$_2$ \cite{steglich}, and the (TMTSF)$_2$PF$_6$, where TMTSF represents a molecule of tetramethyltetraselenafulvalene, organic molecular metal \cite{Jerome}, for instance. The BCS theory of conventional superconductors reveals how to achieve high $T_c$ combining high-frequency phonons, strong electron-phonon coupling, and high-density of states \cite{ginzburg}. Such conditions are satisfied for compounds dominated by hydrogen \cite{ashcroft1,ashcroft2}, predicting compounds with $T_c$ in the range of 50$-$235\,K for numerous hydrides \cite{wang}, since the hydrogen mass is low and it would increase the phonon frequency $\omega$ and thus the critical temperature $T_c$, as shown in the famous BCS theory relation \cite{bcs,annett}:
\begin{equation}
k_BT_c = \frac{\hbar\,\omega_D}{1.45}\cdot \textmd{exp}\left(-\frac{1.04(1+\lambda)}{\lambda - \mu^{\ast}(1+0.62\lambda)}\right),
\end{equation}
\noindent where $\hbar$ is the Planck's constant divided by 2$\pi$, $\omega_D$ the Debye's frequency, $\lambda$ the electron-phonon coupling constant, and $\mu^{\ast}$ the Coulomb pseudopotential, which takes into account the Coulomb repulsion between electrons. In 2015 the hydrogen sulfide compound H$_2$S was discovered to superconduct at $T_c$ $\sim$ 203\,K under 100\,GPa \cite{h2s}. Yet, in 2019 it was reported that a Lanthanum superhydride LaH$_{10\pm x}$ showed signatures of superconductivity at $T_c$ above 260\,K under pressures of 180-200\,GPa \cite{prllanthanum,pnaslanthanum}. Until the writing of this Thesis, there is a report of near-ambient superconductivity in a N-doped lutetium hydride \cite{rangadias}. However, such results are still under high criticism and no consensus has yet been reached. In this context, it is clear that the goal to achieve high-temperature superconductivity remains a challenge. The pressurized Fabre-salt (TMTSF)$_2$PF$_6$ show superconductivity at $T_c = 0.9$\,K under 12\,kbar \cite{Jerome}, playing a crucial role in the frame of superconductivity to fundamentally investigate correlated phenomena, more specifically the relation between the charge-ordered and the superconducting phases that take place in these systems, aiming to discover key-factors for achieving high-temperature superconductivity. In an article published in 1964, it was predicted by W.\,A.\,Little that low-dimensional systems would be strong candidates in exhibiting superconductivity at high temperatures \cite{Little}.
\begin{figure}[!htb]
  \centering
  \includegraphics[scale=0.3]{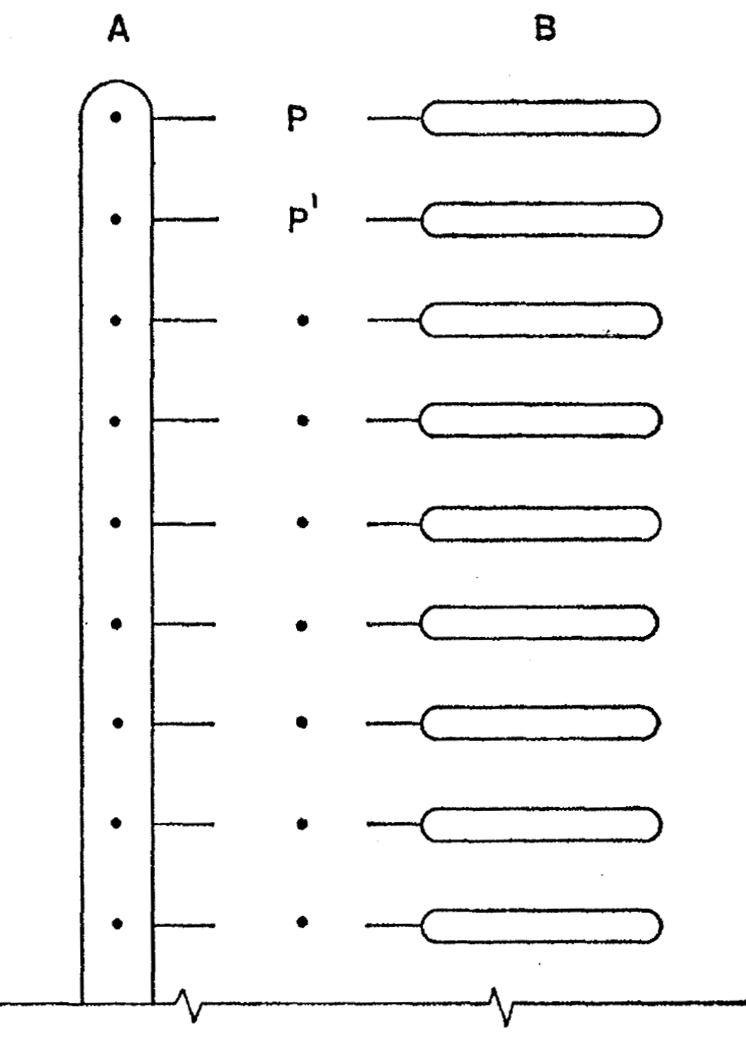}
  \caption{\footnotesize Little proposed model of a superconducting organic molecule, where the molecule A is defined as the ``spine'' and B are the side chains attached to the spine at points P, P', and so on. Picture extracted from Ref.\,\cite{Little}.}\label{figlittle}
\end{figure}
More specifically, the proposed model suggests that in such systems a long chain called the ``spine'' can be observed in which electrons fill the various states and a series of arms or side chains attached to the spine (Fig.\,\ref{figlittle}), making the system easily polarized. In this case, the electric dipoles are formed between the structure A and B (Fig.\,\ref{figlittle}) and, upon appropriate choosing the molecules of the side chains, it would be possible to observe an interaction between electrons in the spine due to the charge oscillations of the side chains, and with such an attractive interaction a superconducting state can emerge. The molecular conductors discussed in this proposal present such characteristics and this is one of the main reasons that make such systems to be intensively explored by the scientific community in the last decades \cite{reviewpouget,reviewlorentz}. In this context, molecular conductors have been recognized as an appropriate playground for the exploration of correlated phenomena in low-dimensions \cite{reviewpouget,reviewlorentz}. Among molecular conductors, the systems of particular interest are the (TMTTF)$_2$$X$, where TMTTF is the base molecule tetramethyltetrathiafulvalene and $X$ a monovalent center-symmetric counter-anion, such as PF$_6$, AsF$_6$, or SbF$_6$. Such systems enable the investigation of contemporary aspects in the field of correlated phenomena, like the Mott metal-insulator transition \cite{reviewlorentz} and the charge-ordered phase \cite{kanoda1997}. It is possible to tune the physical properties of such systems either by employing chemical substitution of the counter-anions or by applying external hydrostatic pressure \cite{dressel,Monceau2} (Fig.\,\ref{jacko}). In such systems, the TMTTF molecules are organized and stacked in chains in the $a$-axis (Fig.\,\ref{tmttf}) and are dimerized, resembling the model proposed by Little \cite{Little}.
Theoretically, the Physics of the molecular conductors of the TMTTF family can be described by the dimerized one-dimensional extended Hubbard model \cite{dimerized}, represented by the Hamiltonian:
\begin{eqnarray}\scalemath{0.88}{
\emph{H}=t_1\sum_{i\,\textmd{even},\,
\sigma}(a_{i\,\sigma}^{\dag}a_{i+1\, \sigma} + H.c.) + t_2\sum_{i\,\textmd{odd},\, \sigma}(a_{i\,\sigma}^{\dag}a_{i+1\, \sigma} +
H.c.) +  U\sum_i{n_{i\uparrow}n_{i\downarrow}} +
V\sum_{i}n_in_{i+1}\label{ExtendedHubbard},}
\end{eqnarray}
\noindent where $t_1$ = $t (1 - \delta_d)$  and  $t_2$ = $t (1 + \delta_d)$  are, respectively, the inter- and intra-dimer hopping terms, $\delta_d$ the dimerization degree, $\sigma$ is the spin index $\uparrow$ or $\downarrow$, $a^{\dag}_{i\sigma}$, $a_{i\sigma}$, and $n_{i\sigma}$ the creation, annihilator, and the number operators for one electron with spin $\sigma$  at the $i$$^{\textmd{th}}$ site, represented by $n_i$ = $n_{i\uparrow}$ + $n_{i\downarrow}$, $U$ and $V$ are the on-site and inter-site nearest-neighbor Coulomb interaction, respectively. Upon analysing the electrical resistivity \emph{versus} temperature data of the TMTTF family salts, a metallic behaviour is observed at room temperature \cite{1}. However, upon decreasing temperature, $U$ increases upon lowering the temperature and splits the energy bands creating a gap. In such a temperature, a Mott-insulating phase \cite{bookmott} takes place, where a partially-filled energy band configuration coexists with the presence of a gap (Fig.\,\ref{jacko}). Such a phase only occurs due to the strong electronic correlation present in such systems.
\noindent A charge-ordered phase can be achieved in the TMTTF-based molecular systems at low temperatures (typically between $\sim$ 50$-$150\,K) (Fig.\,\ref{molecules22}), coexisting with a ferroelectric one \cite{nad}, in the so-called Mott-Hubbard ferroelectric phase. Such phenomena can also be described by employing the extended Hubbard model \cite{hubbard}. In this case, when $U/W \approx 1$ a Mott insulator transition takes place \cite{reviewpouget}, where $W$ is the bandwidth. However, as the inter-site $V$ term increases, a critical value of it can be found at a particular charge-ordering temperature $T_{co}$ where the minimal energy configuration is not a homogeneous charge distribution, but a charge disproportionated one \cite{hubbard}. Above $T_{co}$, electrons are localized without the formation of electric dipoles, while below $T_{co}$ there is a finite charge disproportionation \cite{dressel2} between the molecules (above a critical value of the inter-site Coulomb repulsion $V_c$, a charge disproportion takes place \cite{hubbard}), creating electric dipoles and thus the emergence of a finite electrical polarization below $T_{co}$. From such electrical configuration, the ferroelectric phase emerges, concomitant with the charge-ordering one. Another strong evidence of ferroelectricity is an anomaly (maximum) in the dimensionless real part of the dielectric constant $\varepsilon'$ measurements for such systems reported in the literature \cite{nad2,prbrioclaro}. Since the dielectric constant is directly associated with the polarizability of the system through the Clausius-Mossoti relation \cite{kittel}, such anomaly reveals a finite polarization of the system at $T_{co}$, where a ferroelectric phase develops. In other words, the maximum in $\varepsilon'$ is associated with the free-energy variation between the two phases, namely paraelectric and ferroelectric phases, which is maximized at $T_{co}$ vanishing above and below $T_{co}$.
The latter can be understood upon writing the free-energy density $\hat{F}$ in terms of the order parameter, in this case the polarization $P$, and the relation between $\varepsilon'$ and $P$, namely, $\hat{F} = -EP + g_0 + 1/2g_2 P^2 + ...$ \cite{kittel}, where the $g_n$ are temperature-dependent coefficients, and $P = E(\varepsilon' - 1)$ \cite{kittel}. Hence, it becomes evident that $\hat{F}$ depends on $\varepsilon'$. Recently, we have reported on quasi-static dielectric constant measurements in the $c^*$-axis for the fully-hydrogenated PF$_6$, 97.5\% deuterated PF$_6$, SbF$_6$, and AsF$_6$ systems, showing a systematic comparison between a normal ferroelectric behaviour of the partially deuterated PF$_6$ system and the relaxor-like ferroelectric behaviour of the fully-hydrogenated PF$_6$ one \cite{prbrioclaro}, as well as the transition from a normal to a relaxor ferroelectric behaviour on the AsF$_6$ upon increasing the X-ray irradiation exposure of the samples \cite{prbrioclaro}.
Currently, systems where ferromagnetism and ferroelectricity coexist (multiferroic/magnetoelectric) have been intensively explored by the scientific community \cite{fiebig,cheong,eerestein}, since it is a rare phenomenon and, by the fundamental point of view, very interesting to investigate the possible mechanisms that lead to a multiferroic/magnetoelectric behaviour. Furthermore, multiferroicity has been proposed to be present in the quasi-one-dimensional organic systems of the (TMTTF)$_2$X family \cite{multi}. Note that it is not yet clear wether such systems present a multiferroic or a magnetoelectric behaviour. On one hand, in a multiferroic material there is a coupling between the electric and magnetic order parameters \cite{eerenstein}, i.e., both electric magnetic phases coexist. On the other hand, a magnetoelectric system refers to the change of an electric phase under magnetic field or vice-versa \cite{khomskii}. Hence, the possible multiferroic/magnetoelectric character of the TMTTF salts is one of the key points explored in this Thesis. Another extense investigation on the Fabre-salts has been related to the vibrational modes of the TMTTF molecules. There are several representations of the vibrational character of such a molecule: $b_{1u}$, $b_{2u}$, and $b_{3u}$, which are infrared active and the $a_g$, $b_{1g}$, $b_{2g}$, and $b_{3g}$, which are Raman active \cite{dresselraman}, just to mention a few. It is evident that the charge distribution of the TMTTF molecule directly affects the existent vibrational modes, as can be seen in Fig.\,\ref{ramanmolecules}.
 \newline
Besides that, below $T_{co}$ the charge distribution is inequivalent due to the inter-site Coulomb repulsion $V$ \cite{hubbard,shunsuke} and a splitting of the pattern of the vibrational mode is observed by infrared and Raman optical investigations (Fig.\,\ref{splitting}).
The charge disproportionation $\delta$ observed in Fig.\,\ref{splitting} is related to the difference $\Delta\nu$ between the resonant frequencies and can be computed by the following expression \cite{dresselraman}:
\begin{equation}
2\delta = \frac{\Delta\nu}{80\,\textmd{cm}^{-1}/e}.
\end{equation}
At this point, it is worth mentioning that the lock down that lasted about 2 years due to the Covid-19 pandemic severely affected the realization of the planned experiments in the frame of this Thesis. However, during the lock down period, theoretical works proposed by Prof.\,Dr.\,Mariano de Souza were developed culminating in nice publications, cf.\,Section\,\ref{resultssection}. Hence, this is the reason this Thesis provides more theoretical results than experimental ones. The proposed target experiments will be continued in the frame of a post-doc research project.

\section{Experimental aspects}

Before discussing the obtained experimental results obtained, the basic principles of work and maintenance for an optimized and proper functioning of the Teslatron-PT cryostat are provided.

\subsection{Principle of work of the Teslatron-PT cryostat}

\begin{figure}[!htb]
\centering
\includegraphics[width=0.9\textwidth]{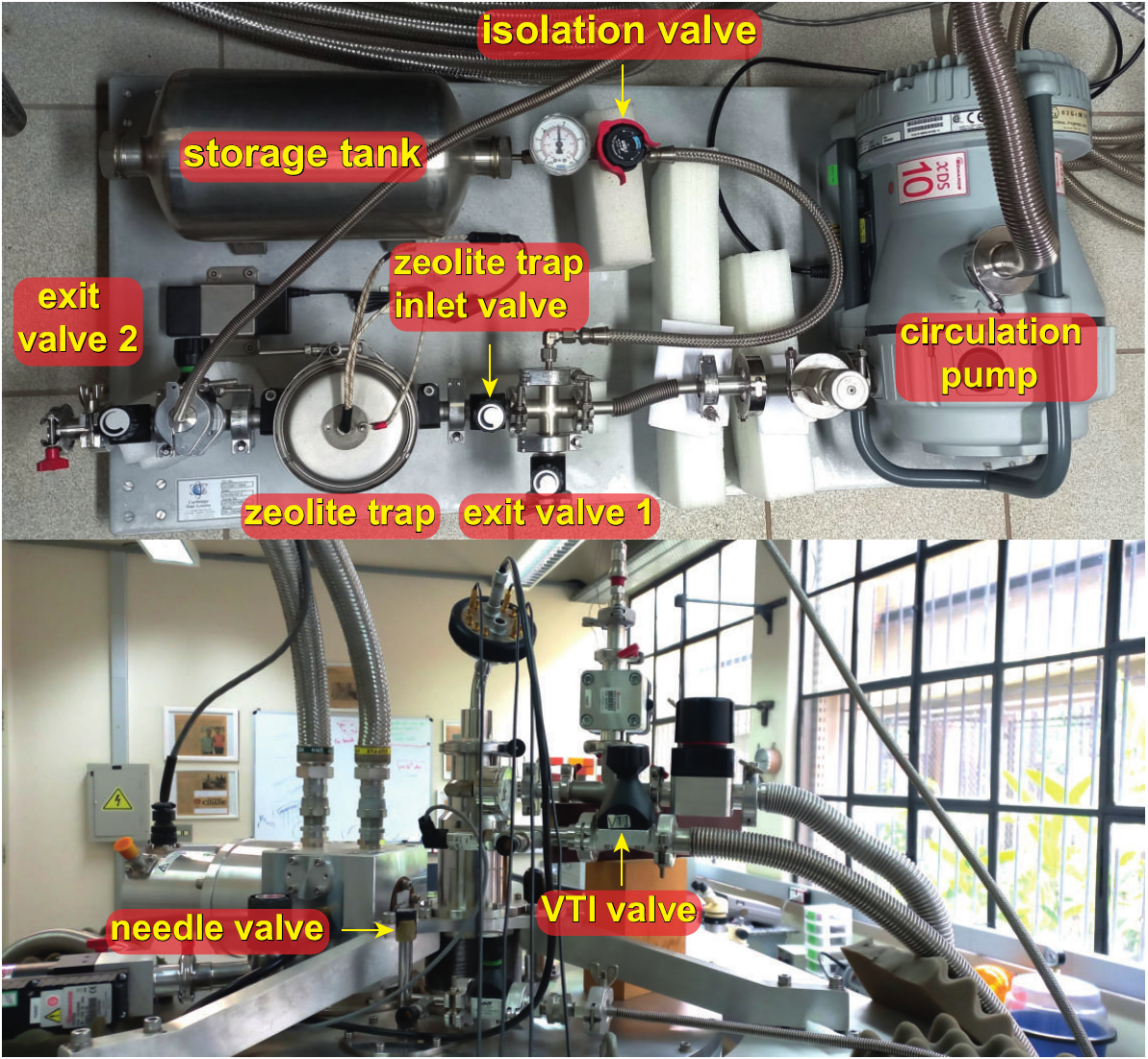}
\caption{\footnotesize Upper panel: circulation system showing the storage tank containing $\sim$\,0.55\,bar of $^4$He, the isolation valve, zeolite trap, zeolite trap  inlet valve, XDS 10 circulation pump, and exit valves 1 and 2. Lower panel: top view of the Teslatron PT cryostat showing the needle valve, which controls the cooling power, and the variable temperature insert (VTI) valve.}
\label{heliumcirculation}
\end{figure}

\begin{figure}[!htb]
\centering
\includegraphics[width=0.8\textwidth]{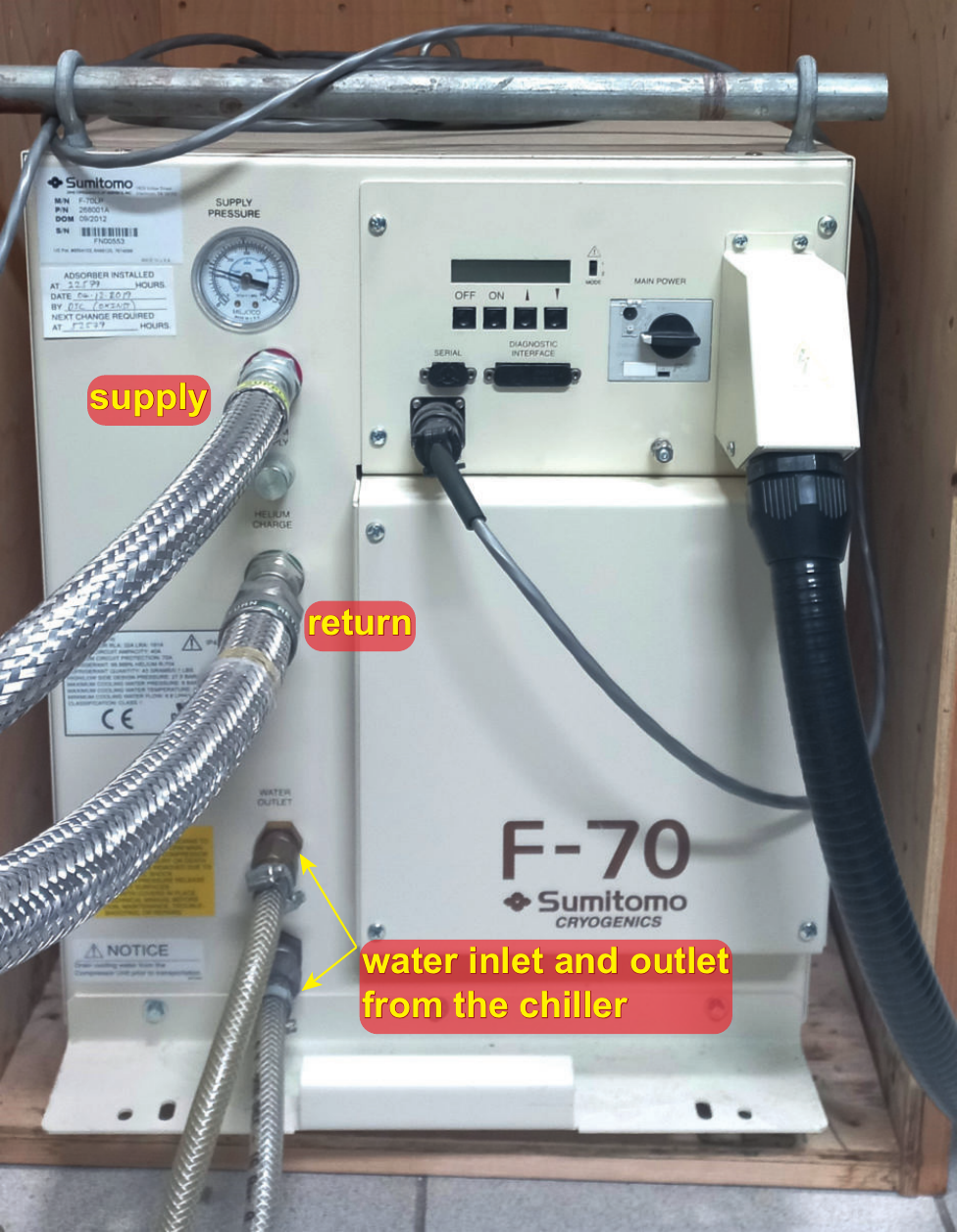}
\caption{\footnotesize F-70 Sumitomo compressor showing the supply and returns $^4$He lines and the water inlet and outlet lines from the chiller to cool down the compressor during its functioning.}
\label{sumitomo}
\end{figure}

The Teslatron-PT cryostat relies in a $^4$He closed cycle where a circulation circuit is attached to it in order to continuously cool down the interior of the cryostat. Such a cooling is achieved through a F-70 Sumitomo compressor (Fig.\,\ref{sumitomo}), which compresses and expands $^4$He. In order for the $^4$He to cool down, first the gas is compressed and then it does work in the so-called compression chamber in order to expand, which decreases its temperature. Such a cold $^4$He is then circulated through the cryostat exchanging heat with it, returning warmer to the compressor, and the cycle is continuously repeated. The $^4$He flow/pressure is controlled by a so-called needle valve (Fig.\,\ref{heliumcirculation}) that uses a tapered needle to regulate the flow of $^4$He. The needle is connected to a handle that enables a fine adjustment of the $^4$He flow/pressure. As the needle is turned, it moves up or down within the valve body, altering the size of the opening through which the $^4$He can pass. When the needle is completely closed, $^4$He cannot flow through the valve and its pressure goes to zero. As the needle is gradually opened, the pressure increases. Such a needle valve is crucial to the cooling process of the cryostat, since the pressure of the $^4$He flow must be precisely adjusted during the cooling steps in order to achieve base temperature, which is $T \approx 1.4$\,K. Now, in order to mathematically understand the Teslatron-PT cooling process, the first-law of Thermodynamics is recalled \cite{ralph}:
\begin{equation}
dQ = dU - dW.
\end{equation}
Considering that $dW = pdv$, the first-law can be rewritten as:
\begin{equation}
dQ = dU - pdv.
\label{principleofwork}
\end{equation}
Note that the sign of $dW$ will depend if the work is being done into the system ($dW < 0$) or by the system ($dW > 0$). Such a simple analysis enables to infer that if work is done on the gas, it increases its internal energy and, as a consequence, its amount of heat, which is the case of the gas being initially compressed. On the other hand, when the gas freely expands after being compressed it performs work at expense of its internal energy and heat amount, which cools the $^4$He. Such a continuous cooling process after $\sim$\,30$-$40\,h enables the achievement of a base temperature close to 1.4\,K. It is worth mentioning that the achievement of the base temperature is not a plug and play process. In order to do so, the following procedures must be followed according to Oxford Instruments operation manual listed in what follows in more details:\\

\textbf{1) Starting of the $^4$He circulation system}

The $^4$He circulation system lies in a room next to the Solid State Physics Laboratory to avoid excessive noise when it is in operation. Essentially, it is composed by a $^4$He storage tank, the so-called zeolite trap, which works as a ``strainer'' to confine moisture, and the XDS 10 pump (supplied by Edwards company) that circulates the $^4$He into the cryostat (Fig.\,\ref{heliumcirculation}). In order to initiate the cool down of the cryostat, first the $^4$He circulation system must be turned on so that the $^4$He is released from the storage tank and thus circulates for about 12\,h before starting the compressor. Such an initial process aims to maximize the homogeneity of the $^4$He circulation and, as a consequence, the corresponding efficiency of the cooling. The recommended $^4$He pressure inside the storage tanks is between $\sim$\,0.45$-$0.60\,bar, cf.\,Fig.\,\ref{heliumcirculation}. In order to start the $^4$He circulation system, the following steps must be carried out (see Fig.\,\ref{heliumcirculation} for guidance): \newline
\emph{i)} Open the zeolite trap inlet valve;\newline
\emph{ii)} Turn on the XDS 10 circulation pump; \newline
\emph{iii)} Open the NW25 VTI pumping port valve to start the circulation of $^4$He; \newline
\emph{iv)} Evacuate the outer vacuum chamber (OVC) and leave the $^4$He to circulate for about $\sim$\,12\,h. Usually, this process is performed in the afternoon of the day before the cool down starts.

\vspace{0.5cm}
\textbf{2) Turning on the compressor}

\begin{figure}[!htb]
\centering
\includegraphics[width=0.8\textwidth]{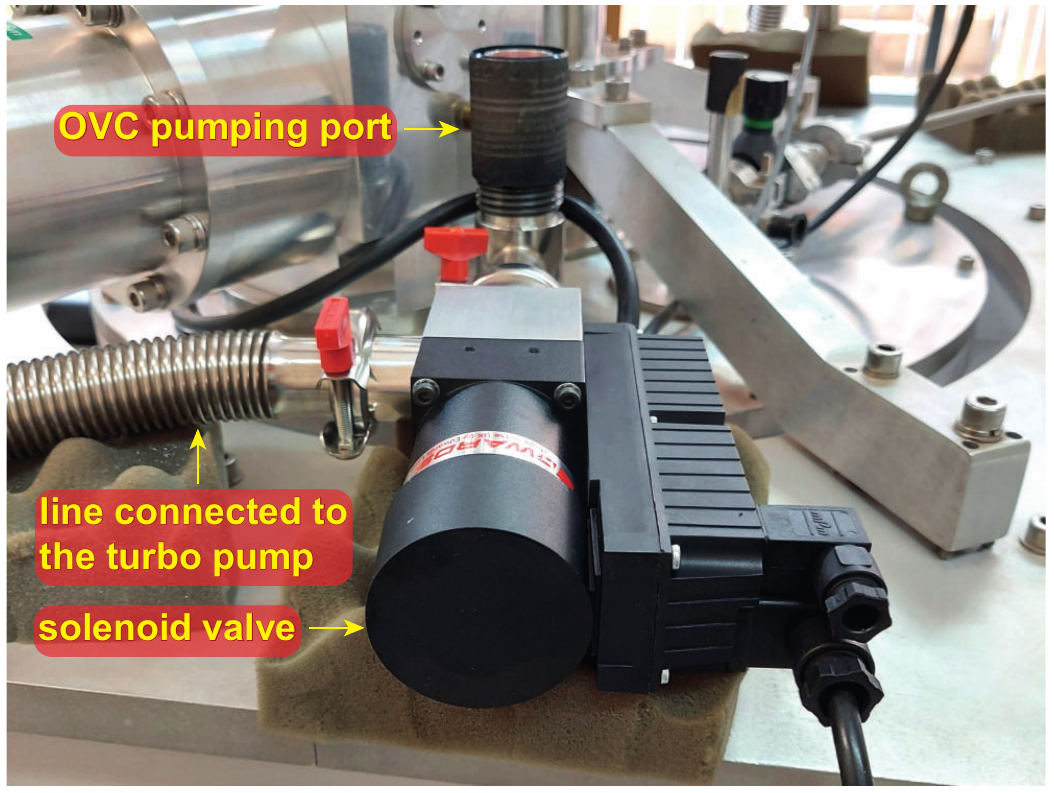}
\caption{\footnotesize Solenoid valve installed between the turbo pump and the OVC pumping port. When a power loss occurs, it becomes de-energized and closes, protecting the OVC from the ventilation of the turbo pump. When power is re-established, it reopens and the pumping of the OVC is continued.}
\label{solenoidvalve}
\end{figure}

After step 1), the needle valve needs to be open half a turn from its fully closed position and the compressor must be started.
Before doing so, it is worth mentioning that, in order to cool down the compressor due to its power dissipated when it is in operation and to avoid overheating, a chiller is attached to it operating with water at about $17\,^\circ\mathrm{\textmd{C}}$. Hence, before turning on the compressor, the chiller must be turned on first for about 15\,minutes. Afterwards, it is finally the time to switch on the compressor (Fig.\,\ref{sumitomo}) to actually initiate the cool down of the cryostat. After the compressor turned on, the OVC must be continuously pumped until the so-called pulse tube refrigerator second state (PT2) temperature sensor reads below $T = $ 5\,K, which takes between 30$-$40\,h. The room where the compressor and the $^4$He circulation system are installed is constantly refrigerated by an air-conditioner fixed at $23\,^\circ\mathrm{C}$ during the whole time the system is turned on. At this point, it is worth mentioning that, in the case of a power loss during this period, the turbo pump connected to the OVC would ventilate and thus contaminate the interior of the OVC, which could lead to the freezing of moisture inside it severely compromising the vacuum and, as a consequence, the achievement of the base temperature of the cryostat. During the last several years, all the Solid State Physics group members have taken turns during this delicate period in order to always have someone at the laboratory in the eventual case of a power loss, which happens very frequently, cf.\,Appendix Section\,\ref{powerloss}. However, in January/2023 it was acquired a solenoid valve that was installed between the turbo pump and the OVC pumping port (Fig.\,\ref{solenoidvalve}).

\begin{figure}[!htb]
\centering
\includegraphics[width=\textwidth]{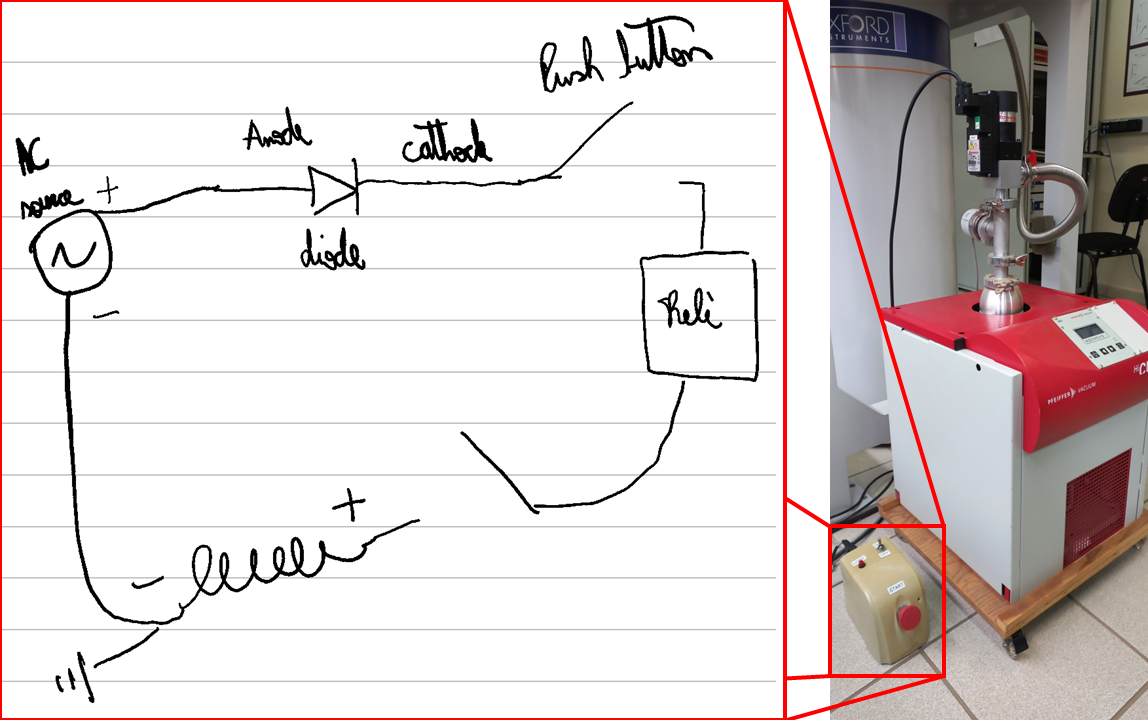}
\caption{\footnotesize Electrical circuit drawn by us to prevent that the solenoid valve opens when power comes back on and the moisture from the ventilation of the turbo pump enters into the OVC. The circuit is composed by a diode, a push button, and a relay. In the case of power being interrupted and then normalized, the solenoid valve is only opened when the push button is manually pressed.}
\label{circuitvalve}
\end{figure}
However, only the installation of the solenoid valve is not enough to fully protect the OVC from being contaminated in the case of a power loss. It turns out that if the power supply is turned on, the solenoid valve is opened and the turbo pump is working fine. If the power goes down, the solenoid valve closes and the turbo pump ventilates in the pumping lines, which connects the turbo pump, the solenoid valve and the OVC pumping valve. Hence, if the power comes back again, the solenoid valve can be opened more quickly than the turbo pump initiates the pumping process. In such a case, the moisture from the ventilation of the turbo pump in the pumping lines can enter into the OVC, since the pumping lines will be at ambient pressure and the OVC in a lower pressure. Then, in order to prevent this to happen and to guarantee that the OVC will not be contaminated, me and Prof.\,Dr.\,Mariano de Souza developed a simple circuit to intermediate the functioning of the solenoid valve. Essentially, such an electrical circuit, cf.\,depicted in Fig.\,\ref{circuitvalve} is composed by a diode, a relay, and a push button. The diode is inserted into the circuit to prevent the solenoid valve to be continuously opened and closed to the instabilities in the power source. When power goes down, both the relay and push button will be de-energized and the electrical contact is opened. When power goes down, the relay is only energized if the push button is pressed. Hence, we guarantee that when power goes down, the solenoid valve is only opened when the push button is manually pressed so that the pumping lines are pumped before opening the solenoid valve again. This simple circuit guarantees that the OVC will never be contaminated by moisture in the case of a power loss followed by its subsequent returning.\newline
\vspace{0.5cm}
\textbf{3) Adjusting the needle valve}

\begin{figure}[!htb]
\centering
\includegraphics[width=0.9\textwidth]{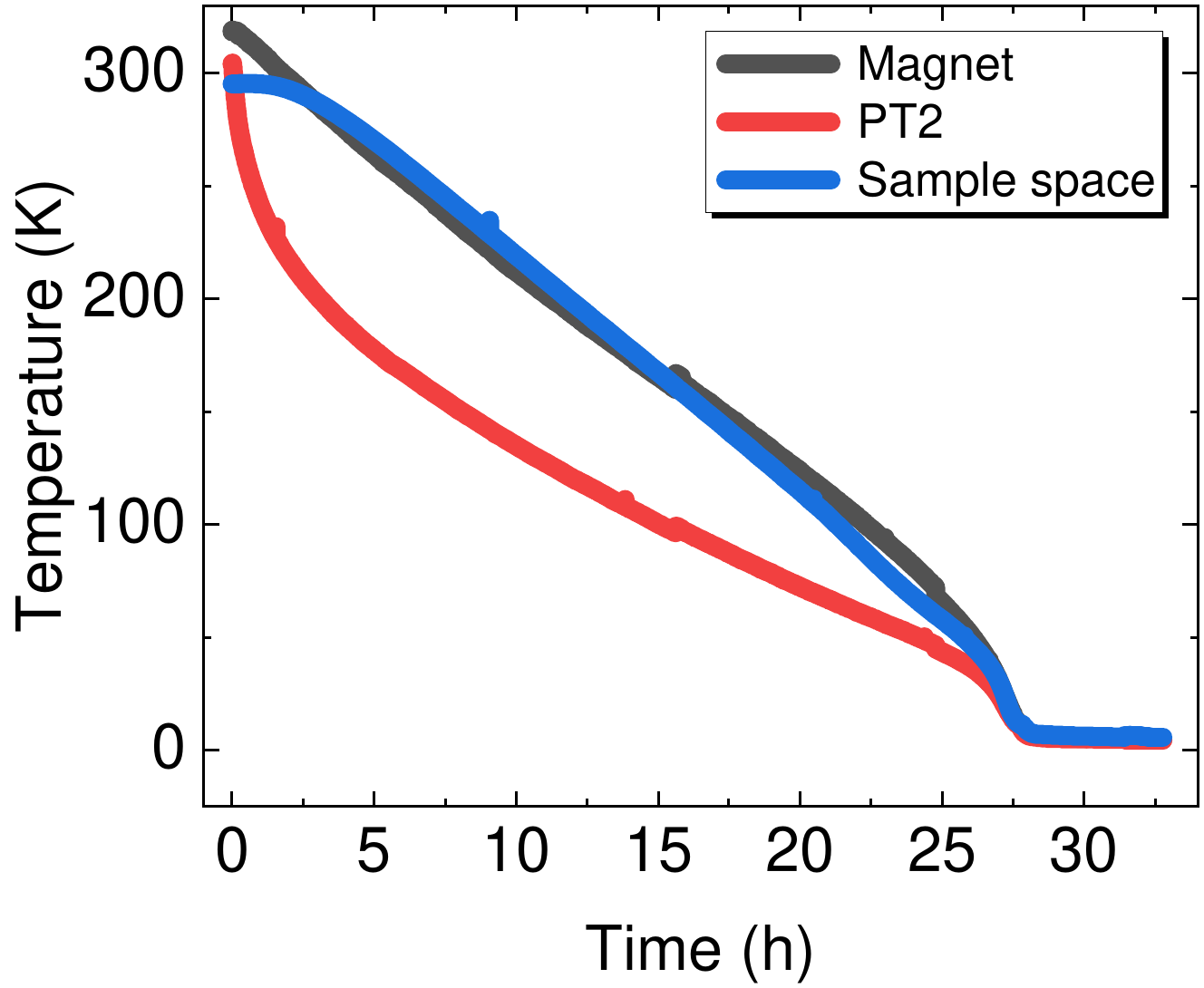}
\caption{\footnotesize Typical behaviour of the temperature \emph{versus} time for the magnet (gray), PT2 (red), and sample space (blue). Note that it takes about 28\,h for all temperatures to reach equilibrium but the cool down is not finished. The last step (not shown) is to adjust the needle valve pressure and wait for all temperatures to further decrease until the base-temperature is achieved, which usually takes several hours.}
\label{cooldownplot}
\end{figure}

Several hours after the compressor is turned on, the temperatures of the PT2, sample space, and the magnet are decreased. According to Oxford Instrument's manual, the pressure of the needle valve should be adjusted to 10\,mbar when all temperatures converge and stop decreasing. Such an adjustment of the needle valve pressure must be continuously repeated until all temperatures falls below 4\,K. Only at this point, the OVC can be closed and the turbo pump turned off, highlighting the importance of the solenoid valve to avoid contamination inside the vacuum chamber. Hence, the needle valve pressure should be adjusted to 5\,mbar in order to achieve base-temperature ($T \backsimeq 1.4$\,K). A typical behaviour of $T$ as a function of time for the PT2, sample space, and the magnet is shown in Fig.\,\ref{cooldownplot}.

\subsection{Cleaning up of the zeolite trap}

\begin{figure}[!htb]
\centering
\includegraphics[width=0.7\textwidth]{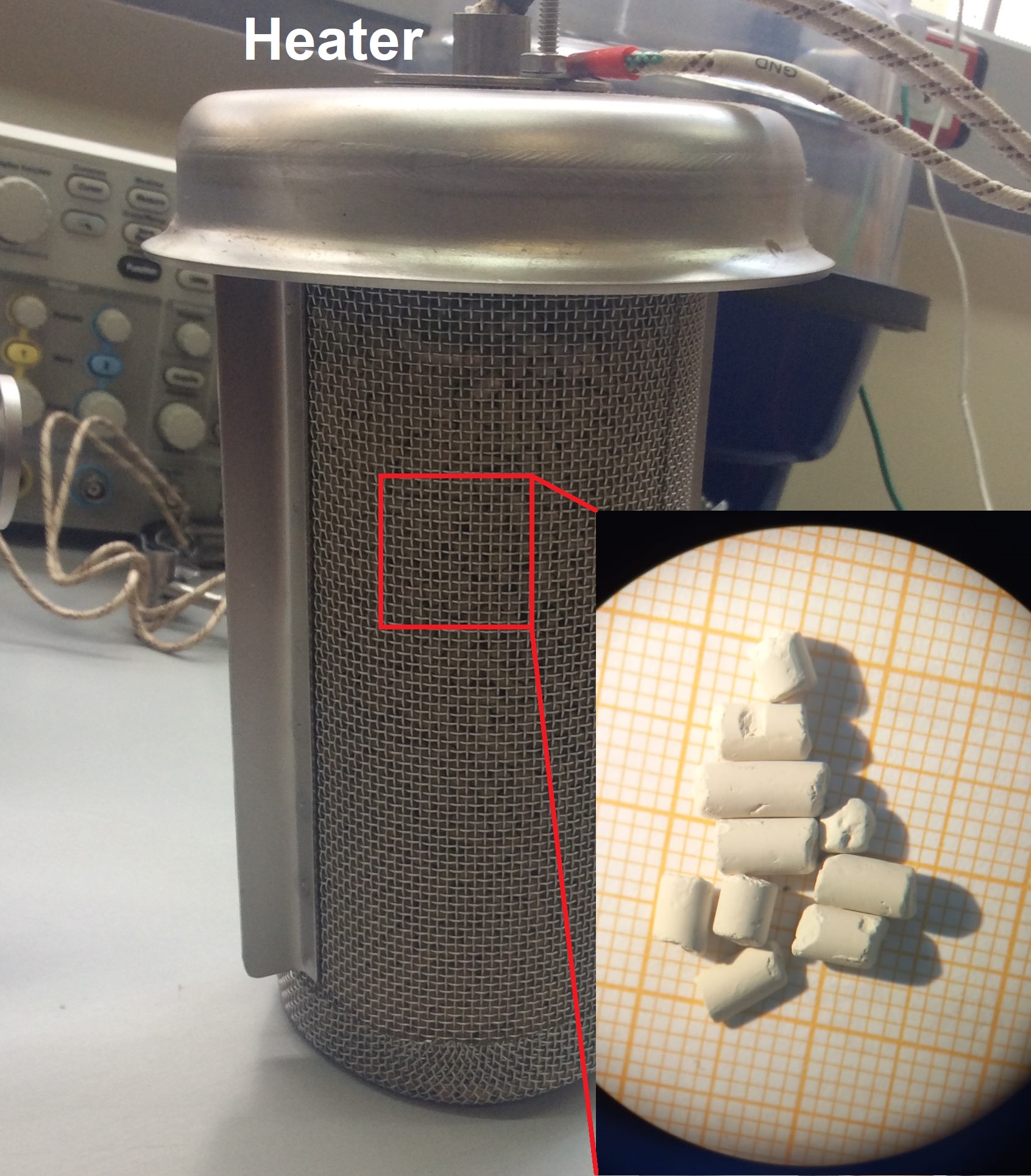}
\caption{\footnotesize Interior of the zeolite trap and the microporous aluminosilicate minerals are show in the inset. At the top of the zeolite trap, it is shown a heater that is used in its cleaning process.}
\label{zeolitetrap}
\end{figure}

A zeolite trap is a device that employs the unique properties of the so-called zeolites, a family of microporous aluminosilicate minerals, to trap and remove impurities from gases \cite{zeolitemolecular}, cf.\,Fig.\,\ref{zeolitetrap}. Such a trap works by adsorption, a process in which impurities are physically adsorbed onto the surface of the zeolite crystals. The pores of the zeolite crystals are precisely sized to selectively adsorb specific molecules, such as water vapor or hydrocarbons, while allowing the passage of other gases. The efficiency of the zeolite trap can be enhanced by heating the zeolite, which desorbs the impurities, allowing for their removal and regeneration of the trap. Zeolite traps have been extensively studied and used in a variety of industrial applications such as air-conditioning and helium circulation systems.

According to Oxford Instruments' manual regarding the operation of the Teslatron PT cryostat, it is recommended that the zeolite trap (Fig.\,\ref{heliumcirculation}) is cleaned after each warmup of the system to room-temperature. In order to do so, the following procedure must be performed:
\begin{itemize}
\item Connect the turbo pump to the exit valve 2 (Fig.\,\ref{heliumcirculation}) but do not open it yet. Note that, at this point following the warm up procedure, the zeolite trap inlet and the VTI valves must be closed, see Fig.\,\ref{heliumcirculation} for guidance;
\item In the mean time, the needle valve must be connected to the power supply for at least one hour, so it heats up and the moisture is easily removed. After one hour, the exit valve, with the turbo pump already turned on, must be slowly opened so that the cleaning starts. At this point, the needle valve pressure is the key indicator to verify the achievement of an appropriate cleaning of the zeolite trap, since, according to Oxford Instruments' manual of operation, the cleaning is completed when the needle valve reaches pressures lower than 0.5\,mbar. Such a process usually takes several hours ($\approx 8$\,h) to be completed;
\item After the needle valve pressure is stabilized close to 0.5\,mbar, the exit valve 2 is closed and the zeolite trap is turned off in order to cool down to room-temperature and the process is then completed.
\end{itemize}

\begin{figure}[!htb]
\centering
\includegraphics[width=0.65\textwidth]{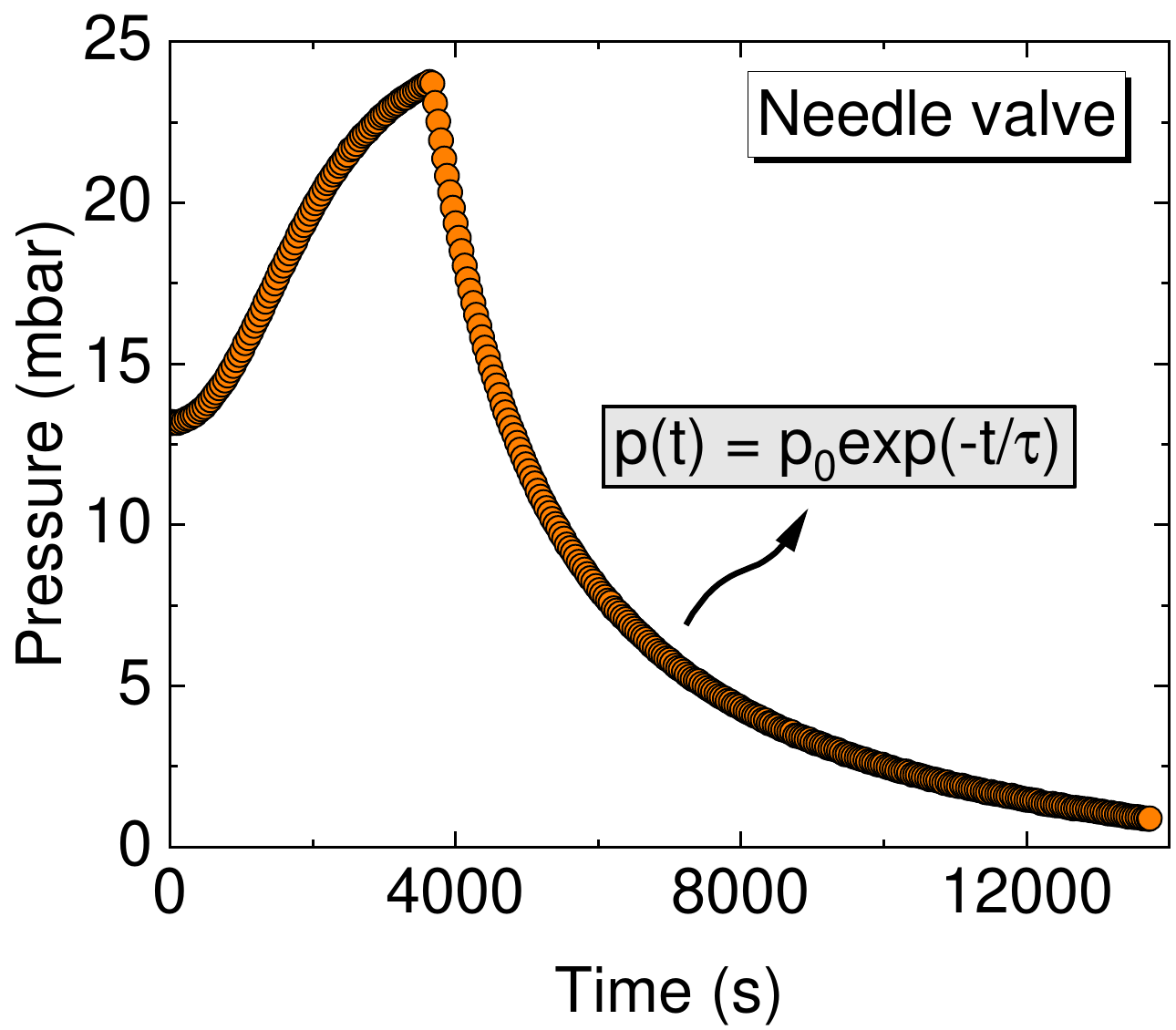}
\caption{\footnotesize Needle valve pressure as a function of time. Note that the pressure is exponentially decreased when the exit valve 2 is opened following a mathematical relation $p(t) = p_0\exp{(-t/\tau)}$, where $p_0$ is the pressure right before the exit valve 2 (see Fig.\,\ref{heliumcirculation}) is opened, $t$ is time, and $\tau$ is the characteristic time or time constant associated with the pumping speed. The initial increase of pressure is due to the heater being turned on and, in order to optimize the cleaning of the zeolite trap, the pumping is performed with the heater still on.}
\label{zeolite}
\end{figure}

A typical behaviour of the needle valve pressure as a function of time can be seen in Fig.\,\ref{zeolite}. Interestingly, an exponential decay of the needle valve pressure is observed when the exit valve 2 is opened. This is not merely a coincidence. In fact, this is predicted by the so-called continuity equation regarding pumping processes, given by \cite{ultravacuum}:
\begin{equation}
vdp = Q_{tot}dt - S'pdt,
\label{continuity}
\end{equation}
where $Q_{tot}$ is the total amount of gas loads entering the vessel with unit of [Js$^{-1}$] and $S'$ is the pumping speed with unit of [m$^3$s$^{-1}$]. The pressure variation $dp$ in a container of fixed volume $v$ is dictated by Eq.\,\ref{continuity}, which takes into account the total amount of gas entering the vessel in a time interval $dt$ and the quantity of gas being pumped out in the same time interval at a given pumping speed $S'$. Since the zeolite trap is completely sealed, i.e., there is no particle influx from the outside, $Q_{tot}$ is equal to zero. Therefore, Equation\,\ref{continuity} can be rewritten as:
\begin{equation}
v\frac{dp}{dt} = -S'p,
\end{equation}
which in turn reads:
\begin{equation}
\frac{1}{p}dp = -\frac{S'}{v}dt.
\label{continuity2}
\end{equation}
Integrating both sides of Eq.\,\ref{continuity2}:
\begin{equation}
\int_{p_0}^{p}\frac{1}{p}dp = \int_{t_0}^{t}-\frac{S'}{v}dt,
\end{equation}
the following relation is achieved assuming $t_0 = 0$:
\begin{equation}
\ln{(p)} - \ln{(p_0)} = -\frac{S'}{v}t,
\end{equation}
\begin{equation}
\ln{p} = \ln{p_0} -\frac{S'}{v}t,
\end{equation}
\begin{equation}
e^{\ln{p}} = e^{\ln{p_0} -\frac{S'}{v}t} = e^{\ln{p_0}}e^{-\frac{S'}{v}t},
\end{equation}
\begin{equation}
p = p_0e^{-\frac{S'}{v}t}.
\label{continuity3}
\end{equation}
Assuming that a characteristic time $\tau$ can be defined as $\tau = v/S'$, Eq.\,\ref{continuity3} reads:
\begin{equation}
\boxed{p = p_0e^{-\frac{t}{\tau}}}.
\label{continuity4}
\end{equation}
Equation\,\ref{continuity4} describes the exponential decay of pressure over time due to pumping, which is the case of the zeolite trap cleaning shown in Fig.\,\ref{zeolite}.

\subsection{Pump and purge of the $^4$He circulation lines}

As discussed in the last Section, although a zeolite trap is inserted into the circulation lines in order to minimize the moisture in the circulating $^4$He, it is strongly suggested that the $^4$He circulation lines are completely cleaned periodically. Such a process is described in the following:
\begin{itemize}
\item First, the turbo pump is connected to the exit valve 1 (Fig.\,\ref{heliumcirculation}) together with the $^4$He line connected to the $^4$He tank. The needle and the VTI valves are completely opened, the circulation pump is turned on and the $^4$He is released from the storage tank. At this point, the exit valve 1 is slowly opened so that all the $^4$He is pumped out from the circulation system. The heater of the zeolite trap can be turned on to optimize this process;
\item As a second step, a pump and purge of the system is carried out systematically, i.e., about 0.5\,bar of $^4$He is inserted and pumped out a few times in order to guarantee that all moisture is removed;
\item As a final step, 0.6\,bar of 5N high-purity (99.999\%) $^4$He is inserted and the storage tank is closed. Then, all $^4$He is pumped out from the circulation lines for a few hours and the cleaning process is completed.
\end{itemize}
After following the steps above, the system is ready for the initiation of a subsequent cool down of the Teslatron-PT system.

\section{Experimental results}

\subsection{Dielectric constant}
\begin{figure}[!htb]
\centering
\includegraphics[width=0.8\textwidth]{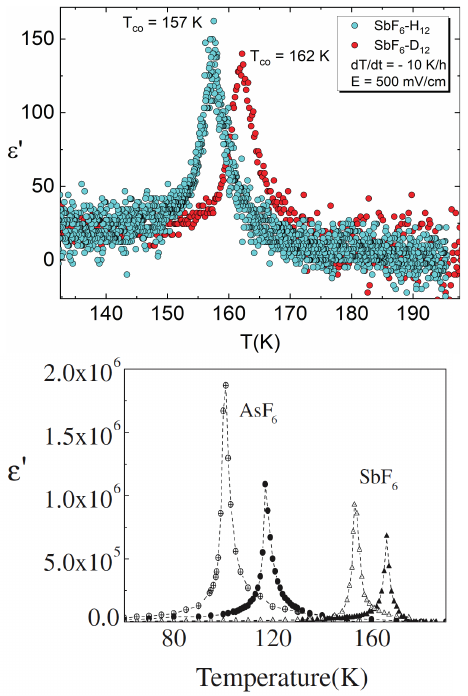}
\caption{\footnotesize Top panel: dielectric constant $\varepsilon'$ measurements as a function of temperature $T$ for the hydrogenated (blue circles) and deuterated (red circles) variant of the (TMTTF)$_2$SbF$_6$ systems for a fixed frequency $f = 1$\,kHz and electric field $E$ = 500 mV/cm. Their corresponding charge-ordering temperature $T_{co}$ are shown. Bottom panel: dielectric constant $\varepsilon'$ measurements as a function of temperature $T$ at a frequency $f$ = 100\,kHz for the hydrogenated (crossed circles) and deuterated (full circles) of (TMTTF)$_2$AsF$_6$ and hydrogenated (white triangles) and deuterated (full triangles) of (TMTTF)$_2$SbF$_6$. Figure extracted from Ref.\,\cite{nad2}.}
\label{dielectricconstantmeasurements}
\end{figure}

Systematic dielectric constant measurements were carried out at both zero and under various magnetic fields values for the Fabre salts. The obtained results reproduce literature results nicely, cf.\,Fig.\,\ref{dielectricconstantmeasurements}. The experiments were affected by the Covid-19 pandemic and shall be continued in the frame of future projects.

The samples were prepared upon painting the surfaces associated with the $c^*$ axis (Fig.\,\ref{tmttf}) with carbon paint in order to emulate a parallel plates capacitor aiming to explore the ionic contribution to the dielectric constant \cite{prbrioclaro}. Next, tempered golden wires with 25\,\textmd{$\mu$}m diameter were attached to the surfaces painted with carbon paste. The sample was then attached to an insulating mask and the golden wires were connected to the mask employing silver paste. Then, such a mask with the contacted sample is attached to the sample holder of the cryostat that goes inside the sample chamber, which is then electrically connected to an Andeen-Hagerling capacitance bridge for the electrical capacitance measurements as a function of temperature.

\subsection{Raman spectra and fluorescence background of the TMTTF salts}

The optical investigations on the TMTTF-based salts were performed in the laboratory of Prof.\,Dr.\,Marcio Daldin Teodoro at the Federal University of S\~ao Carlos (Fapesp process number 2018/06328-1). Systematic Raman spectroscopy measurements as a function of both $T$ and $\vec{B}$ were carried out employing the following available experimental setup:

\begin{itemize}

\item Solid State Diode Cobolt 08-01 Series laser with emission in $\lambda$ = 532\,nm (green) and high-stability with width line lower than 1\,MHz and power up to 20\,mW. Also, this laser has a so-called integrated optical insulator, making it immune to the destabilization by the reflections of light through the microscope.

\item 3 Semrock filters attached to the experimental setup in order to achieve Raman laser lines up to 100\,cm$^{-1}$ ($\lambda = 100\,\mu$m).

\item Confocal attocube microscope of the attoCFM I model designed in titanium aiming to provide high mechanical and thermal stability.

\item Andor Shamrock 75\,cm spectrometer with Peltier cooling system, employing a diffraction grid of 1,200\,lines/mm, since the intensity of the Raman signal is low. In the detection system, together with the spectrometer, a CCD silicon detector is employed, providing a resolution of about 2\,cm$^{-1}$.

\item Attocube attoDRY1000 $^4$He closed-cycle cryostat, reaching the temperature range of 3.8\,K $<$ $T$ $<$ 300\,K and external magnetic fields of $\vec{B}$ $<$ 9\,T, with mechanical vibrations in the region of the sample lower than 1\,nm. In the cooling process, no liquid helium is employed and the sample is not in direct contact with $^4$He that cools the confocal microscope.
\end{itemize}

\begin{figure}[!htb]
\centering
\includegraphics[width=0.5\textwidth]{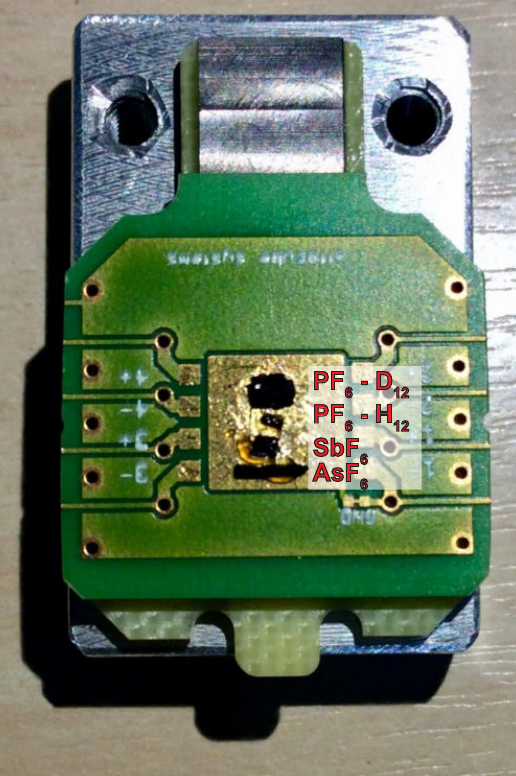}
\caption{\footnotesize Samples attached to the sample holder. From top to bottom the samples are: deuterated variant of (TMTTF)$_2$PF$_6$, hydrogenated variant of (TMTTF)$_2$PF$_6$, (TMTTF)$_2$SbF$_6$, and (TMTTF)$_2$AsF$_6$. The area of the samples were about 2\,mm$^2$ and their thickness of about 1\,mm.}
\label{1.jpg}
\end{figure}

In Fig.\,\ref{1.jpg} it is shown the investigated samples attached to the sample holder. A systematic set of Raman spectra measurements were carried out as a function of temperature and, at each fixed $T$, the external magnetic field was then applied ranging from 0 to 3 to 6\,T, being completely removed before the temperature was varied again. Following such experimental procedure, we have measured the Raman intensity as a function of Raman shift for $T$ = 200, 87, 30, and 4\,K.

The motivation of such experiments under $\vec{B}$ was to detect evidences of multiferroicity/magnetoelectricity in the TMTTF-based systems. Hence, it was possible to check the Raman spectra results from the literature to compare our results enabling us to recover several intensity peaks associated with particular vibrational modes of the TMTTF molecule.
\begin{figure}[!htb]
\centering
\includegraphics[width=\textwidth]{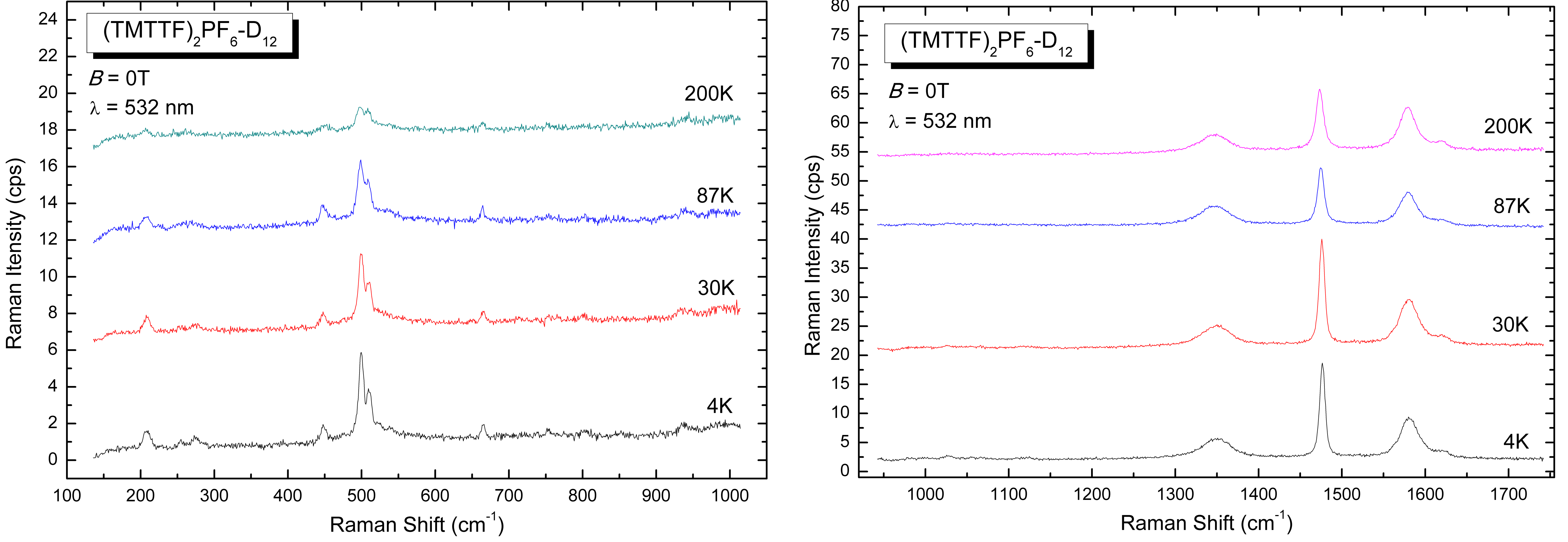}
\caption{\footnotesize Raman intensity \emph{versus} intensity at $\lambda$ = 532\,nm without external magnetic field ($B = 0$\,T) for the deuterated variant of (TMTTF)$_2$PF$_6$ at $T$ = 200, 87, 30, and 4\,K.}
\label{0T}
\end{figure}
\begin{figure}[!htb]
\centering
\includegraphics[width=\textwidth]{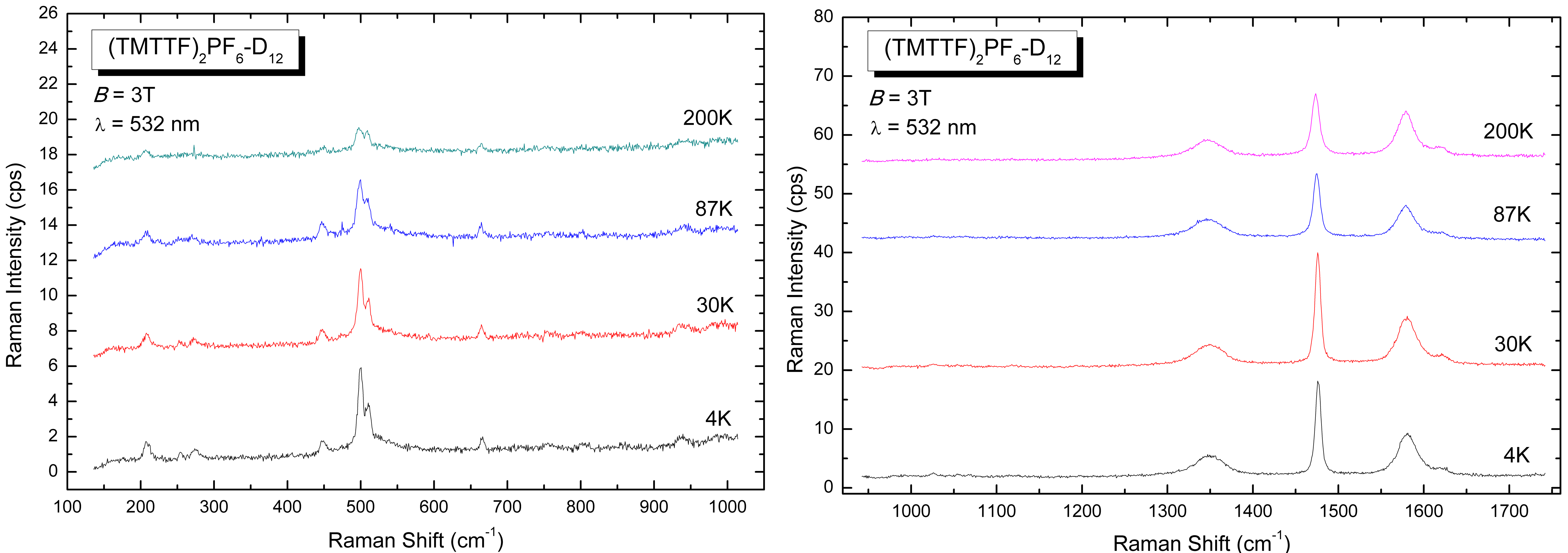}
\caption{\footnotesize Raman intensity \emph{versus} intensity at $\lambda$ = 532\,nm under $B = 3$\,T for the deuterated variant of (TMTTF)$_2$PF$_6$ at $T$ = 200, 87, 30, and 4\,K.}
\label{3T}
\end{figure}
\begin{figure}[!htb]
\centering
\includegraphics[width=\textwidth]{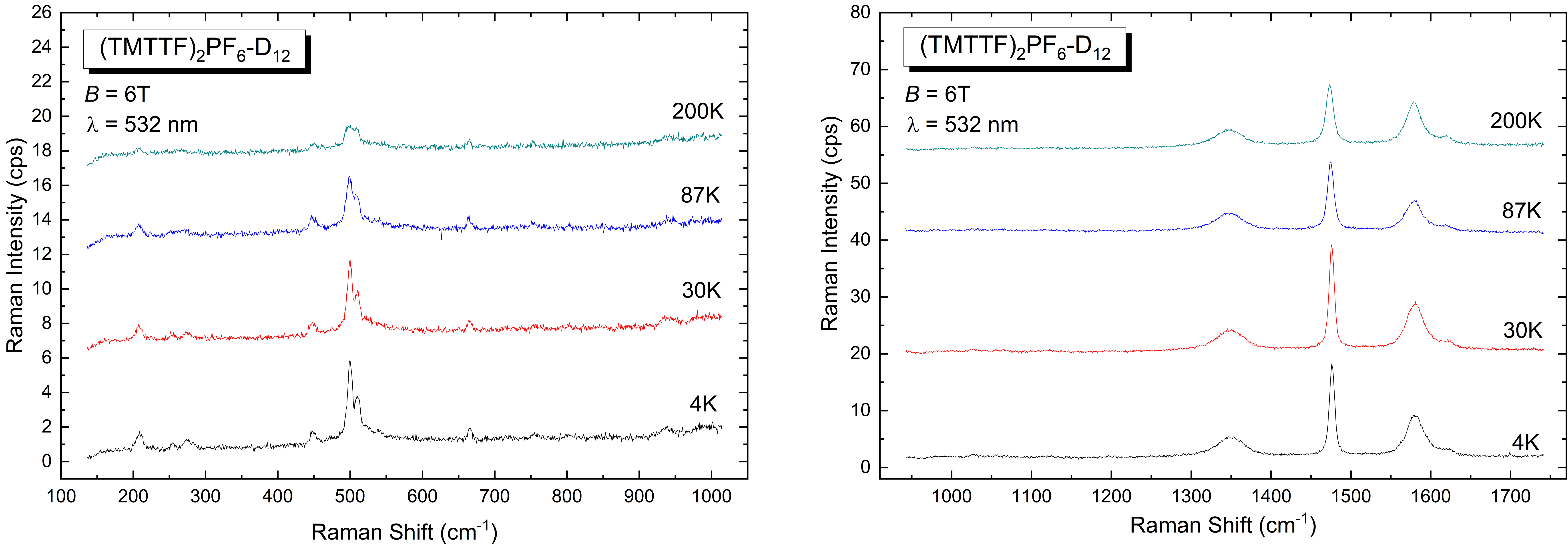}
\caption{\footnotesize Raman intensity \emph{versus} intensity at $\lambda$ = 532\,nm under $B = 6$\,T for the deuterated variant of (TMTTF)$_2$PF$_6$ at $T$ = 200, 87, 30, and 4\,K.}
\label{6T}
\end{figure}
\begin{figure}[!htb]
\centering
\includegraphics[width=0.9\textwidth]{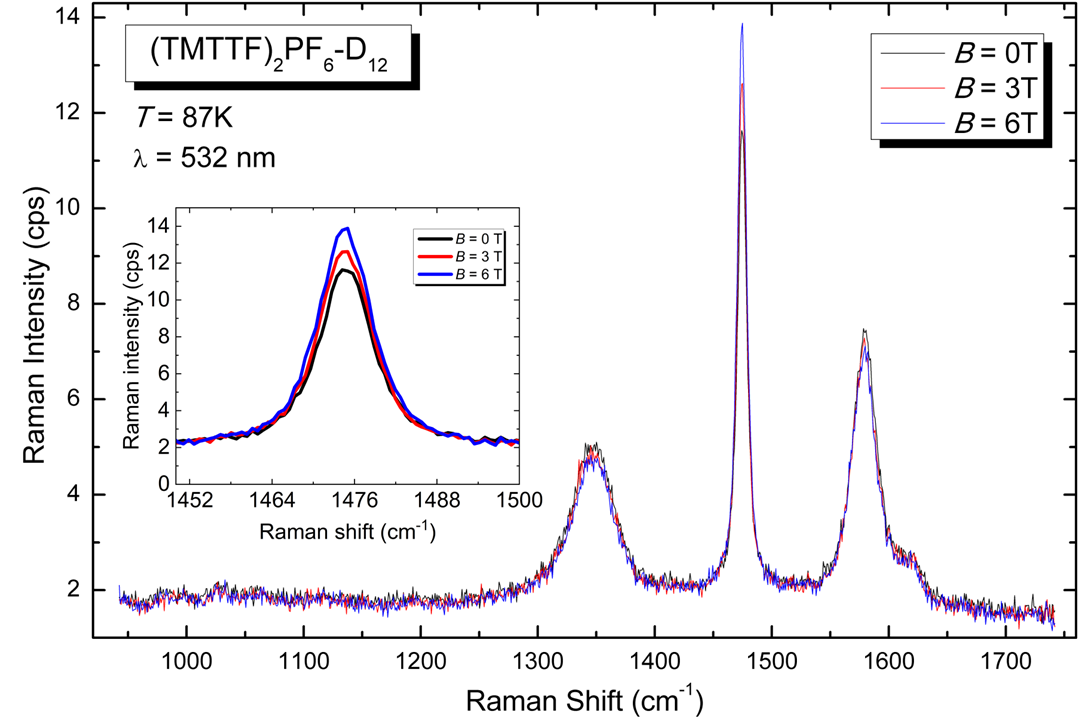}
\caption{\footnotesize Main panel: Raman intensity as a function of Raman shift for the deuterated variant of the (TMTTF)$_2$PF$_6$ system under $B = 0$\,T (black line), $B = 3$\,T (red line), and $B = 6$\,T (blue line). Inset: zoomed region around the Raman shift 1476\,cm$^{-1}$ to indicate that the increase of $B$ reduces the Raman intensity.}
\label{6T2}
\end{figure}
\begin{figure}[!htb]
\centering
\includegraphics[width=0.6\textwidth]{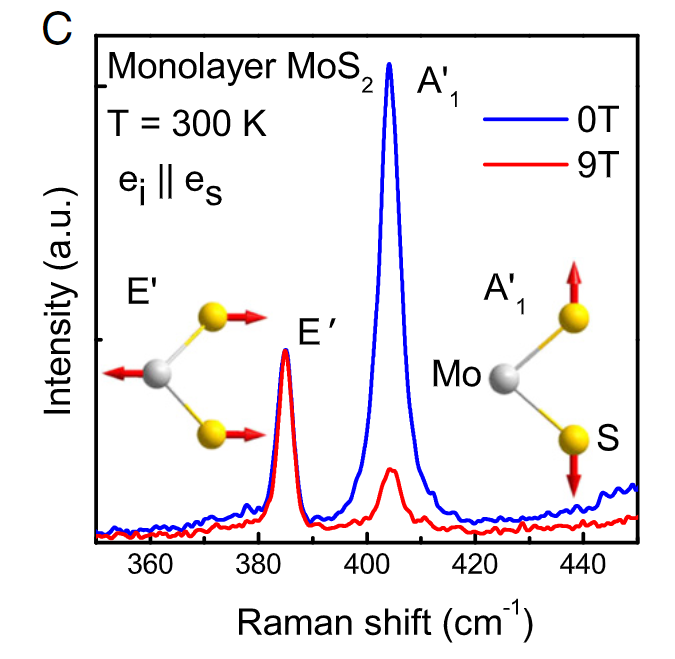}
\caption{\footnotesize Intensity \emph{versus} Raman shift for monolayer MoS$_2$ at $T = 300$\,K showing that the intensity of the Raman shift $\approx 405$\,cm$^{-1}$ is reduced under $\vec{B} = 9$\,T. The vibrational patterns of the corresponding Raman modes are also illustrated. Figure extracted from Ref.\,\cite{mos2}.}
\label{mos2}
\end{figure}
Two key unprecedented results were extracted from such measurements. The first one is regarded to the $\vec{B}$-dependence of the Raman intensity peak at 1476\,cm$^{-1}$ (Fig.\,\ref{6T2}) at $T_{co}$ = 87\,K for the deuterated variant of the (TMTTF)$_2$PF$_6$ salt, which is primordially associated with the methyl-end groups vibrational modes \cite{dresselcrystals}. We have observed that, upon increasing $\vec{B}$ the Raman intensity peak at 1476\,cm$^{-1}$ was decreased by about 27\,\% from 0 to 6\,T, which in turn can be interpreted as a magneto-optical effect similarly to that observed for layered MoS$_2$ (Fig.\,\ref{mos2}) \cite{mos2}. It is worth mentioning that, until the writing of this Thesis, such an effect was not reported for the TMTTF-based salts yet.
\begin{figure}[t]
\centering
\includegraphics[width=0.8\textwidth]{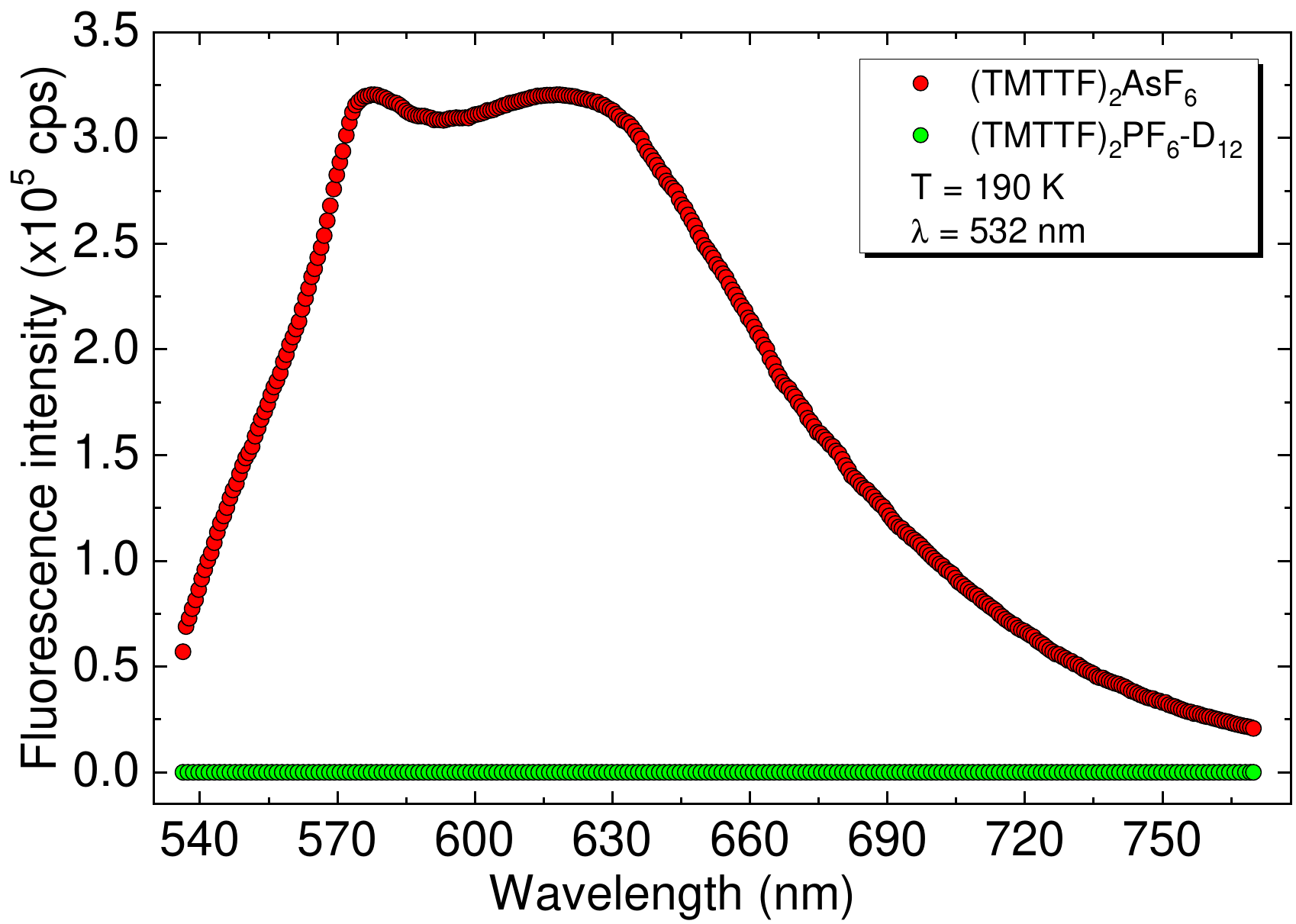}
\caption{\footnotesize Fluorescence intensity \emph{versus} wavelength at $T = 190$\,K and $\lambda = 532$\,nm for (TMTTF)$_2$AsF$_6$ (red circles) and the deuterated variant of (TMTTF)$_2$PF$_6$ (green circles).}
\label{fluorescencebackground}
\end{figure}
Another key result was related to the fluorescence background of the TMTTF-based salts, cf.\,Fig.\,\ref{fluorescencebackground}. We have observed that the fluorescence background of the hydrogenated system (TMTTF)$_2$AsF$_6$ was about five orders of magnitude higher than the one associated with the deuterated variant of the (TMTTF)$_2$PF$_6$ system, showing that deuteration might affect the fluorescence background of the TMTTF molecules. Unfortunately, due to the dilated time of the pandemic such experiments could not be properly continued in order to check the reproducibility of the data and possible sample-to-sample dependence. However, such experimental results are unprecedent in the literature and are planned to be further explored after the finishing of my Ph.D. in the frame of a post-doc project. 
\begin{savequote}[8cm]
``You think I know the first thing about how hard your life has been, how you feel, who you are because I read Oliver Twist? Does that encapsulate you? Personally… I don’t give a damn about all that, because you know what, I can’t learn anything from you I can’t read in some book. Unless you want to talk about you, who you are, then I’m fascinated. I’m in. But you don’t want to do that, do you sport? You’re terrified of what you might say. Your move, chief.''
  \qauthor{--- Good Will Hunting (1997)}
\end{savequote}

\chapter{Summary and conclusions}

This Ph.D. Thesis represents a symbiosis between theory and experiments and it is worth mentioning that the theoretical investigated topics were only possible due to the experimental expertise of the Solid State Physics Group. The theoretical results obtained in the frame of this Thesis employing the various definitions of the Gr\"uneisen parameter covered the fields of magnetism at ultra-low temperatures, caloric effects and critical phenomena, adiabatic magnetization in real paramagnetic systems, and the expansion of the universe, which is a central problem in modern Cosmology. The Brillouin-like paramagnet was first explored and $\Gamma_{mag} = 1/B$ was obtained, being further explored taking into account the mutual interactions between neighbouring magnetic moments. Hence, it was demonstrated that the conditions for probing a genuine zero-field quantum phase transition in terms of $\Gamma_{mag}$ are not fulfilled in real paramagnetic systems when finite $B_{loc}$ is taken into account. Also, a connection between $\Gamma_{mag}$ and the canonical definition of temperature was achieved and the so-called adiabatic magnetization without external magnetic field was proposed by only manipulating the mutual interactions. Experimental setups were proposed to attain adiabatic temperature changes and to detect such a subtle effect of the adiabatic magnetization. Also, the Gr\"uneisen ratio was employed to propose key-ingredients in terms of entropy arguments to design materials that can attain giant caloric effects and can be broadly used in technological applications, as well as that caloric effects are enhanced close to any finite-temperature critical end point. Yet, $\Gamma_{eff}$ was identified as the equation of state parameter $\omega$ in the frame of Cosmology, being recognized to be embedded in Einstein field equations enabling to describe the expansion of the universe under the light of condensed matter Physics and to propose that $\Gamma_{ec}$ can be employed to investigate the anisotropic expansion of the universe. The theoretical results obtained in the frame of this Thesis significantly contributed to the advance of each field. It is remarkable that the various forms of the Gr\"uneisen parameter are embedded in many different areas of knowledge, making one wonders which other fields of research remains unexplored until now in terms of the Gr\"uneisen parameter.

Regarding the experimental part of this Thesis, experiments were carried out in the Solid State Physics Laboratory in Rio Claro, SP and also at the laboratory from Prof.\,Dr.\,Marcio Daldin at UFSCAR in São Carlos, SP. Among the experimental results, dielectric constant and Raman spectra measurements were carried out during the Ph.D. The goal was to probe a possible multiferroic/magnetoelectric character of the TMTTF-based salts. Unfortunately, due to the lock down of the Covid-19 pandemic that lasted 2 years, the planned experiments were severely affected and are planned to be further explored in the frame of a post-doc project. The Raman spectra measurements revealed a possible magneto-optical effect in the intensity peak 1476\,cm$^{-1}$ of the deuterated variant of the (TMTTF)$_2$PF$_6$ system, which deserves further exploration. Also, a huge fluorescence background was observed in (TMTTF)$_2$AsF$_6$, being 5 orders of magnitude higher than in the deuterated variant of the (TMTTF)$_2$PF$_6$ salt. These investigations will be continued after the Ph.D. aiming to clarify the possible multiferroic/magnetoelectric and the magneto-optical behaviours in the TMTTF-based systems.

\begin{savequote}[8cm]
``To know that we know, and that we do not know what we do not know, that is true knowledge.''
  \qauthor{--- Confuncius (551 - 479 BCE)}
\end{savequote}

\chapter{Perspectives and outlook}

The research topics investigated in this Ph.D. Thesis will continue to be explored after the Ph.D. The results obtained for the Gr\"uneisen parameters, namely $\Gamma$, $\Gamma_{eff}$, $\Gamma_{E}$, $\Gamma_{mag}$, and $\Gamma_{ec}$, will enable to extend such analysis to other physical scenarios regarding, for instance, exotic magnetic phases and even to biological systems. Also, there are some well-established projects in this regard already ongoing that will continue to be developed after the Ph.D. Also, planned experiments for a near future are systematic dielectric constant and electric polarization measurements as a function of temperature and external magnetic field to probe the possible multiferroic/magnetoelectric character of the Fabre-salts. High-resolution thermal expansion and magnetostriction measurements are also planned for the various salts. Yet, Raman spectra measurements as a function of temperature and external magnetic field are also planned in a cooperation between Prof.\,Dr.\,Mariano de Souza, Prof.\,Dr.\,Marcio Daldin, and Orsay's team to probe a possible magneto-optical behavior in these salts, as well as a systematic exploration of the fluorescence background of the TMTTF salts and a possible suppression of such a background upon deuterating the methyl-end groups. All these theoretical and experimental topics, among others, are planned to be continued in the frame of a post-doc research project.

\clearpage\begin{savequote}[8cm]
"Rather than love, than money, than fame, give me truth."
  \qauthor{--- Henry David Thoreau (1817 - 1862)}
\end{savequote}

\chapter{\label{app:1-cardiophys}Appendices}

\section{Absence of a perfect alignment between $\vec{\mu}$ and $\vec{B}$}\label{perfect}

\begin{figure}[h!]
\centering
\includegraphics[width=\textwidth]{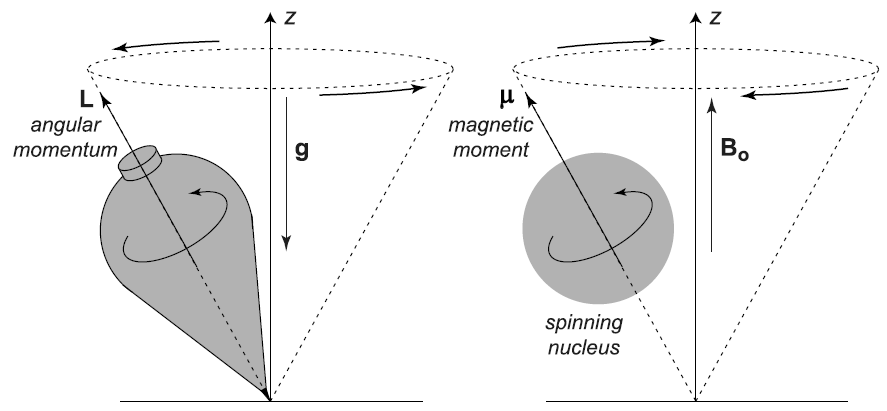}
\caption{\footnotesize Schematic representation of the analogy between the precession of the angular momentum vector $\vec{L}$ of a peon around the $\vec{z}$ axis due to the gravitational field $g$ and the precession of the magnetic moment vector $\vec{\mu}$ of a spinning nucleus around the $\vec{z}$ axis due to an applied magnetic field $\vec{B}$. Figure extracted from Ref.\,\cite{nmr}.}
\label{nmr}
\end{figure}

The magnetic moment vector $\vec{\mu}$ can be analogous to the angular momentum vector $\vec{L}$ of a spinning peon, cf.\,Fig.\,\ref{nmr}. Essentially, when a magnetic field $\vec{B}$ is applied, for instance, in the $z$ direction, there will be a precession of $\vec{\mu}$ around the $z$ axis, which is called Larmor precession \cite{nmr}. In an analogous way, when the peon is spined, $\vec{L}$ precesses around the $z$ axis as well due to the presence of gravity.
The well-known magnetic energy $E_{mag}$ of a magnetic moment under an applied $\vec{B}$ is given by \cite{ralph}:
\begin{equation}
E_{mag} = -\vec{\mu}\cdot\vec{B} = -\mu B\cos{(\phi)},
\end{equation}
where $\phi$ is the angle between $\vec{\mu}$ and $\vec{B}$. It turns out that the minimum energy configuration between $\vec{\mu}$ and $\vec{B}$ refers to a parallel configuration between them, i.e., $E_{min} = -\mu B$ and $\vec{\mu}$ is aligned in the direction of $\vec{B}$, while the maximum magnetic energy $E_{max} = +\mu B$ refers to an anti-parallel alignment between $\vec{\mu}$ and $\vec{B}$.
\begin{figure}[h!]
\centering
\includegraphics[width=0.6\textwidth]{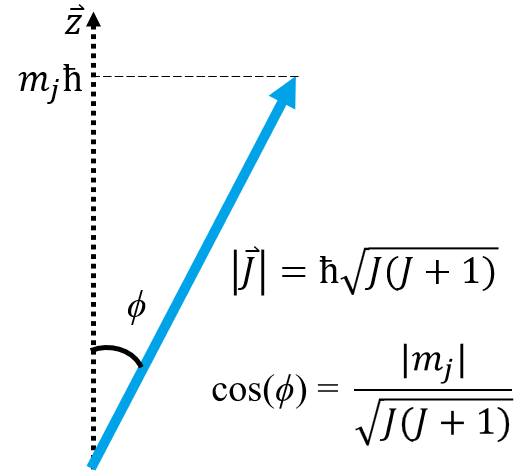}
\caption{\footnotesize Schematic representation of the total angular momentum vector $\vec{J}$ that makes an angle $\phi$ in the $\vec{z}$ direction. The modulus of $|\vec{J}| = \hbar\sqrt{J(J+1)}$ \cite{guimaraes} and its projection in the $\vec{z}$ axis is $m_{j}\hbar$, where $m_j$ is the magnetic quantum number.}
\label{magneticmoment}
\end{figure}
It is possible to write $E_{mag}$ in terms of the magnetic quantum number $m_j = -J, -J+1, ..., J-1, J$ \cite{guimaraes}, which describes all the possible orientations $\vec{\mu}$ can have, being $J$ the total angular momentum quantum number. The vector $\vec{B}$ is along the $\vec{z}$ direction and $\vec{\mu}$ and the total angular momentum vector $\vec{J}$ have the very same direction. Upon considering the projection of $|\vec{J}| = \hbar\sqrt{J(J+1)}$ \cite{guimaraes} in the $\vec{z}$ axis as $m_j\hbar$ \cite{guimaraes}, it is possible to rewrite $\cos{(\phi)}$ between $\vec{\mu}$ and $\vec{B}$ by (Fig.\,\ref{magneticmoment}) \cite{guimaraes}:
\begin{equation}
E_{mag} = -\mu B\cos{(\phi)} = -\mu B\frac{m_j\hbar}{\hbar\sqrt{J(J+1)}} = -\mu B\frac{m_j}{\sqrt{J(J+1)}}.
\label{projection}
\end{equation}
Note from Eq.\,\ref{projection} that a perfect alignment between $\vec{\mu}$ and $\vec{B}$ means that either $\phi$ is 0$^\circ$ or 180$^\circ$, so that $\cos{(\phi)}$ is +1 or -1. This means that both numerator and denominator of the fraction $m_j/\sqrt{J(J+1)}$ in Eq.\,\ref{projection} must be equal. However, upon analysing Fig.\,\ref{magneticmoment}, the projection of a vector along a particular axis will never have the same length as the original vector. Hence, $m_j$ is always lower than $\sqrt{J(J+1)}$ so that a perfect alignment, either parallel or anti-parallel, between $\vec{\mu}$ and $\vec{B}$ is prevented by Quantum Mechanics.

\section{The impossibility of connecting adiabatic deformations and the concept of negative temperatures}\label{stressnegativetemperatures}

As an intellectual exercise, the fundamental concepts of the Purcell and Pound's experiment \cite{purcellandpound} are recalled in order to make a connection with the concept of negative temperatures and the adiabatic stress application. Then, although such analysis looks correct at first, a discussion in terms of the requirements to attain negative temperatures is made to show that such a connection is not sustainable.

Following the canonical definition of temperature, the sign of $T$ is ruled by either a positive or negative $(\partial S/\partial E)_B$, where $E$ is the average magnetic energy \cite{ralph}. Essentially, when a magnetic field is applied in a particular direction, the magnetic moments align in the direction of the field and then $(\partial S/\partial E) > 0$ indicating a positive temperature. However, if the magnetic field is reversed very quickly in the time-scale of micro seconds, i.e., faster than the typical spin-relaxation time, then the magnetic moments will be in an anti-parallel configuration with respect to the magnetic field and thus $(\partial S/\partial E) < 0$ leading to the concept of negative absolute temperature. Now, we present an analogy between $(\partial S/\partial E)_B$ dictating the sign of $T$ and $(\partial S/\partial U)$ in the case of adiabatic strain with a change of temperature \cite{Landau}. First, we recall the entropy associated with the deformation of a system due to a temperature change \cite{Landau}:
\begin{equation}
S = S_0 + K\alpha_p u_{ll},
\label{entropydeformation}
\end{equation}
where $S_0$ is the entropy before the deformation, $K$ is the bulk modulus, and $u_{ll} = (v-v_0)/v_0$ refers to the relative volume change, being $v_0$ the initial volume of the specimen before the deformation. Since we are interested in the adiabatic case, the adiabatic bulk modulus $K_{ad}$ must be employed and its relation with $K$ is given by \cite{Landau}:
\begin{equation}
\frac{1}{K_{ad}} = \frac{1}{K} - \frac{T\alpha_p^2}{c_p},
\end{equation}
which in turn can be rewritten as:
\begin{equation}
K = \frac{K_{ad}c_p}{(c_p + K_{ad}T\alpha_p^2)}.
\label{kad}
\end{equation}
Replacing Eq.\,\ref{kad} in Eq.\,\ref{entropydeformation}, we obtain:
\begin{equation}
S = S_0 + \frac{\alpha_p u_{ll}K_{ad}c_p}{(c_p + K_{ad}T\alpha_p^2)}.
\label{entropydeformation2}
\end{equation}
Now, we make use of the internal energy $U$ per unit of volume associated with an adiabatic deformation \cite{Landau}:
\begin{equation}
U = \frac{K_{ad}{u_{ll}}^2}{2} + \mu'\left(u_{ij} - \frac{1}{3}u_{ll}\delta_{ij}\right)^2,
\label{internalenergypervolume}
\end{equation}
where $\mu'$ is the shear modulus, also called modulus of rigidity, $u_{ij}$ is the strain tensor, and $\delta_{ij}$ is the unit tensor or Kronecker delta. Solving Eq.\,\ref{internalenergypervolume} for $K_{ad}$ and replacing it into Eq.\,\ref{entropydeformation2}, we obtain the expression for $S$ in terms of $U$:%
\begin{equation}
S = S_0 + \frac{2\alpha c_p\left[U-\mu'\left(u_{ij}-\frac{\delta_{ij}u_{ll}}{3}\right)^2\right]}{u_{ll}\left\{c_p+\frac{2\alpha^2 T\left[U-\mu'\left(u_{ij}-\frac{\delta_{ij}u_{ll}}{3}\right)^2\right]}{{u_{ll}}^2}\right\}}.
\end{equation}
Making the derivative of $S$ in respect to $U$ and employing the relation $T = (u_{ll}+\alpha T_0)/\alpha$ \cite{Landau}, where $T_0$ is the starting temperature, we obtain:
\begin{equation}
\left(\frac{\partial S}{\partial U}\right) = \frac{162 \alpha_p  {c_p}^2 {u_{ll}}^3}{\left\{2
   \alpha_p  (\alpha_p  T_0+u_{ll}) \left[9 U-\mu'
   (\delta_{ij} u_{ll}-3 u_{ik})^2\right]+9
   c_p {u_{ll}}^2\right\}^2} = \frac{1}{T}.
   \label{dSdU}
\end{equation}
Note that the denominator of Eq.\,\ref{dSdU} is always positive since it is squared. Assuming that $\alpha_p$ is positive and $c_p$ is always positive, the sign change of Eq.\,\ref{dSdU} is ruled by the term ${u_{ll}}^3$. Note that the consideration that $\alpha_p$ is constant lies on the adiabatic character of the uniaxial stress application, since $\alpha_p = -1/v(\partial S/\partial p)_T$ \cite{ralph}. Thus, if an adiabatic contraction takes place in a time-scale lower than the thermal relaxation time of the system (Fig.\,\ref{pzt}), ${u_{ll}}^3$ is negative since $v < v_0$ and thus $T < 0$ as well. However, in an adiabatic expansion, ${u_{ll}}^3$ is positive since $v > v_0$ and thus $T > 0$. Therefore, the sign change of $T$ is governed by the sign of the relative volume change in an adiabatic deformation. The latter is analogous to the sign of $T$ ruled by the direction of the magnetic moments with respect to $B$ in the Purcell and Pound's experiment \cite{purcellandpound}.  Essentially, the inversion of the voltage applied to the PZT stacks at a given frequency (Fig.\,\ref{pzt}), and consequently, the applied electric field, could be seen as a corresponding situation to the fast inversion of the magnetic field in the seminal experiment proposed by Purcell and Pound's.

Next, a discussion is performed about the non-validity of this approach based on the essential requirements for a thermodynamical system to be capable of negative temperatures, which reads \cite{Ramsey1956}:

\begin{quote}
“(…) (1) The elements of the thermodynamical system must be in thermodynamical equilibrium among themselves in order that the system can be described by a temperature at all; (2) there must be an upper limit to the possible energy of the allowed states of the system; and (3) the system must be thermally isolated from all systems which do not satisfy both of the above conditions, i.e., the thermal equilibrium time among the elements of the system must be short compared to the time during which appreciable energy is lost to or gained from other systems. (…)”
\end{quote}

In the experiment performed by Purcell and Pound \cite{purcellandpound}, the nuclear spin system lies in thermodynamical equilibrium, which means that the temperature is kept constant during the fast inversion of the applied magnetic field, changing the regime of positive $(\partial S/\partial U)_B$ to a negative one. During the adiabatic volume change discussed in terms of a uniaxial stress application, such a deformation is always accompanied by an adiabatic temperature change to hold the entropy constant, thus violating the previously mentioned requirement (1). Also, there is no upper limit in the elastic energy associated with the adiabatic volume variation, which is at odds with requirement (2). Following such arguments, it is straightforward to conclude that requirement (3) is also not fulfilled. Hence, the connection between adiabatic deformations and negative temperatures in connection with the Purcell and Pound’s experiment is not sustainable.

\section{Power down during experiments}\label{powerloss}
During the Ph.D. period, several power losses took place, which affected our experiments since the cryostat had to be warmed up to room temperature before cooling it down again. This takes several days to finish and, by summing up the almost 40 hours to cool down the system again, significantly impacted our experiments. Some of the power loss logs can be seen in the records written in the laboratory notebook below.

\includepdf[pages=-]{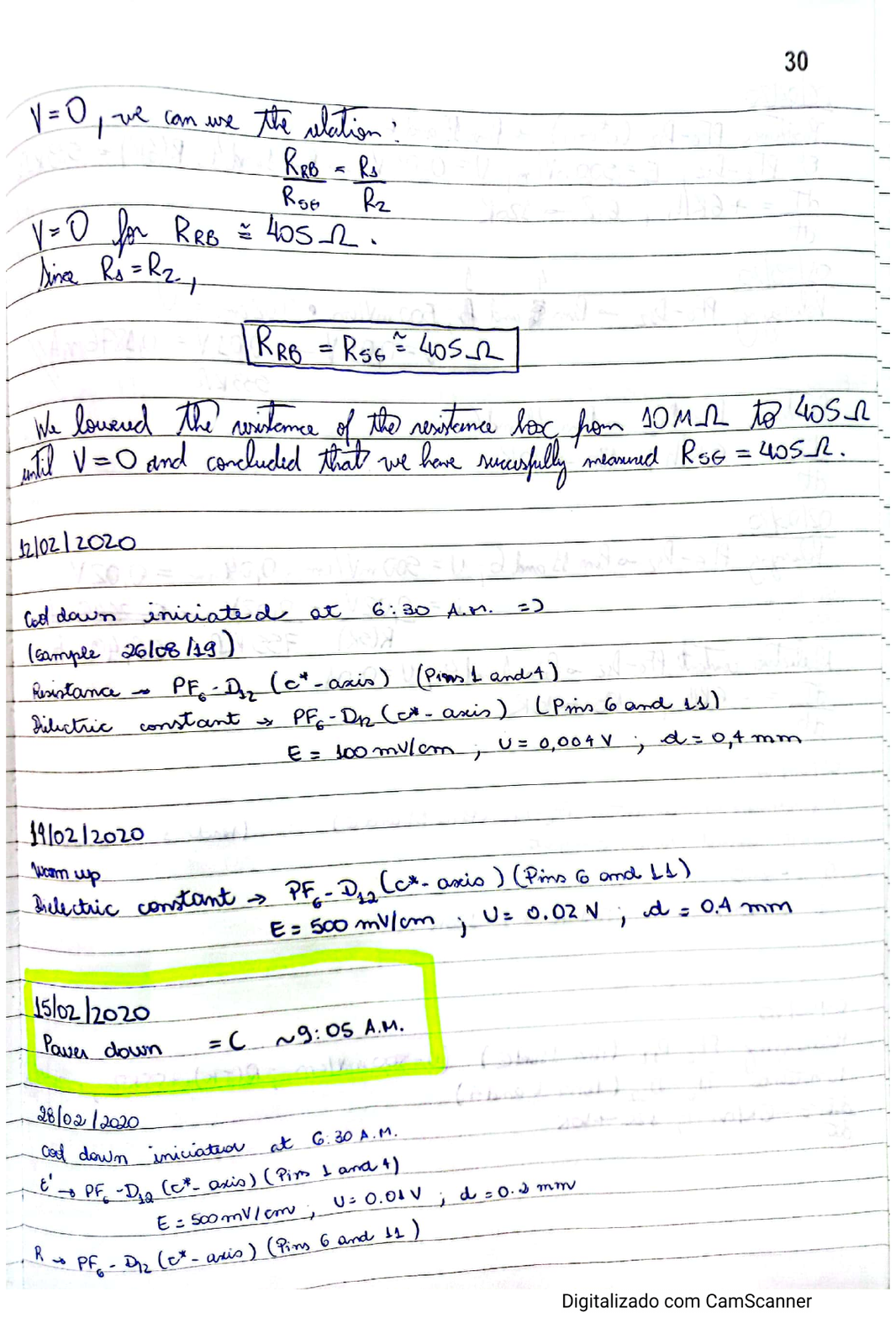}

\section{Technical aspects}

\begin{figure}[h!]
\centering
\includegraphics[width=\textwidth]{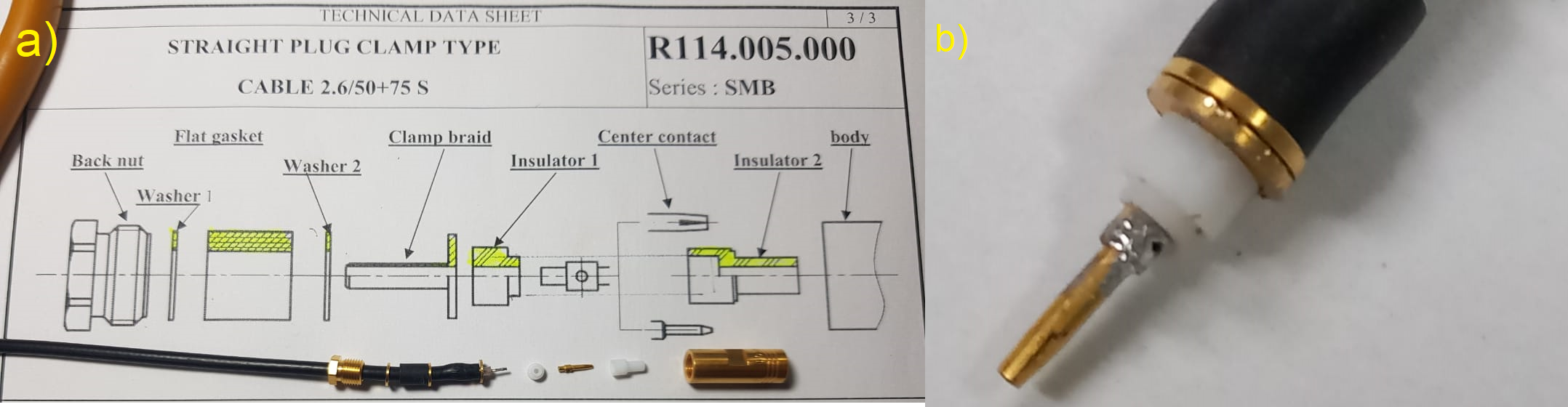}
\caption{\footnotesize a) All the individual parts of the SMB connector together with its datasheet. b) The golden tip of the SMB connector is already soldered in the coaxial cable. Special thanks to Leandro Xavier who performed such a delicate repair.}
\label{coaxialcable}
\end{figure}

During the period of my Ph.D. I had the opportunity to spectate the fixing of a coaxial cable by soldering its tip on a R114.005.000 SMB connector, cf.\,Fig.\ref{coaxialcable}.

\section{Joint reports}

During the period of my Ph.D. degree, Prof. Dr. Mariano de Souza provided me to the opportunity of acting as a co-reviewer in various manuscripts he had being requested to provide a report. The journals I have participated as a co-reviewer are the following:
\begin{itemize}
\item Physical Review B (12 manuscripts);
\item Brazilian Journal of Physics (1 manuscript);
\item AIMS Mathematics (2 manuscripts);
\item Mathematical Biosciences and Engineering (2 manuscript);
\item Physica A: Statistical Mechanics and Its Applications (1 manuscript);
\item Physical Review Materials (2 manuscripts);
\item PLOS ONE (2 manuscripts);
\item Physics Letters A (1 manuscript).
\end{itemize}
Due to these great opportunities Prof. Mariano de Souza has given me, I became an active referee of Physical Review B and Physical Review Research.

\section{Tutorships}

As part of my Ph.D. degree, I acted as a tutor in the following courses ministered by Prof.\,Dr.\,Mariano\,de\,Souza:
\begin{itemize}
\item Physics II (First semester/2019 - volunteer);
\item Solid State Physics (Second semester/2020);
\item Electromagnetism (First semester/2021);
\item Electromagnetism (Second semester/2021 - volunteer).
\end{itemize}

\section{Participation in scientific events}

\begin{itemize}
\item Autumn Meeting of the Brazilian Physical Society - Ouro Preto, MG (2023).
\textbf{Poster presented:} Adiabatic magnetization in paramagnetic salts without applied magnetic field. \\
\textbf{Talk:} Materials design for eco-friendly refrigeration employing entropy arguments.
\item First Iberoamerican Meeting on Quantum Materials and Electronic Structures: Theory and Experiments - UFES, Vitória, ES (2023). \\
\textbf{Poster presented:} Elastocaloric-efect-induced adiabatic magnetization in paramagnetic salts due to the mutual interactions
\item II Condensed Matter Theory in the Metropolis - ICTP-SAIFR/IFT-UNESP, São Paulo, SP (2022). \\
\textbf{Poster presented:} Zero-field quantum criticality and the role played by the mutual interactions in paramagnets. \\
\textbf{Poster presented:} Giant caloric effects close to \emph{any} critical end point.
\item School on disordered elastic systems - ICTP-SAIFR/IFT-UNESP, São Paulo, SP (2022);
\item Autumn meeting of the Brazilian physical society - University of São Paulo (USP), São Paulo, SP (2022). \\
\textbf{Poster presented:} Giant caloric effects close to \emph{any} critical end point. \\
\textbf{Talk:} Elastocaloric-efect-induced adiabatic magnetization in paramagnetic salts due to the mutual interactions.
\item Autumn meeting of the Brazilian physical society - Online (2021). \\
\textbf{Talk:} Unveiling the physics of the mutual interactions in paramagnets.
\item International conference on strongly correlated electron systems - Online (2021).\\
\textbf{Poster presented:} Zero-field quantum criticality and the role played by the mutual interactions in paramagnets.
\item Autumn meeting of the Brazilian physical society - Online (2020).\\
\textbf{Talk:} Magnetic Gr\"uneisen parameter for model systems.
\item Autumn meeting of the Brazilian physical society - Aracaju, SE (2019). \\
\textbf{Talk:} Exploring the ionic dielectric constant contribution in the Mott-Hubbard phase of molecular metals.
\item Second Brazilian synchrotron school: fundamentals and applications (2018).
\end{itemize}

\section{Prizes and awards}

\begin{itemize}
\item Honorable mention of poster entitled ``Giant caloric effects close to \emph{any} critical end point'' presented at the Autumn meeting of the Brazilian physical society (2022);
\item Best poster award of poster entitled ``Zero-field quantum criticality and the role played by the mutual interactions in paramagnets'' presented at the international conference on strongly correlated electron systems (2022);
\item Best poster award of poster entitled ``Elastocaloric-effect-induces adiabatic magnetization in paramagnetic salts due to the mutual interactions'' presented at the First Iberoamerican Meeting on Quantum Materials and Electronic Structures: Theory and Experiments (2023).
\item Honorable mention in the section ``Dynamical Magnetization'' of poster entitled ``Adiabatic magnetization in paramagnetic salts without applied magnetic field'' presented at the Autumn Meeting of  the Brazilian Physical Society (2023).
\item Institute of Physics (IOP) best poster award (2023) - 1$^{\textmd{st}}$ place - of poster entitled ``Adiabatic magnetization in paramagnetic salts without applied magnetic field'' presented at the Autumn Meeting of  the Brazilian Physical Society (2023).

\end{itemize}

\section{Graduate student representative}

Based on the suggestion made by Prof. Dr. Mariano de Souza, I had the opportunity to act as the Physics graduate student representative in the Physics graduate program at Unesp, Rio Claro, SP, together with my colleague Ricardo Brandolt Júnior. During this very important period, we had the opportunity to participate and to bring the student's vision/opinion in the graduate board meetings. We participated in numerous activities, some of which were the elaboration of normative instructions, which are the rules associated with the Physics graduate course. We also spectated the approvals of students research projects, tutorships requests, and participated in many other academic discussions that contributed a lot to our scientific careers in a broader way. In addition, we spectated 2 masters and doctoral selection processes in the Physics graduate program, which were unique experiences to learn many academic and bureaucratic aspects. Hence, it is worth mentioning that, for the first time in the history of the Physics graduate program at Unesp Rio Claro, SP, there was an active graduate student representation.

\section{Published manuscripts as a co-author}

Apart from the topics investigated in the frame of this Thesis, I had the opportunity to participate as a co-author in the following articles:
\begin{itemize}
\item I.F. Mello, L. Squillante, G.O. Gomes, A.C. Seridonio, Mariano de Souza, Griffiths-like phase close to the Mott transition, Journal of Applied Physics \textbf{128}, 225102 (2020).
  \item I.F. Mello, L. Squillante, G.O. Gomes, A.C. Seridonio, Mariano de Souza, Epidemics, the Ising-model and percolation theory: A comprehensive review focused on Covid-19, Physica A: Statistical Mechanics and Its Applications \textbf{573}, 125963 (2021).
  \item L. Squillante, Mariano de Souza, Review of: ``High-pressure thermal conductivity and compressional velocity of NaCl in B1 and B2 phase'', Qeios ID: ZWB69H, \url{https://doi.org/10.32388/ZWB69H} (2021).
  \item Isys F. Mello, Lucas Squillante, Roberto E. Lagos-Monaco, Antonio C. Seridonio, Mariano de Souza, Gr\"uneisen parameter, the specific heat ratio, and phases coexistence region (submitted) (2023).
\end{itemize}










\end{document}